%% file: main.tex
\documentclass[12pt]{scrreprt}
\usepackage[utf8]{inputenc}
\usepackage[english]{babel}
\usepackage{amsmath}
\usepackage{amssymb}
\usepackage{float}
\usepackage{graphicx}
\usepackage[a4paper,left=3cm, right=3cm, top=3cm, bottom=3cm]{geometry}
\usepackage{subfigure}
\usepackage[colorlinks=true,linkcolor=blue,citecolor =blue]{hyperref}
\usepackage{multirow}
\usepackage{url}
\usepackage{subfigure}
\usepackage{makecell}
\usepackage{mathptmx}
\usepackage{dsfont}

\usepackage{tcolorbox}

\tcbset{
    mydefinition/.style={
        colback=white,    
        colframe=blue,  
        fonttitle=\bfseries, 
        boxrule=0.5mm,     
        coltitle=black,    
        enhanced,          
        attach boxed title to top left={xshift=2mm, yshift=-2mm}, 
        boxed title style={colback=blue, colframe=blue}
    }
}

\DeclareUnicodeCharacter{223C}{\textasciitilde}

\usepackage[headsepline]{scrlayer-scrpage}
\clearpairofpagestyles
\ohead{\pagemark}
\ihead{\headmark}
\automark{section}
\automark*{chapter}

\usepackage{xcolor}
\usepackage[labelfont={color=blue,bf},format=hang]{caption}
\usepackage[style=alphabetic,sorting=nyt,maxnames =20]{biblatex}
\usepackage{csquotes}

\begin{document}
\begin{titlepage}
\centering
\includegraphics[width=0.3\textwidth]{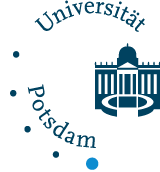}\hspace{2cm}
\includegraphics[width=0.3\textwidth]{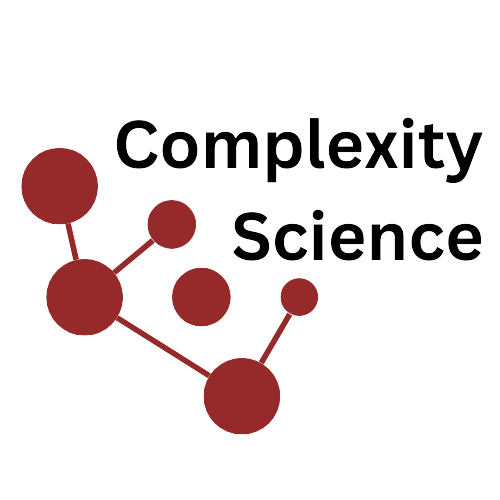}
\\
\vspace{1cm}
{\Huge Study of the \textit{Diffusion Map} Method in the Context of Social Science Data Sets \\}
\vspace{0.5cm}
{\large As an Example for Spectral Dimensionality Reduction Methods \\}

\vspace{2cm}
{\Large Masterarbeit \\}
\vspace{0.5cm}
{im Studiengang \\}
\vspace{0.5cm}
{\Large Physik \\}
{aus den Bereichen der Soziophysik und der Komplexitätswissenschaft\\}
\vspace{2cm}
{\Large Sönke Beier \\}
{Matrikelnummer: 790592 \\
soenbeier@uni-potsdam.de\\}
\vspace{2cm}
{Erstbegutachterin: Prof. Dr. Karoline Wiesner\\}
{Zweitbegutachter: Dr. Eckehard Olbrich\\}
{Wissenschaftliche Sparringpartnerin: Paula Pirker-Díaz\\}
\vfill
{Potsdam, 12.03.2025}
\end{titlepage}

\chapter*{Abstract}
\thispagestyle{empty}

The \textit{Diffusion Map} is a nonlinear dimensionality reduction technique used to analyze high-dimensional data, with recent applications extending to datasets from the social sciences. Previous research has given little attention to how the specific characteristics of these datasets might influence the results of the \textit{Diffusion Map} and what conditions must be met for the \textit{Diffusion Map} to yield meaningful and interpretable results. Moreover, there is a lack of clear, comprehensive explanations of the fundamental principles, which has led to misunderstandings in the literature.\\

This work first addresses the fundamental principles of the \textit{Diffusion Map} and compares them with other spectral methods. It investigates the impact of the \textit{Diffusion Map} parameters as well as the structure of the underlying data on the results. The V-Dem democracy dataset, British census data, and data on German urban and rural districts are then analyzed, considering their possible natural parameters. A focus is placed on the benefits of the \textit{Diffusion Map} in comparison to the established linear principal component analysis (PCA).\\

The analysis shows that the time parameter $t$ of the \textit{Diffusion Map} framework has no significant influence on the analysis. In contrast, discrete and redundant variables, as well as the scaling and normalization of the data, have a substantial impact. Unlike PCA, the \textit{Diffusion Map} eigenspectrum does not provide a clear indication of which components are important. Therefore, typical polynomial patterns related to one-dimensional datasets within the \textit{Diffusion Map} framework are explored.\\

The thesis presents insights suggesting that several underconsidered effects need further examination, and emphasizes the need for a framework to accurately analyze complex datasets using the \textit{Diffusion Map}.

\setcounter{secnumdepth}{4}
\setcounter{tocdepth}{4}
\addtocontents{toc}{\protect\thispagestyle{empty}}
\tableofcontents

\thispagestyle{empty}

\chapter{Introduction}
\setcounter{page}{1}
\input{Chapter/1Introduction}

\chapter{Theoretical background of the \textit{Diffusion Map}}
\label{1theory}
\input{Chapter/2Theory}

\chapter{Understanding principal effects by exploring classic examples}
\label{2classic_examples}
\input{Chapter/classic_examples}

\chapter{Results from the analysis of social science data sets}
\label{3results_social_datasets}

In this chapter, we will analyze three real-world datasets from the field of social sciences to evaluate the ability of the \textit{Diffusion Map} to effectively describe such data and how its findings align with existing social science research. We aim to uncover meaningful insights, identify potential natural parameters, and investigate whether the effects discussed in Section \ref{classical_examples} are reflected in the results.

In the context of these data sets, we will investigate how individual parameters influence the outcome and whether typical polynomial structures emerge, which could indicate the intrinsic dimensionality of the data. Furthermore, we will explore the impact of redundant variables on the mapping, evaluate whether the sorting capability of the \textit{Diffusion Map} is consistently maintained or if limitations arise, and will see if cluster formations occur. A crucial aspect of this analysis is the comparison with non-linear PCA to determine whether the observed effects are genuinely specific to the \textit{Diffusion Map} or if they can already be explained by simpler methods.

\include{Chapter/41_vdem}

\include{Chapter/43_census_britain}
\include{Chapter/44_germany}

\chapter{Discussion and outlook}
\input{Chapter/5Outlook}

\chapter{Appendix}
\label{counterexample_normalisation}
\clearpairofpagestyles
\ohead{\pagemark}
\input{Chapter/6Appendix}

\cleardoublepage
\addcontentsline{toc}{section}{Bibliography}
\printbibliography

\chapter*{Selbstständigkeitserklärung}
\addcontentsline{toc}{section}{Selbstständigkeitserklärung}
\thispagestyle{empty}
Hiermit versichere ich, dass ich die vorliegende Arbeit ohne Hilfe Dritter und ohne Zuhilfenahme anderer als der angegebenen Quellen und Hilfsmittel angefertigt habe. Die den benutzten Quellen wörtlich oder inhaltlich entnommenen Stellen sind als solche kenntlich gemacht. Für Übersetzungen wurde auf KI-Anwendungen wie DeepL und ChatGPT zurückgegriffen.\\

Die „Richtlinie zur Sicherung guter wissenschaftlicher Praxis für Studierende an der Universität Potsdam (Plagiatsrichtlinie), habe ich zur Kenntnis genommen. Während des 'Introductory projects' zur Masterarbeit habe ich mich mit dem Thema \textit{Diffusion Map} beschäftigt und kurze Abschnitte zur Theorie in Kapitel \ref{probabilistic_interpretation}, zur Arbeit mit klassischen Beispielen aus Kapitel \ref{classical_examples}, zur Erstellung der \textit{Diffusion Map} in Kapitel \ref{vdem_section} und zu Kapitel \ref{counterexample_normalisation} in diese Arbeit übernommen. 

\vspace{1.5cm}

Potsdam, den 12.03.2025\\

\vspace{1.5cm}

Sönke Beier\\

\chapter*{Zusammenfassung}
\addcontentsline{toc}{section}{Zusammenfassung}
\thispagestyle{empty}

In der heutigen Zeit werden zunehmend hochdimensionale Datensätze erzeugt und analysiert. Die \textit{Diffusion Map} ist eine nicht-lineare Dimensionsreduktionsmethode, die dazu dient, solche Datensätze greifbar zu machen. In den letzten Jahren fand diese Methode auch Anwendung bei der Analyse von Datensätzen aus den Sozialwissenschaften. Bisherige Arbeiten haben jedoch wenig untersucht, welchen Einfluss die besonderen Eigenschaften dieser Datensätze auf die Ergebnisse der \textit{Diffusion Map} haben können und welche Voraussetzungen erfüllt sein müssen, damit die \textit{Diffusion Map} sinnvolle und interpretierbare Ergebnisse liefert. Zudem mangelt es an klaren, zusammenfassenden Erklärungen der Grundprinzipien, was zu Missverständnissen in der Literatur führt.\\

Diese Arbeit setzt sich zunächst mit den Grundprinzipien der \textit{Diffusion Map} auseinander und vergleicht diese mit anderen spektralen Methoden. Es wird untersucht, welche Einflüsse die Parameter der \textit{Diffusion Map} sowie die Struktur der zugrunde liegenden Daten auf die Ergebnisse haben. Anschließend werden unter Berücksichtigung dieses Wissens der V-Dem Demokratiedatensatz, britische Census-Daten und Daten über deutsche Städte und Landkreise analysiert. Ein besonderer Fokus liegt auf dem Nutzen der \textit{Diffusion Map} im Vergleich zur etablierten linearen Principal Component Analysis (PCA).\\

Die Analyse zeigt, dass der Zeitparameter $t$ keinen nennenswerten Einfluss auf das Ergebnis hat. Hingegen haben diskrete und redundante Variablen sowie Skalierungen und Normalisierungen der Daten einen erheblichen Einfluss. Im Gegensatz zur PCA bieten die \textit{Diffusion Map} Eigenwerte keine klare Auskunft darüber, welche Dimensionen wichtig sind. Es werden die möglichen natürlichen Parameter der drei untersuchten Datensätze untersucht, typische Formen der \textit{Diffusion Map} identifiziert und gezeigt, dass auch die PCA die Daten sinnvoll anordnen kann.\\

Die Arbeit kommt zu dem Schluss, dass verschiedene Effekte der \textit{Diffusion Map} noch nicht ausreichend erforscht sind. Es wird ein Framework angeregt, das es ermöglicht, Datensätze mit der \textit{Diffusion Map} korrekt zu analysieren und Interpretationsfehler zu vermeiden. Diese Arbeit liefert erste Impulse für die Entwicklung eines solchen Ansatzes.

\chapter*{Danksagung}
\addcontentsline{toc}{section}{Danksagung}
\thispagestyle{empty}

Danke an Lisa, Karoline, Paula, Kolja, Lea, Julian, Sylke, Stefan, Sina, Hans, Luca, Alexander, Lisa, Rick, Flo, Elli, Lilly.\\



\end{document}

%% file: Chapter/1Introduction.tex
In today's highly connected and digital world, the amount of complex and high-dimensional data is continuously increasing. Dimensionality reduction techniques are a key approach for effectively analyzing and utilizing data, as they provide interpretable low-dimensional representations and enable visualization. In this thesis, we focus on one such technique: the \textit{Diffusion Map} method. This spectral, nonlinear dimensionality reduction method has been widely used since its introduction at the beginning of the 21st century \cite{lafon2004, Nadler2005, Coifman2006}. 

In addition to simpler datasets, such as initial tests on image data \cite{lafon2004} and basic nonlinear datasets like the Swiss Roll \cite{Nadler2007}, this method has been applied to a wide range of complex scientific analyses. These include sorting bacterial genera based on genomic data \cite{Fahimipour2020}, analyzing gene expression data for cancer detection \cite{Xu2010}, studying molecular folding pathways and chain dynamics \cite{Ferguson2010, Kim2015}, investigating crash data analysis \cite{Schwartz2014}, addressing the \textit{wireless localization matching problem} by analyzing Wi-Fi signal strength data \cite{Ghafourian2020}, and distinguishing between earthquakes and explosions for nuclear test and seismic event monitoring \cite{Rabin2016}. Furthermore, social science datasets have also been analyzed using \textit{Diffusion Maps}, including census data \cite{Barter2019, Xiu2023}, mobility and movement data \cite{Levin2021, Xiu2024, Zeng2024}, and the voting pattern of the General Assembly of the UN \cite{Le2013}.\\

A key observation is that the fundamental principles and mechanics of the \textit{Diffusion Map} method are often not explicitly stated or are described inconsistently. As a result, this thesis presents an overview of the theoretical foundations and historical development of the method (in section \ref{1theory}).

Furthermore, there is limited discussion regarding how the characteristics of different datasets influence the results of the \textit{Diffusion Map} method. This issue is particularly relevant for social science datasets, where data collection and preprocessing play a crucial role. Key questions include how normalization and scaling of variables affect results, whether the \textit{Diffusion Map} method performs differently for discrete versus continuous variables, and the impact of redundant variables in the dataset. Additionally, fundamental aspects of \textit{Diffusion Map} results remain underexplored: How many dimensions should be considered to accurately represent the data and can an accurate statement be made at all? Are the leading dimensions always the most significant, and can there be complex relationships between them? What conditions must be met for the \textit{Diffusion Map} to correctly structure the data? Furthermore, the effect of the \textit{Diffusion Map} parameters and the right selection is a crucial aspect that directly influences the quality and interpretability of results, but is rarely discussed.\\

To address these questions and to also resolve misunderstandings in this regard, we will first examine classical examples such as the Swiss Roll dataset in section \ref{2classic_examples}. Followed by an analysis of three complex social science datasets (section \ref{3results_social_datasets}): the V-Dem dataset, which describes democracies (section \ref{vdem_section}), a census dataset from Bristol (section \ref{census_britain_section}), and a dataset characterizing various districts in Germany (section \ref{regionaldata_germany_section}). From these case studies, we aim to assess what insights can be gained from \textit{Diffusion Map} embeddings and how well this method captures the structures of these social science data sets.\\

Since this is a method-based work that also questions and analyzes fundamental aspects of \textit{Diffusion Maps}, we will omit a strict separation of theory, results, and discussion in favor of maintaining a clear argumentation flow.

%% file: Chapter/2Theory.tex
In this chapter, we will explain the theoretical background of the \textit{Diffusion Map}. To this end, we will first present the principal ideas behind the \textit{Diffusion Map}, followed by an overview of the algorithm to obtain a \textit{Diffusion Map} (section \ref{idea_algorithm}). Then, we will look in detail at the probabilistic interpretation of the \textit{Diffusion Map}, which gives the method its name (section \ref{probabilistic_interpretation}). Finally, we will look at which (historical) related methods exist and what distinguishes them from the \textit{Diffusion Map} (section \ref{related_methods_and_history}). In particular, we will also introduce methods that are equivalent to the \textit{Diffusion Map} and make the connection to the widely used Principal Component Analysis (PCA) (section \ref{equivalence_pca}).

\section{Main ideas and algorithm}

\label{idea_algorithm}
\subsection{Aim of the algorithm: manifold learning}
\begin{figure}[h]
    \centering
    \includegraphics[width=0.5\linewidth]{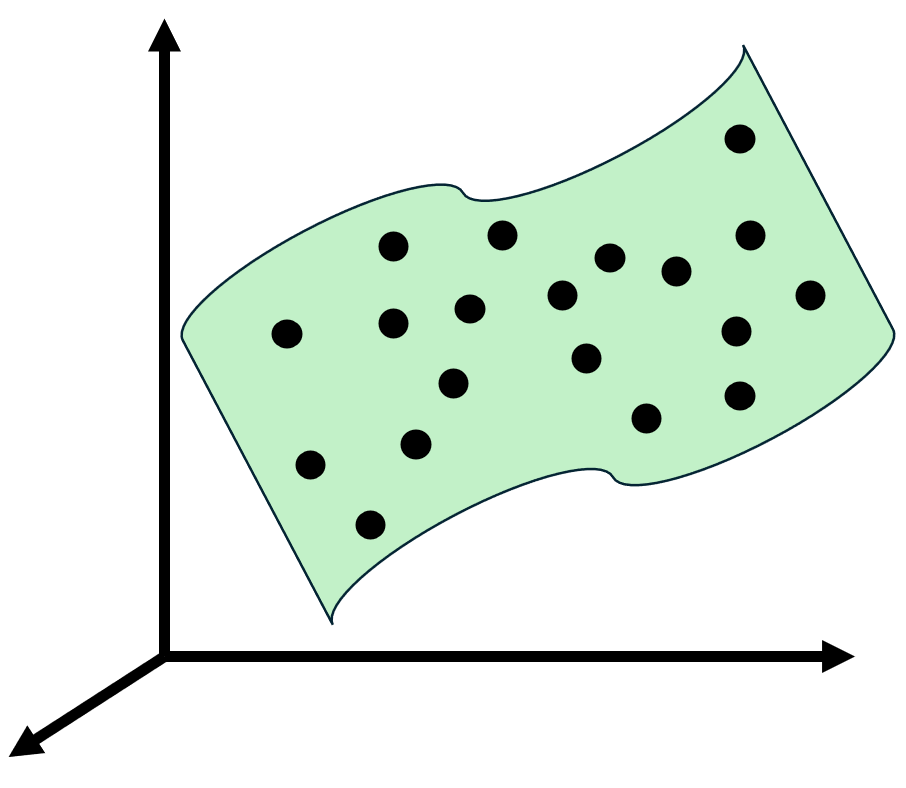}
    \caption{\textbf{Illustrating the manifold hypothesis:} Three-dimensional data points do not fill the entire three-dimensional space, but lie on a two-dimensional manifold.}
    \label{fig:illustrating_manifold_hypothesis}
\end{figure}

\begin{figure}[h]
    \centering
    \includegraphics[width=0.9\linewidth]{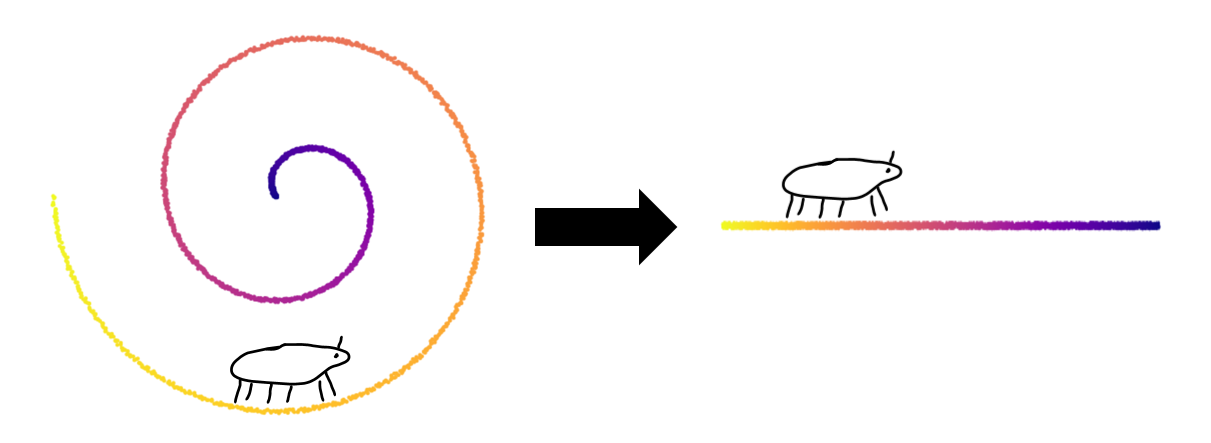}
    \caption{\textbf{Illustrating the idea of nonlinear dimensional reduction:} On the left the data points lie on a one dimensional manifold (spiral) in a 2 dimensional feature space. The idea of nonlinear dimensional reduction is to find the underlying intrinsic dimension. The procedure can be compared to a small insect located at one data point, that only recognizes its immediate environment, but not the global appearance of the data in a higher-dimensional sense. If it moves along the manifold it can put together the local information to create the low-dimensional representation (right). The color can be seen as the natural parameter of the manifold. The illustration is inspired by \cite{Ghojogh2023}.}
    \label{fig:illustrating_insect}
\end{figure}

As a start, we consider a dataset $X$ that consists of $n$ data points $X = \{x_1,x_2,...,x_n\}$. These data points lie in a $D$-dimensional feature space $\mathbb{R}^D$, where every dimension represents a feature or variable of the dataset. The aim of dimensionality reduction methods like the \textit{Diffusion Map} algorithm is to represent the data points of the high-dimensional feature space in the best possible way in fewer dimensions, preserving as much information as possible. Typically, this transformation is performed into two or three dimensions, enabling effective visualization of the data. Fortunately, according to the manifold hypothesis, natural data usually lie on a low-dimensional manifold with dimension (illustrated in figure \ref{fig:illustrating_manifold_hypothesis}). So it has an intrinsic dimension $k<D$. A good representation of the dataset would be to describe the data along this manifold. The data points could therefore be represented with only $k$ parameters.  We will call these meaningful parameters the natural parameters of the dataset. This (non-linear) dimensionality reduction with the aim of finding the low-dimensional structure of the data is also called manifold learning. An illustration of the procedure of the dimension-reducing methods can be seen in figure \ref{fig:illustrating_insect}\\

Now we will take a step-by-step look at how the algorithm achieves this aim. Unlike in the usual literature, we will first ignore the probabilistic interpretation\cite{lafon2004,Nadler2005} of the \textit{Diffusion Map} and we come back to it later in section \ref{probabilistic_interpretation}. The effects of dimensional reduction can also be achieved without this probabilistic explanation and the main ideas already existed before this interpretation (more on this in section \ref{related_methods_and_history}).


\subsection{Constructing neighborhood graph}
Most dimensional reduction methods start by finding the neighborhood information of the data \cite{Meila2024}. This process is best understood as constructing a neighborhood graph, where the nodes represent the individual data points and the edge weights represent their similarities. As a result, the graph encodes the local geometric and topological properties of the data \cite{Meila2024}.

\subsection{Defining similarity measure}
For that reason we need the information on how similar the different data points are to each other. Therefore, we first need a fitting similarity measure for our data. Similarity can be represented using a suitable distance measure: a small distance means high similarity, and large distances correspond to low similarity. Although other distance measures are thinkable, the Euclidean distance $||x_i-x_j||$ between the points is typically calculated for this purpose, resulting in the distance matrix $\mathcal{D}$ with the following entries:
\begin{equation}
    \mathcal{D}_{ij} = ||x_i-x_j||
\end{equation}
 As we only provide the similarity of the data points in the algorithm, and not their locations, the result is also independent of rotations, translations or reflecting of the original data \cite{Nadler2007}. On the other hand, changing the range of certain features, for instance, through normalization or standardization, will result in a change in the distance and potentially impact the resulting \textit{Diffusion Map}. For some datasets, it may be useful to normalize or standardize the data beforehand (see therefore section \ref{normalization}).  This can be particularly useful for social science datasets, since the individual dimensions of the data points are often different types of variables(different range, discrete or continuous variables), which can lead to one variable having too much influence (such effects can be seen in section \ref{vdem_suffrage}).

\subsection{Adding locality}
\label{adding_locality}
In order to ensure the accurate recognition of manifolds, it is essential to adjust the similarity measure such that only data points within a local neighborhood are connected. This adjustment is necessary to ensure that the method can effectively follow the respective manifold. To illustrate this concept, one may consider an insect that considers where it can place the next step (by only recognizing the local environment) and follows the respective manifold, combining what it has seen into a global representation that reflects the intrinsic dimension, as already shown in figure \ref{fig:illustrating_insect}. The local information is aggregated and later reassembled into a global representation \cite{lafon2004,ham2004} which happens as an effect of the global operation of spectral decomposition \cite{Saul2003,Wu2018}. One also speaks here of the global structure being preserved by local fitting, and this is something that many non-linear methods have in common \cite{Ghojogh2023}.\\

But why is that important? If two points are close to each other, the Euclidean distance gives us useful information, the global ‘long-range’ similarities given by the Euclidean distance are not important and trustworthy \cite{lafon2004,Luxburg2007,Porte2008} and should not be preserved. This becomes clearer in the following examples:

\begin{itemize}
    \item \cite{lafon2004, Porte2008} describe a case where the goal is to define an appropriate similarity measure for images of numbers that are slightly rotated by a given angle. The feature space has a dimension of $m \times n$, where $n$ is the width and $m$ the height of the image. While the human eye can easily recognize the same number across different rotations, the Euclidean distance struggles with this task. 
    The problem with using the Euclidean distance is that there are only two cases if we only look at one specific number: 
    Either the images have a small angle of rotation and the distance is correspondingly very small and there is the useful information that the two images are very similar. 
    Or the angle of rotation is so large (e.g. if the specific number is upside down) that the distance is very large and there is no more useful information, because it makes no difference whether it would be the inverted number or simply a different number. 
    Therefore the approach is: You only trust the small distances between very similar data points and hope that with enough data points the local information will agglomerate such that it will infer the global structure of the dataset \cite{lafon2004}. We will discuss the sorting ability of images again in section \ref{sorting_ability}. 
    \item Another example is the measurement of distances between different molecular configurations, where each atom's position in the molecule represents a variable/dimension (e.g., as shown in \cite{Meila2024,Kim2015}). For very similar configurations, the Euclidean distance provides useful information, indicating their similarity. However, for larger distances, it becomes less meaningful since many configurations are physically impossible, and transitioning from one configuration to another may require passing through several distinct intermediate states.
\end{itemize}

To summarize: we now want to create a local similarity graph, in order to group similar data points and therefore retain the local topology of the data in the dimensionality reduction process. This graph can be represented by a similarity matrix. \\

The locality is the crucial property of the method to be able to recognize non-linear global structures. This is because the crucial difference to linear spectral methods such as PCA is that these do not use local, but global distances and similarity matrices. We will discuss this in more detail in section \ref{equivalence_pca}.\\

\begin{figure}[h]
    \centering
    \includegraphics[width=0.5\linewidth]{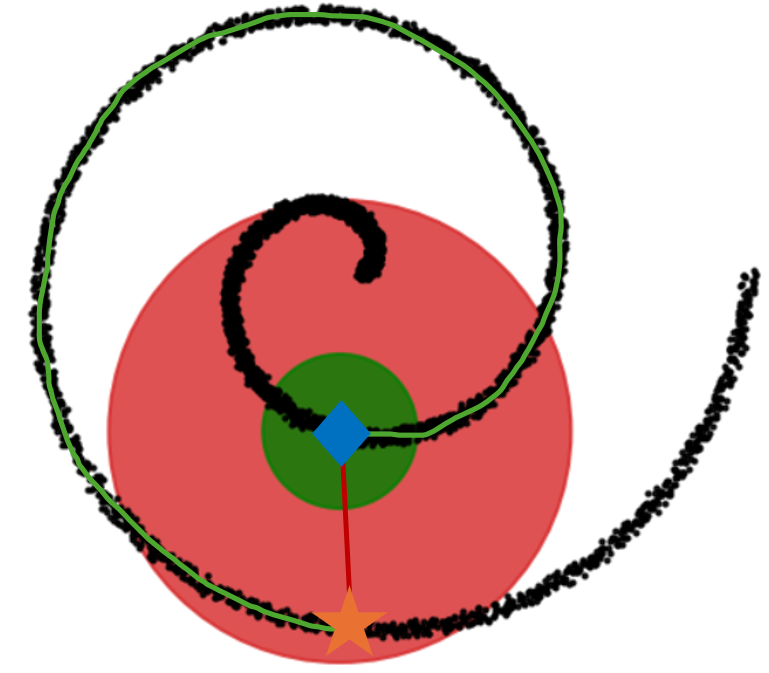}
    \caption{\textbf{Illustrating the importance of locality:} The neighborhood has to be small enough to detect the true underlying manifold. Here, data points are shown on a spiral. If we want to know the distance along the manifold, for example from the blue diamond to the orange star, the neighborhood must not be so large that the points are connected. If the neighborhood is too large (red), then the algorithm finds the wrong shortcut across the gap. If the neighborhood is chosen correctly (green), We can move along the manifold. Inspired by \cite{Roozemond2021}.}
    \label{fig:illustrating_locality}
\end{figure}

But how do we practically create the local similarity graph out of the distance matrix $\mathcal{D}$? There are various possibilities for this, which are used in dimensionality reduction methods\cite{Luxburg2007,Meila2024}:

\begin{itemize}
    \item \textbf{Radius-neighbor graph}: Less commonly used is the creation of a neighborhood graph in which only points with pairwise distances smaller than a radius $r$ are connected.
    \item \textbf{Fully connected graph with a kernel}: A widely used option is to use a Gaussian neighborhood, which is determined by the width parameter $\epsilon$. The kernel matrix, which stores the local similarities, is given by: 
    \begin{equation}
        \label{kernel_definition}
        K_{ij} = \exp{(-\mathcal{D}_{ij}^2/\epsilon)}
    \end{equation}
    Here, it is important not to choose a neighborhood that is sufficient small, such that there are no unintended shortcuts (see figure \ref{fig:illustrating_locality}). To ensure that the graphic along the manifold is not disconnected, it must not be too small either. These effects are also examined in more detail in section \ref{finding_parameter}. It should also be noted that with this defined graph, the associated matrix is not a sparse matrix, because the graph remains fully connected.
    It is also generally possible to use other symmetric and positivity preserving kernel functions \cite{lafon2004,Coifman2006}. 
    \item \textbf{Nearest neighbor graph}: 
    An alternative approach is to construct a neighborhood graph where each data point is only connected to its $N$ nearest neighbors. The similarity entries of these connected data points can either be obtained from the distance matrix, a kernel matrix, or set to 1 if one does not want to use weights. 
    In practice, the resulting matrix is often symmetrized to get an undirected neighborhood graph \cite{Meila2024}. This is done by keeping the similarity entries between points $i$ and $j$ if either point $i$ or $j$ is among the nearest neighbors of the other. In this case, each point has at least $N$ connected points.    
\end{itemize}

In the earliest works on \textit{Diffusion Maps}, only the Gaussian kernel was used \cite{lafon2004, Nadler2005, Coifman2006}. Note that in some publications different pre-factors are used and the width parameter $\epsilon$ has to be converted to compare the results (for example in \cite{pydiffmap} or \cite{Lee2018} the width parameter is multiplied by $4$). 
To enforce locality, later studies introduced an additional constraint and considered only nearest neighbors in combination with the Gaussian kernel \cite{Izenman2012,Barter2019,Fahimipour2020,Xiu2023}. This approach is also implemented in the Python package \cite{pydiffmap}. The primary motivation for this is that using the nearest neighbors results in a sparse matrix, which is computationally more efficient to handle \cite{DeSilva2002, Luxburg2007, VanDerMaaten2009a, Barter2019, Meila2024}. 
Moreover, the $N$-nearest neighbors approach is recommended as a starting point because it is easier to choose a suitable parameter \cite{Luxburg2007}.\\

In this work, we use the kernel matrix $K$ as defined in formula \ref{kernel_definition}, and for certain datasets, we optionally set the entries $K_{ij}$ to zero, if point $i$ is not among the $N$ nearest neighbors of $j$ or vice versa.
Finding the appropriate neighborhood parameters $\epsilon$ and $N$ is generally non-trivial, as discussed in more detail in section \ref{finding_parameter}, because the optimal value of $\epsilon$ depends on the dimension, curvature, volume, and number of the manifold \cite{Singer2006}.

\subsection{Different normalizations}
\label{different_normalizations}

We are following \cite{Nadler2005} and calculating the matrix
\begin{equation}
    M = D^{-1}K
\end{equation}
which is a Markovian transition matrix, which comes from the probabilistic interpretation from section \ref{probabilistic_interpretation}. To improve the performance for the spectral decomposition \cite{Ng2001} we symmetrize it:
\begin{equation}
    M_s = D^{-1/2}KD^{-1/2}
\end{equation}
Here $D$ is the diagonal normalization matrix with the diagonal entries $D_{ii}=\sum_iK_{ij}$, while all other entries are zero.
In the literature, there are different definitions regarding which of the mentioned matrices should be considered the normalized Laplacian matrix: \cite{Nadler2005} says $K$ is the normalized Laplacian. According to \cite{Ghojogh2023} it is $D^{-1/2}KD^{-1/2}$. \cite{Izenman2012} defines it as $D^{-1/2}(D-K)D^{-1/2}$. \cite{Luxburg2007} and \cite{Meila2024} say it is $\mathds{1} - D^{-1/2}KD^{-1/2}$, which would fit to \cite{Weiss1999}, who names $D^{-1/2}KD^{-1/2}$ as the normalized adjacency matrix. Due to these different definitions, we will not attempt to name the matrix, but note that there may be slight differences in the result.

\subsection{Spectral decomposition}
\label{spectral_decomposition}

The local information stored in the neighborhood graph, and therefore in the matrix $M$ or the symmetrized version $M_s$, is now aggregated by the spectral decomposition. This is a global operation that couples all local information to discover the intrinsic global structure of the manifold \cite{Saul2003}.\\

$M$ and $M_s$ have the same eigenvalues $\{\lambda_j\}_{j=0}^{n-1}$. The eigenvectors of the symmetric matrix $M_s$ with the length $1$ are denoted as $\{v_j\}$.\\ 

In contrast to \cite{Nadler2005}, we motivate in \ref{counterexample_normalisation} that the left eigenvectors $\{\phi_j\}$ and the right eigenvectors $\{\psi_j$\} can be defined as follows:
\begin{align}
    \phi_j &= \frac{1}{\sqrt{tr(D)}} v_jD^{1/2}\\
    \psi_j &= \sqrt{tr(D)} v_jD^{-1/2}
\end{align}
For a fully connected graph, the eigenvalues are bound from above by $1 = \lambda_0 > \lambda_1 \geq \lambda_1 \geq \lambda_2 \geq...$ \cite{Coifman2006}. Now, we have up to $n$ eigenvalues and eigenvectors, which can be used to embed the $n$ data points.\\

Note that the eigenvectors are independent of multiplications by 
$-1$ because eigenvectors multiplied by a scalar are still eigenvectors. However, in this algorithm, the length of the eigenvectors is fixed. This leads to the effect that when performing the \textit{Diffusion Map} algorithms multiple times, it can occur mirrored results, depending on the run.
\\

In the first papers on \textit{Diffusion Maps}, the additional parameter $t$ is often introduced, with which the matrix $M^t$ can be calculated. When we are doing the spectral decomposition of this matrix, the eigenvectors do not change, but the eigenvalues are changed by the power of $t$ \cite{proof_eigenvalues_matrix_powers}.
The effect will be discussed in section \ref{probabilistic_interpretation} and \ref{time_parameter}.

\subsection{Dimensional reduction}

To achieve a dimension reduction, we only use the first $k$ most important eigenvectors. These are the eigenvectors with the largest eigenvalues. For this, we will ignore the eigenvector with $\lambda_0 = 1$, since $\psi_0(x) = 1$ is constant. 

We call the new embedding the \textit{Diffusion Map} and it is given by: 
\begin{equation}
    \label{equation:diffusion_map}
    \Psi(x) = (\lambda_1^t\psi_1(x), \lambda_2^t\psi_2(x), ..., \lambda_k^t\psi_k(x))    
\end{equation}

The space defined by the \textit{Diffusion Map} is now also referred to as diffusion space and the components as diffusion components.

\subsection{Summary of the algorithm}
\label{algorithm_summary}
The algorithm used to generate the results of this work is summarized in this section.
\begin{description}
    \item [optional]{Normalize or standardize the data beforehand} 
    \item [Step 1]{\textbf{Calculate distance matrix $\mathcal{D}$}}
        \begin{itemize}
            \item Using the euclidean distance $\mathcal{D}_{ij} = ||x_i - x_j||$
            \item Other distance measures are possible
        \end{itemize}
    \item [Step 2]{\textbf{Calculate a kernel matrix $K$}}
        \begin{itemize}
            \item Choose width of the neighborhood $\epsilon$.
            \item The matrix is given by $K_{ij} = exp(-\mathcal{D}_{ij}^2/\epsilon)$. Other kernel functions are possible.
            \item Optionally choose a number of considered nearest neighbors $N$. Entries $K_{ij}$ are kept, if point $i$ is among the $N$ nearest neighbors of $j$ and vice versa. All other entries are set to zero.
        \end{itemize}
    \item [Step 3]{\textbf{Create a symmetrized stochastic matrix $M_s$}}
        \begin{itemize}
            \item By normalizing each row of $K$.
            \item Use for example a diagonal normalization matrix $D_{ii} = \sum_j K_{ij}$. Then the stochastic matrix is $M = D^{-1}K$. 
            \item Symmetrize $M_s = D^{1/2}MD^{-1/2} = D^{-1/2}KD^{-1/2}$
        \end{itemize}
    \item [Step 4]{\textbf{Calculate eigenvalues $\lambda_k$ and eigenvectors $\psi_k$}}
        \begin{itemize}
            \item To keep the \textit{Diffusion Map} property, the eigenvectors must be normalized as follows: $\psi_i = \sqrt{tr(N)} v_i N^{-1/2}$, when $v_i$ are the eigenvectors with length 1. 
        \end{itemize}
    \item [Step 5]{\textbf{Build \textit{Diffusion Map} $\Psi(x)$}}
        \begin{itemize}
            \item The \textit{Diffusion Map} is defined as \\$\Psi(x) = (\lambda_1^t\psi_1(x), \lambda_2^t\psi_2(x), ..., \lambda_k^t\psi_k(x))$. 
        \end{itemize}
\end{description}

The Python code implementation, which was used for this master thesis, can be found on \url{https://github.com/SoenBeier/diffusion_map}. \\

Aside from this code, there are other implementations of the method, as \cite{pydiffmap} or \cite{mdanalysis}.

\subsection{Local justification: solution of an optimization problem}
\label{justification_optimization}

In this section, we want to show that \textit{Diffusion Map} embedding is a reasonable embedding in a local sense: 
when developing the Laplacian eigenmap, \cite{Belkin2001} take into consideration that the local information of the data should be optimally preserved (similar to other spectral methods like Locally Linear Embedding \cite{roweis2000}). Because the Laplacian eigenmap \cite{Belkin2001} differs from the \textit{Diffusion Map} primarily in the use of different Laplacian matrices \cite{Meila2024}, this justification also applies to the \textit{Diffusion Map} \cite{Nadler2005,Bah2008}.

The idea is that if points $x_i$ and $x_j$ that are close to each other in the feature space remain close to each other in the new embedding. For this purpose, \cite{Belkin2001} has formulated the following minimization problem:
\begin{equation}
    minimize~ \sum_{ij}(\psi_i-\psi_j)^2K_{ij}
\end{equation}
where $\psi_i$ represents the new embedding of the data points. $K$ is the kernel matrix, which indicates the similarity of the points to each other. If $x_i$ and $x_j$ are close, $K_{ij}$ is large. Therefore, minimizing the expression results in close $\psi_i$ and $\psi_j$. If $x_i$ and $x_j$ are far away, $K_{ij}$ is small (or zero when using only nearest neighbors) and the overall expression is already small \cite{Ghojogh2023}. This also shows that long Euclidean distances are not as important when examining the \textit{Diffusion Map}. 
Now we look at how to find the optimal embedding given the assumption. Reformulating the expression yields the following result (see also \cite{Ghojogh2023} Proposition 9.2):
\begin{align*}
    \sum_{ij}(\psi_i-\psi_j)^2K_{ij} &= \sum_{ij}(\psi_i^2+\psi_j^2-2\psi_i\psi_j)^2K_{ij}\\
    &= \sum_i \psi_i^2 D_{ii} + \sum_j \psi_j^2 D_{jj} - 2\sum_{i,j}\psi_i\psi_j K_{ij} \\
    &= 2(\sum_i \psi_i^2 D_{i} - \sum_{ij} \psi_i \psi_j K_{ij} ) \\
    &= 2(\psi^TD\psi - \psi^TK\psi) = 2\psi^T(D-K)\psi\\
    &= 2\psi^TL^{un}\psi
\end{align*}
with the diagonal normalization matrix $D_{ii} = \sum_j W_{ij}$ and the unnormalized Laplacian $L^{un} = D-K$.

Acording to \cite{Ghojogh2023}, the minimization problem $minimize ~~\psi^TL^{un}\psi$ under the constraint $\psi^T\psi = 1$ can be solved with a Lagrangian:
\begin{equation}
    \mathcal{L} = \psi^TL^{un}\psi-\lambda (\psi^T\psi-1)
\end{equation}
Considering the derivative, this results into the eigenvalue problem for the Laplacian matrix:
\begin{align}
    \frac{\mathcal{\partial L}}{\partial \psi} &= 2L^{un}\psi-2\lambda\psi \stackrel{\text{!}}{=} 0 \\
    &\Rightarrow L^{un}\psi = \lambda \psi
\end{align}

Thus, we have demonstrated that the embedding given by the eigenvectors, i.e., the \textit{Diffusion Map}, preserves the local neighborhood.
It depends on the normalization of the Laplacian if the eigenvectors should be sorted from the corresponding largest to the smallest eigenvalues (for $L^{nor}$) or visa versa (for $L^{un}$)\cite{Weiss1999}. The eigenvectors with eigenvalues $1$ respectively $0$ are ignored to build the embedding. As described in section \ref{different_normalizations}, the \textit{Diffusion Map} uses the normalized Laplacian $L^{nor}$ and the Laplacian eigenmap the unnormalized Laplacian $L^{un}$. According to \cite{Meila2024}, $L^{norm}$ and $L^{un}$ should produce the same embedding.\\

A second difference between the \textit{Diffusion Map} and the Laplacian eigenmap is that for the Laplacian eigenmap the individual components are not multiplied by the corresponding eigenvalues. However, this does not change the statement of the minimization problem, since the eigenvalues only have a scaling effect. In \cite{Ghojogh2023} one can also find the translation of various spectral methods into optimization problems.

\section{The probabilistic interpretation}
\label{probabilistic_interpretation}

\subsection{The algorithm from a probabilistic point of view}

The idea behind the publications on the \textit{Diffusion Map} (\cite{lafon2004,Nadler2005,Coifman2005}) was to find a probabilistic interpretation of nonlinear spectral dimensional reduction procedures and to establish a connection to diffusion processes, although the method differs only in minor aspects from other methods (see \ref{related_methods_and_history}). \\

The paper by \cite{Nadler2005} states that the \textit{Diffusion Map} is "a diffusion-based probabilistic interpretation of spectral clustering and dimensionality reduction algorithms that use the eigenvectors of the normalized graph Laplacian." Specifically, this normalized Laplacian represents the transition matrix of a discrete-time Markov process. Now we will have a closer look at how the method is motivated by the probabilistic perspective: \\

The idea is to create a random walk on the considered data points \cite{Porte2008}. 
The probability to jump from one point to another is defined by their similarity: Therefore it should be more likely to jump to closer data points, which are more similar. Since only the probability of jumps along the manifold should receive a high probability, locality is established by using a kernel function. So we get the similarity matrix $K_{ij}$ as defined in formula \ref{kernel_definition}. 

The entries of the kernel matrix can now be converted into probabilities. For this purpose, the diagonal normalization matrix $D_{ii} = \sum_j K_{ij}$ can be used. This results in the Markov matrix:
\begin{equation}
    M = D^{-1}K
\end{equation}
And $M_{ij}$ is the probability of jumping from the data point $x_i$ to $x_j$. Because the Markov process is reversible, we can build an associated symmetric matrix \cite{Nadler2005,Coifman2006}
\begin{equation}    
    M_s = D^{1/2}MD^{-1/2} = D^{-1/2}KD^{-1/2}
\end{equation}

We now let the Markov process evolve forward. The jump probabilities after $t$ time steps are given by $M^t$, where the entry $(M^t)_{ij}$ represents the probability of transitioning from state $i$ to state $j$ in $t$ time steps.

As described in section \ref{spectral_decomposition} we can calculate the corresponding eigenvalues and eigenvectors and can define the \textit{Diffusion Map} at time $t$, which is given by the formula \ref{equation:diffusion_map}. We will examine the effect of $t$ in section \ref{time_parameter}.\\

In \cite{Nadler2005, Nadler2007}, the asymptotics of the \textit{Diffusion Map} were analyzed. It was shown that for increasingly smaller neighborhood scales $\epsilon \rightarrow 0$ and an increasing number of data points on the manifold $n \rightarrow \infty$, the operator $M$, which describes a random walk in discrete space with discrete time, converges to the infinitesimal generator of the diffusion process $\mathbf{H}$. This operator is described by the Fokker-Planck equation:
\begin{equation}
    \label{FokkerPlanck}
    \mathbf{H}\psi = \Delta \psi - 2\nabla \psi \cdot \nabla U
\end{equation}
The Fokker-Planck equation describes the time evolution of the probability density function $\psi$ for a diffusion process with drift under a potential $U$, which vanishes if the data is uniformly sampled from the manifold \cite{Nadler2007}.

It is important that the eigenvectors of the matrix $M$ can be interpreted as approximations of the eigenfunctions of $\mathbf{H}$, if there are enough data points on the manifold \cite{Nadler2007}. This knowledge will also be essential for deriving simple typical shapes of the \textit{Diffusion Map} in section \ref{typical_shapes} using the eigenfunctions of the Fokker-Planck operator.



\subsection{Global justification: the diffusion distance}
In \cite{lafon2004, Nadler2005} the \textit{Diffusion Map} is given a further meaning in the form of a conserved global quantity. It is shown that the Euclidean distance in diffusion space is equal to the diffusion distance in Euclidean space. This is also called the \textit{Diffusion Map} property and gives the method a global justification. The diffusion distance is defined as:

\begin{equation}
\label{equation:diffusion_distance}
D_t(x_i,x_j)^2 = \sum_{x_u\in X}|p_t(x_i,x_u)-p_t(x_j,x_u)|
\end{equation}
where $p_t(x_u,x_i)$ is the probability of landing at $x_u$ at time $t$, when starting at $x_i$\cite{Porte2008}. Therefore, we sum up all paths via all the possible intermediate points $x_u$.

According to the \textit{Diffusion Map} property, it applies:
\begin{equation}
    \label{diffusion_property}
    D_t^2(x_0,x_1) = ||\Psi_t(x_0) - \Psi_t(x_1)||^2 = \sum_{j \geq 1} \lambda_j^{2t}(\psi_j(x_0)-\psi_j(x_1))^2
\end{equation}

A detailed proof can be found for example in \cite{Yao2011}.

But why does this property make sense? To answer this, we examine what the diffusion distance reveals about the nature of the data we are analyzing.
The diffusion distance describes the connectivity between two data points \cite{lafon2004}. If two data points are far apart in terms of Euclidian distance, but have a lot of intermediate data points connecting them, then the probability to get from one point to the other is increased. They have a low diffusion distance and are close in the \textit{Diffusion Map}. On the other hand, if points are in two different clusters, they will have a larger diffusion distance and will be further apart in the \textit{Diffusion Map}.  \\

\begin{figure}[H]
    \centering
    \includegraphics[scale=0.6]{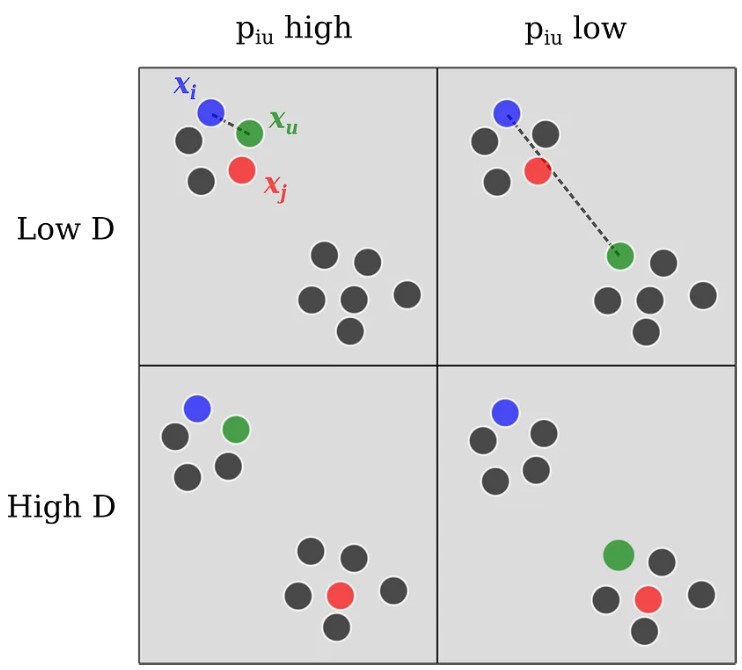}
    \caption{Illustration of the meaning of diffusion distance. Figure originates from \cite{unwrapping_swiss_roll}. $p_{iu} = p_t(x_u,x_i)$ in our notation. D stands for the diffusion distance.}
    \label{fig:ddist} 
\end{figure}

This becomes clear when we consider the illustration in figure \ref{fig:ddist} and look at the summands of the formula \ref{equation:diffusion_distance} following \cite{unwrapping_swiss_roll}:
\begin{itemize}
    \item If $x_i$ and $x_j$ are in the same cluster, then $p_t(x_u,x_i)$ and $p_t(x_u,x_j)$ will have a similar magnitude: If $x_u$ is in the same cluster (top left of the figure) as $x_i$ and $x_j$, both terms are big. If it is in different clusters (top right of the figure), both are small. So both terms always nearly cancel out and $D_t(x_i,x_j)$ is small.
    \item If $x_i$ and $x_j$ are in different clusters, $p_t(x_u,x_i)$ and $p_t(x_u,x_j)$ will have different magnitudes (both cases at the bottom of the figure).
\end{itemize}

The diffusion distance is based on an average of all paths connecting 2 points; the measure is more stable against noise and topological short-circuits than other measures like geodesic distances \cite{lafon2004}.\\

In summary, it can be stated that the diffusion distance not only provides proof that locality is preserved, as shown in section \ref{justification_optimization}, but also ensures that distant points are meaningfully embedded according to the diffusion distance. 

\subsection{Accuracy}
\label{accuracy}

We can find information about the accuracy of \textit{Diffusion Map} in different publications \cite{Coifman2006,Nadler2007,Roozemond2021,Lee2018}.
The accuracy $\delta$ is defined as

$$\delta < \frac{\lambda_k^t}{\lambda_1^t}$$

The accuracy here describes the difference between the Euclidean distance in the $k$ dimensional diffusion space and the diffusion distance in the higher dimensional feature space. So it indicates how precisely the \textit{Diffusion Map} property is fulfilled.\\

It is difficult to say to what extent this accuracy describes whether a result is usable or not. In \cite{Coifman2006,Lee2018} it is indicated that $\delta < 0.1$ is a good value. However, it is not possible to determine the source of this statement. In \cite{Lee2018} the parameter $t$ is also increased to get greater accuracy. This can be seen from the formula, since $\lambda <1$. As we will see in section \ref{time_parameter}, $t$ only scales the axes to different degrees but the shape of the \textit{Diffusion Map} does not actually change. Therefore, the argument cannot be fully understood. 
It seems to be that $\delta$ is an absolute error and so it is clear that the contraction of the \textit{Diffusion Map} (by a larger $t$) leads to a smaller error. But then it would not be a statement about the quality of the dimension reduction. That would contradict the statement that the dimensionality of the correct embedding, so the minimum number of components needed, is dependent on $t$, which is given in \cite{Coifman2006}.\\
Overall it is not clear if a precise accuracy indicates whether the dimension reduction provides a meaningful result and finds the intrinsic low-dimensional representation. It is worth noting that the accuracy is not considered in many applications. 

\section{Related methods}
\label{related_methods_and_history}

\subsection{Linear methods}
\label{equivalence_pca}

Linear dimension reduction methods were developed in the beginning of the 20th century. The most prominent are principal component analysis (PCA) \cite{Pearson1901,Hotelling1933} and the classical multidimensional scaling (MDS) \cite{Torgerson1952} (at least with Euclidean distances). These methods follow a global approach and can therefore only represent the data accurately in low dimensions if the original data lies on a linear manifold \cite{Meila2024}.\\

PCA restructures the data by transforming it into a new coordinate system defined by principal components, which are ranked according to the amount of variance they explain. This can be reached by using the spectral decomposition of the covariance matrix of the data, where the eigenvectors corresponding to the largest eigenvalues represent the principal components that capture the most variance.\\

MDS operates in a similar manner: It involves performing the spectral decomposition of a double-centered distance matrix, where the eigenvectors define the new embedding. It can be shown that MDS and PCA are equivalent when an Euclidean distance matrix is used for MDS, despite the fact that different matrices (covariance or distance matrix) are utilized \cite{Ek2021,Ghojogh2023}.\\

In this work, we will demonstrate for various datasets that the Diffusion Map converges to the PCA embedding for large neighborhoods i.e., for a large width parameter $\epsilon$ (see section \ref{equivalence_pca}, \ref{vdem_PCA}, \ref{census_differentN_rotating}). This can be explained by the fact that locality is the key property of nonlinear dimensionality reduction and the fundamental difference to linear dimensionality reduction methods.

\subsection{Nonlinear methods}
\label{theroy:non-linear-methods}
\paragraph*{Developmental history}~\\
In the 1990s and early 2000s, various so-called non-linear spectral dimension reduction methods were developed. In all of these methods, the spectrum of eigenvectors of differently defined adjacency matrices of the data are calculated in order to obtain a low dimensional representation \cite{Weiss1999}. Therefore, all these methods can be referred to as spectral dimension reduction methods. A good overview of the early developments can be found in \cite{Weiss1999}. One of the earliest methods seems to be published in \cite{Scott1990}, although here also references are made to earlier publications in molecular physics that cannot be further verified. Molecule configurations and their folding pathways can be well described by such methods \cite{Meila2024}. \\

Today the most established spectral methods are the Locally Linear Embedding \cite{roweis2000}, the Isomap \cite{Tenebaum2000}, the Spectral Clustering or Embedding \cite{Ng2001}, kernel PCA \cite{schoelkopf1997}, the Laplacian eigenmap \cite{Belkin2001} and the Diffusion Map \cite{lafon2004,Nadler2005,Coifman2006}. The \textit{Diffusion Map} is therefore not really a new algorithm. Rather the method gives a probabilistic interpretation of the spectral dimension reduction algorithm \cite{Nadler2005}. But the Diffusion Map is not the first probabilistic interpretation of a spectral embedding method. \cite{Meila2000,Meila2001} presented earlier a new view on clustering and segmentation by interpreting pairwise similarities as edge flows in a Markov random walk and by using the eigenvectors of the transition matrix of this random walk. The name \textit{Diffusion Map} first appeared with Lafon's doctoral thesis \cite{lafon2004}. This work was followed by several publications in which connections were drawn between the eigenfunctions of the Fokker-Planck operator, and the \textit{Diffusion Map} was further generalized \cite{Coifman2005,Nadler2005,Coifman2006}. \\

In \cite{Coifman2006} the question of the influence of a non-uniform density of the data points is addressed, and a new parameter $\alpha$ is added, which adjusts the normalization of the graph Laplacian. We will not discuss this generalization because it plays no significant role in the current applications, and there were no improvements in the experiments of this master thesis. \\

In general, there are various modifications of the \textit{Diffusion Map} algorithm. For instance, in \cite{Xiu2023,Xiu2024}, the Spearman rank correlation is employed as a similarity measure. However, this approach could potentially violate the positivity-preserving property of valid kernel functions, as defined in \cite{lafon2004, Coifman2005}. Therefore, caution is required when determining which proven properties of the \textit{Diffusion Map} can be transferred to alternative definitions. Besides, we already know from section \ref{equivalence_pca} that when using linear spectral methods, the PCA using the covariance matrix obtains the same results as the classical MDS using a distance matrix. So this also could motivate the use of such a similarity measure.

\paragraph*{Global vs. local}~\\
In the literature, spectral dimensionality reduction methods are often classified into global and local approaches \cite{DeSilva2002,VanDerMaaten2009a}. This distinction is primarily driven by the introduction of the Isomap method as a global framework, presented in contrast to other methods \cite{Tenebaum2000}, while alternative approaches have been framed as counterexamples \cite{Belkin2001}.\\

However, this classification may be misleading: each of these methods includes a step in which the local neighborhood is analyzed to construct a neighborhood graph. The corresponding adjacency matrices store local information about the underlying manifold, which is crucial for determining its intrinsic dimension and learning its structure. This local information is later integrated through spectral decomposition to discover global structures  \cite{Saul2003,ham2004,Wu2018}.\\

The key difference in methods classified as global, such as the Isomap, lies in an additional intermediate step. In the Isomap algorithm, the local neighborhood graph is first converted into a geodesic distance matrix before spectral decomposition is applied. This ensures that all geodesic distances are taken into account \cite{Belkin2003}. Thus, the integration of local information from the neighborhood graph does not occur solely during spectral decomposition. \\

Interestingly, the \textit{Diffusion Map} is also classified as a global method by \cite{VanDerMaaten2009a}, despite the nearly equivalent Laplacian Eigenmap being categorized as a local method. This distinction does not stem from fundamental algorithmic differences but rather from the fact that the \textit{Diffusion Map} gains a global property through its probabilistic interpretation: the diffusion distance in the feature space corresponds to the Euclidean distance in the diffusion space. However, this should not justify the creation of two separate categories.\\

One could argue that methods labeled as global, such as Isomap and the \textit{Diffusion Map}, aim to preserve not only local geometries but also maintain global structures and distances. In particular, they ensure that points that are far apart in the original space remain distant in the reduced representation \cite{DeSilva2002}. Similar preservation principles likely apply also for methods that emphasize the local approach, as their algorithms differ only slightly. However, these aspects were not the primary focus when these methods were originally developed.\\

As described in section \ref{adding_locality}, there are multiple ways to incorporate locality, and different methods employ distinct strategies. For example, the Laplacian Eigenmap uses either an $\epsilon$ kernel or a $N$ nearest neighbors approach \cite{Belkin2001}, while Locally Linear Embedding applies the nearest neighbors method \cite{roweis2000}. The \textit{Diffusion Map}, in its original formulation, relies on a kernel neighborhood \cite{lafon2004,Nadler2005}, whereas spectral clustering employs either a general kernel \cite{Ng2001} or a kernel in combination with a strict radius \cite{Shi2000}. Isomap, on the other hand, utilizes both the kernel and nearest neighbors approach \cite{Tenebaum2000}. Therefore, locality is an inherent component of all these methods.

Overall \cite{VanDerMaaten2009a} emphasizes that local methods might struggle with datasets, with a high dimensional intrinsic manifold, as the number of data points required to characterize the manifold increases exponentially.

\paragraph*{Unified framework}~\\
Due to the similarities among these methods, the idea of developing a unified framework has existed for a long time \cite{Weiss1999}. Many of these approaches follow the same algorithm, differing only in the definition of the adjacency matrix or the graph Laplacian \cite{Ng2001,Meila2024}. \\

It has been shown that all these methods can be described using Kernel PCA \cite{ham2004,Bengio2004,Strange2014,Ghojogh2023}, which is an extension of PCA to handle nonlinear data by using the kernel trick: Nonlinear data is transformed with a nonlinear kernel function into a higher-dimensional space, where it is linearly analyzable by PCA \cite{schoelkopf1997}. Different kernels were identified for the various methods.
Since the methods are closely related, \cite{Coifman2006} also states that all kernel methods are special cases of the \textit{Diffusion Map} framework and can be interpreted in terms of diffusion processes. One apparent difference is that the \textit{Diffusion Map} introduces a time parameter $t$ \cite{Roozemond2021}. However, as we will demonstrate in section \ref{time_parameter}, this parameter does not significantly alter the embedding.\\

In summary, all non-linear spectral dimensionality reduction methods follow nearly the same steps and are based on very similar principles. The methods only differ in the way they construct different local or global proximity/distance matrices.


%% file: Chapter/classic_examples.tex
\label{classical_examples}
There are various example datasets given in the literature to introduce the functionality of the \textit{Diffusion Map} method. In this section, we will explore several of these examples and use intuitive illustrations to understand what it means to identify the underlying manifold of a dataset. 
To achieve this, we first aim to understand one of the simplest types of datasets: a one-dimensional curve embedded in a higher-dimensional space. For such a dataset, we can determine the resulting structure of the \textit{Diffusion Map} (section \ref{typical_shapes}). After that, we will look in section \ref{swiss_roll} at one of the standard examples used to explain how the \textit{Diffusion Map} works: the Swiss roll. Building on this example, we proceed to examine the influence of the parameters $\epsilon$, $N$, and $t$ of the \textit{Diffusion Map} method (section \ref{finding_parameter}) and analyze the effects of a prior normalization. After that, we discuss how data clustering can be approached (section \ref{clustering}) and demonstrate how the \textit{Diffusion Map} can organize and sort high-dimensional data points, using images as an example (section \ref{sorting_ability}). The insights gained from these simple datasets will help us analyze the more complex social science datasets in section \ref{3results_social_datasets}.

\section{Typical shape of the \textit{\textit{Diffusion Map}}: one dimensional curves}
\label{typical_shapes}

\begin{figure}[H]
\centering
\label{fig:typical_shapes}
\subfigure[One dimensional non-intersecting datasets.]{
   \centering
   \label{fig:one_dimensional_typical_shapes}
   \includegraphics[width=1\linewidth]{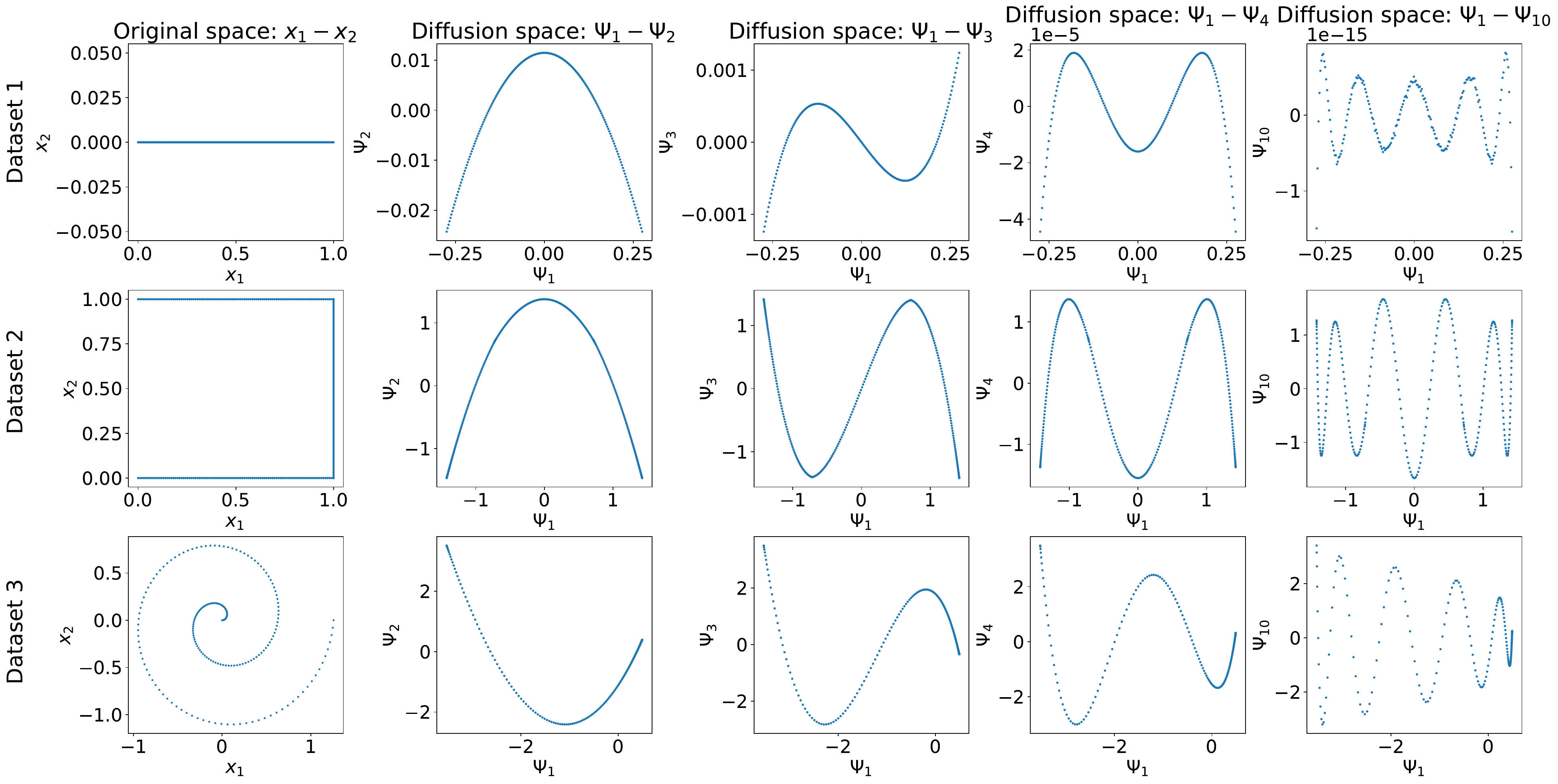}}\\
\subfigure[More complex datasets with small peaks (dataset 4), loop (dataset 5), small gap at one end (dataset 6) and a closed one-dimensional curve (dataset 7).]{
   \centering
   \label{fig:complex_typical_shapes}
   \includegraphics[width=1\linewidth]{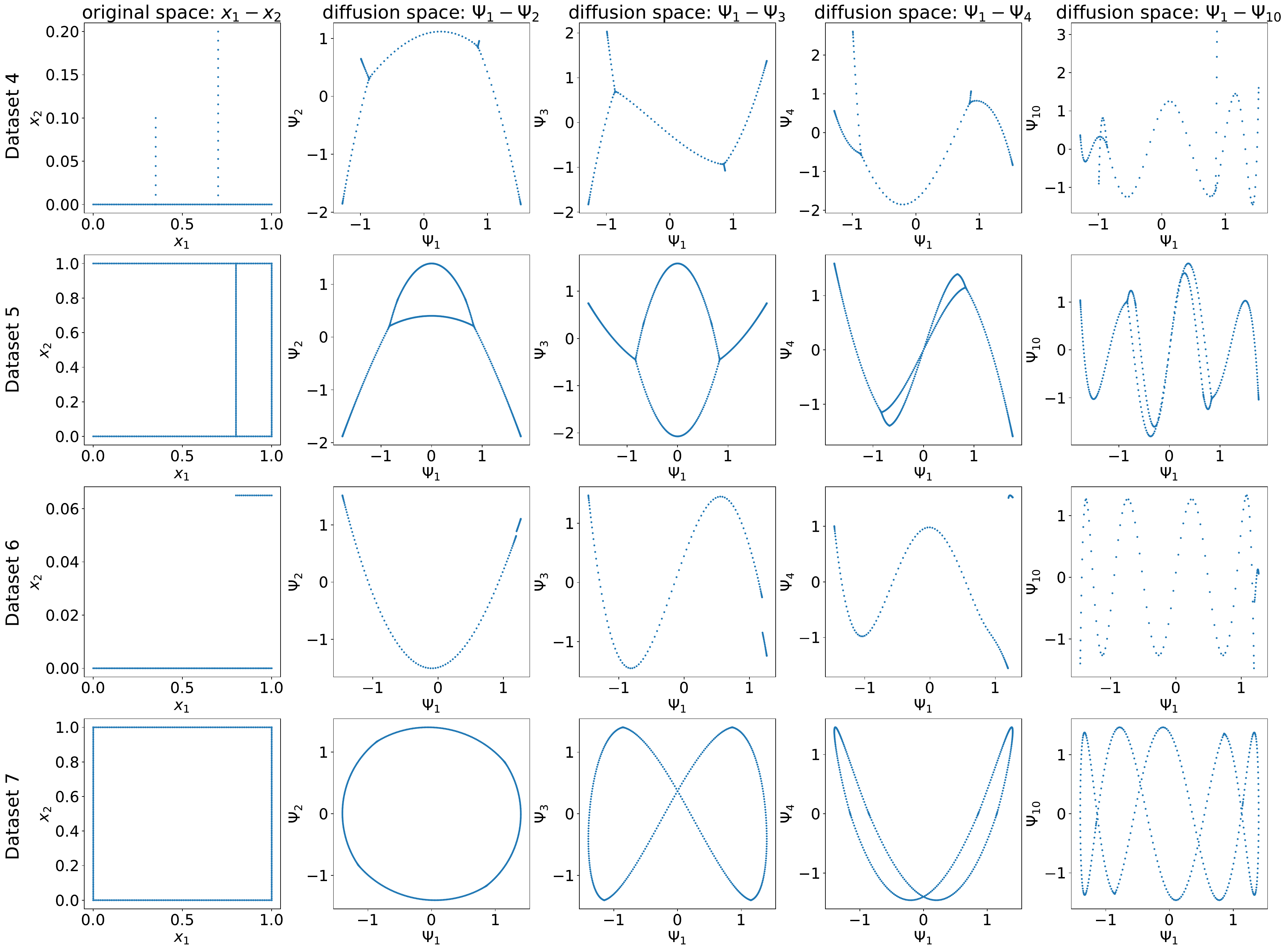}}
   \caption{\textbf{Showing the \textit{Diffusion Map} of different example datasets:} The first column displays the original datasets (consisting each of 100-400 data points) in the feature space. The subsequent columns show different diffusion components in dependence of the first diffusion component $\Psi_1$.}
\end{figure}

\paragraph*{Different Examples}~\\
The simplest example datasets for manifold learning and nonlinear dimensional reduction are one-dimensional curves in a higher-dimensional space \cite{Nadler2007}.
In figure \ref{fig:one_dimensional_typical_shapes}, we can see the diffusion components of different one-dimensional non-intersecting curves as examples. We observe here that the \textit{Diffusion Map} for the first two diffusion components shows a parabolic shape. Higher diffusion components also show a higher-order polynomial relationship. This behavior can be shown for various one-dimensional curves. Note that, as mentioned in \ref{spectral_decomposition}, the orientation of the \textit{Diffusion Map} (in the sense of, for example, the parabola pointing up or down for the first two components) is only a random phenomenon and has nothing to do with the datasets.
In figure \ref{fig:complex_typical_shapes}  smaller features, such as a loop and data points in an additional vertical line, were added to examine the effect of such features. We see that for these modified datasets 4, 5, we also find similar results. The topology of these manifolds is preserved, but the final shape resembles the shapes of the one-dimensional non-intersecting manifold. In dataset 6, we can see that gaps in the dataset also result in gaps in the \textit{Diffusion Map} but depending on the neighborhood parameters (considered in more detail in the section \ref{finding_parameter}) and the size of the gaps. These gaps can also be difficult to detect because they can be molded to the existing structure.

\paragraph*{Polynomial relationships between components}~\\
From \cite{Nadler2007} we know that these typical shapes of the datasets in figure \ref{fig:one_dimensional_typical_shapes} are not a coincidence, but follow from the following proven theorem:
\begin{quote}
    \glqq \textit{Consider data sampled uniformly from a non-intersecting smooth 1-D curve embedded in a high dimensional space. Then, in the limit of a large number of samples and small kernel width the first \textit{Diffusion Map} coordinate gives a one-to-one parametrization of the curve.}\grqq
\end{quote}

Lets try to provide some motivation for this statement:
\cite{Nadler2007} describes that the diffusion components converge to the eigenfunctions of the Fokker-Planck operator $\mathbf{H}$, if the data points are uniformly distributed on the manifold. For the 1-D curve this is $\mathbf{H}\psi = \frac{d^2\psi}{dx^2}$. The eigenfunctions are given by 
\begin{equation}
\label{psi_n}
    \psi_n = cos(n\pi x)\text{,}
\end{equation} if we solve the eigenvalue problem $H\psi = \lambda \psi$ with Neumann boundary conditions with $x=0,1$ at the edges \cite{Nadler2007}. 
Following \cite{angular_formula_wolfram}, the first cases of $cos(nx)$ can be expressed as:
\begin{align*}
    cos(2x) &= -1 + 2cos^2(x)\\
    cos(3x) &= -3cos(x) + 4cos^3(x)\\
    cos(4x) &= 1-8cos^2(x)+8cos^4(x)\\
    cos(5x) &= 5cos(x)-20cos^3(x)+16cos^5(x)\\
    &...
\end{align*}
As $\psi_1 = cos(x)$ the larger diffusion components are directly dependent on the first diffusion component, because they can be expressed as:
\begin{align}
\label{polynomials}
    \psi_2 &= -1 + 2\psi_1^2\\
    \psi_3 &= -3\psi_1 + 4\psi_1^3\\
    \psi_4 &= 1-8\psi_1^2+8\psi_1^4\\
    \psi_5 &= 5\psi_1-20\psi_1^3+16\psi_1^5\\
    &... \nonumber
\end{align}

\cite{Nadler2007} also mentions that if one assumes a closed one-dimensional manifold (as in figure \ref{fig:complex_typical_shapes} dataset 7) then one should use periodic boundary conditions, which would lead to the \textit{Diffusion Map} shapes shown in the figure: The dataset is mapped into a circle in the first two diffusion components.

We can conclude, following \cite{Nadler2007} that:
\begin{enumerate}
    \item If we see such polynomial dependencies of the diffusion components we can say, that the data perhaps lie on a one dimensional manifold in a higher dimensional feature space and can be described by one natural parameter. However, this represents only a necessary condition.
    \item It follows, that we cannot assume that these diffusion components provide any information that is not already given by the first diffusion component. 
\end{enumerate}

\paragraph*{Usability of the spectral gaps to find the intrinsic dimension}~\\

It is important to note that these properties (the presence of polynomial relationships) differ significantly from those observed in methods such as PCA:  In PCA, the first dimensions inherently contain the most important information because they correspond to the directions of greatest variance in the dataset. When working with PCA, one primarily examines the spectrum of eigenvalues. The eigenvalues indicate the proportion of variance explained by each principal component. By analyzing the spectrum and identifying large jumps in the spectrum, so-called spectral gaps, one can determine how many dimensions are needed for a meaningful representation of the data, thereby estimating the intrinsic dimensionality of the dataset. Spectral gaps indicate that the principal components before the gap contribute significantly more to explaining the data than those after the gap.\\

This raises the question: How does this concept apply to the \textit{Diffusion Map}? Is there any indication that we can use the spectrum in a similar way? Several publications attempt to ensure that the relative magnitude of the eigenvalues corresponding to the selected diffusion components remains large \cite{lafon2004, Coifman2006, Nadler2007, Strange2014, Lee2018}. Paying attention to spectral gaps should, in principle, help maintain a high accuracy $\delta$, as defined in section \ref{accuracy}, and ensure that the \textit{Diffusion Map} property (equation \ref{diffusion_property}) is preserved.\\

\begin{figure}[h]
    \centering
    \includegraphics[width=0.9\linewidth]{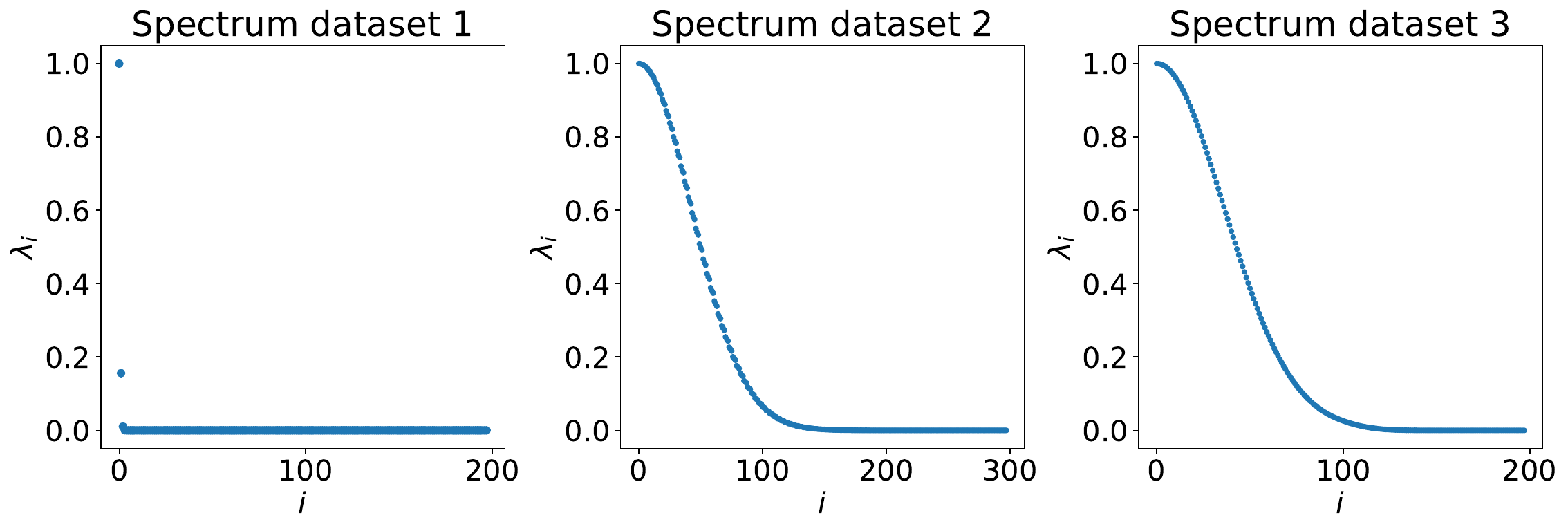}
    \caption{Eigenvalue spectra of the one dimensional non intersecting datasets 1,2,3 form figure \ref{fig:one_dimensional_typical_shapes}.}
    \label{fig:spectra}
\end{figure}

However, when we examine figure \ref{fig:spectra}, which shows the eigenspectra of the first three datasets, we can only identify a clear spectral gap in the first dataset. Yet, even though datasets 2 and 3 are also one-dimensional curves without intersections (just like dataset 1) and can be well represented by the first principal component, with the other principal components simply depending on the first, we do not see a spectral gap. Thus, based on the spectrum, one cannot predict that only the first diffusion component is relevant.

\section{Exploring a two-dimensional manifold: the Swiss roll}
\label{swiss_roll}
\begin{figure}[H]
\centering
\subfigure[Original dataset. The color encodes the length dimension of the sheet, which is rolled up to the Swiss roll. The width of the Swiss roll is along the Y-axis]{
   \centering
   \label{fig:swiss_roll_original_color}
   \includegraphics[width=0.5\linewidth]{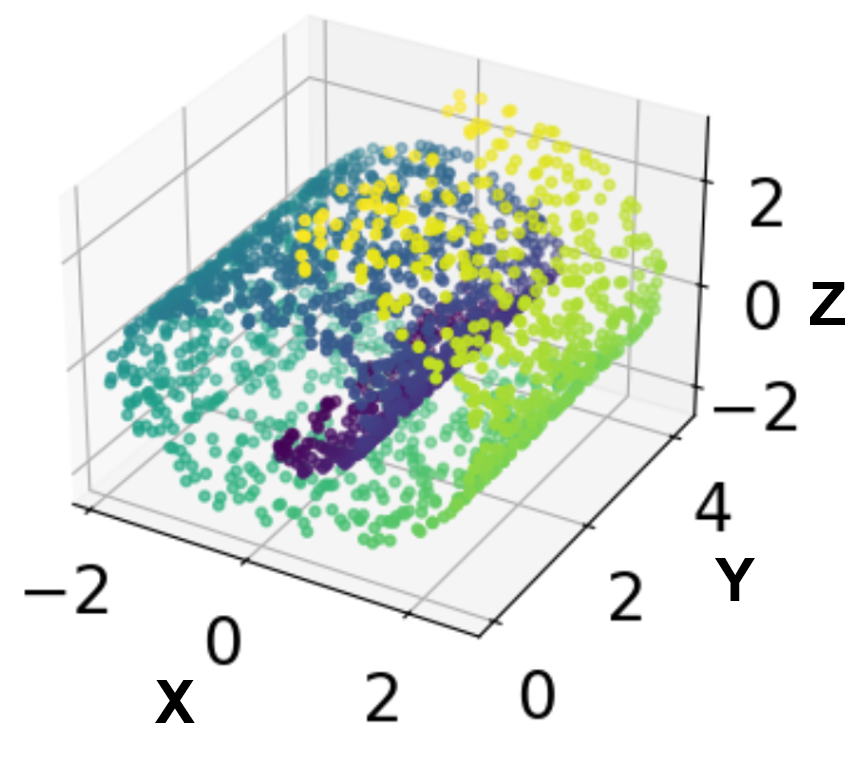}}\\
\subfigure[\textit{Diffusion Map} with the first $5$ components. In the four images on the left, the color describes the length of the rolled-up sheet. In the image on the far right, the color describes the width of the sheet.]{
   \centering
   \label{fig:different_components_swissroll}
   \includegraphics[width=\linewidth]{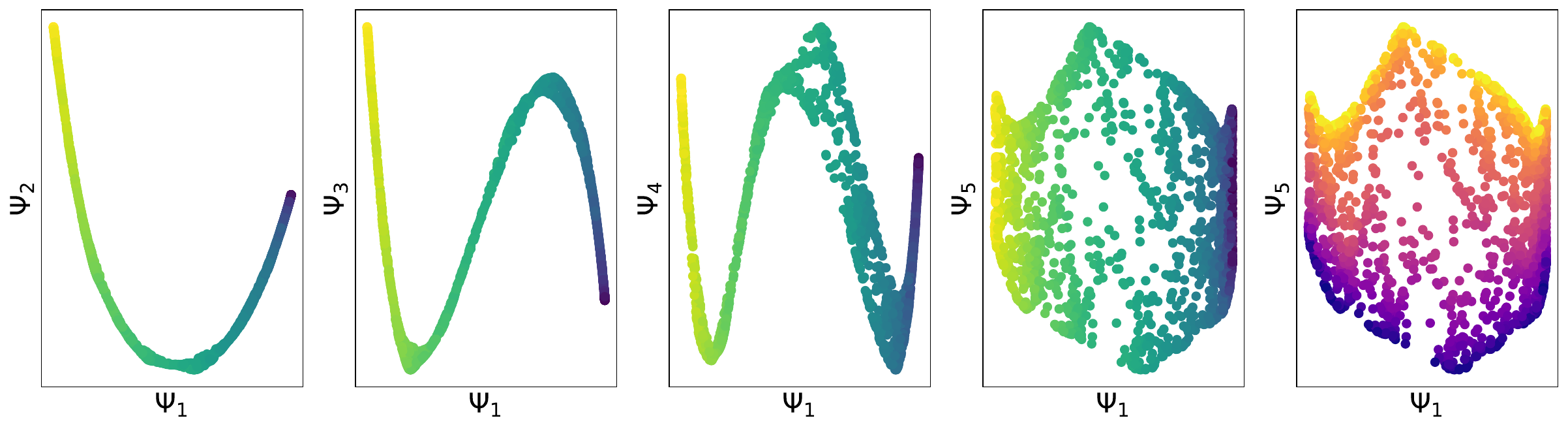}}\\
\subfigure[The first ten eigenvalues of the eigenspectrum which are obtained from the \textit{Diffusion Map} of the Swiss roll.]{
   \centering
   \label{fig:spectrum_diffusion_swissroll}
   \includegraphics[width=0.5\linewidth]{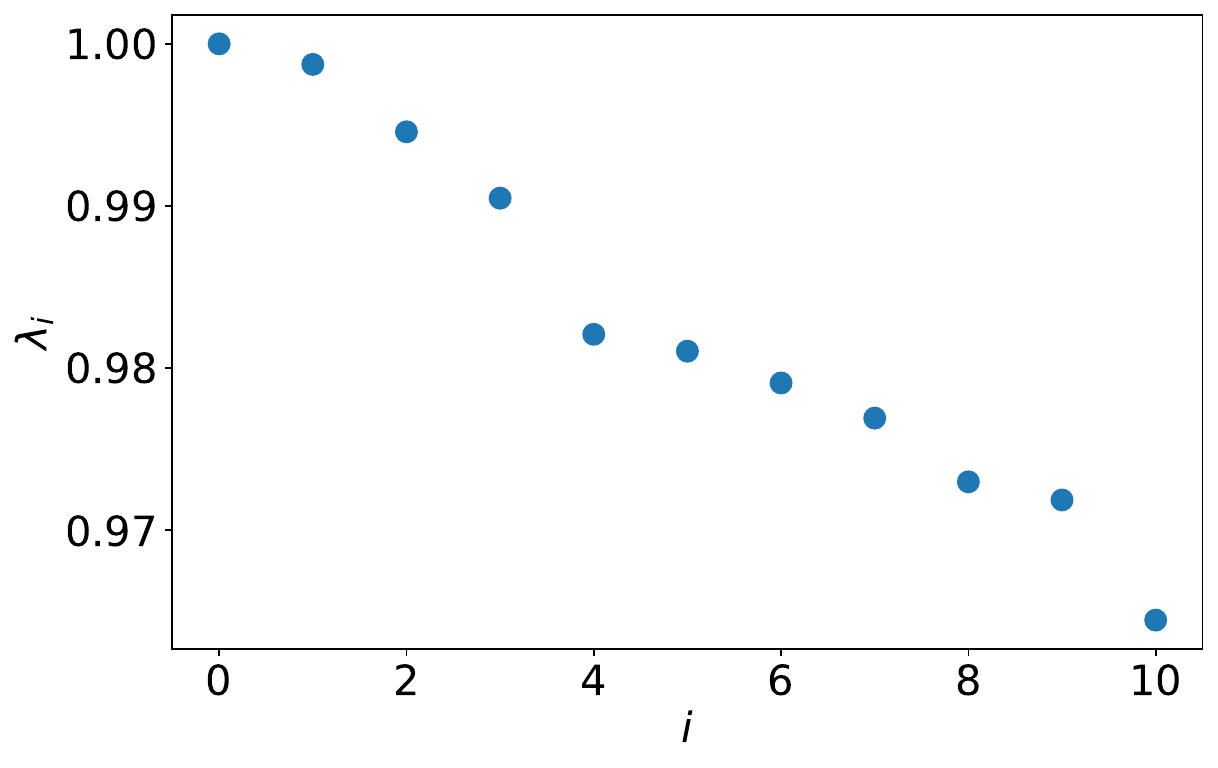}}
   \caption{Exploring the \textit{Diffusion Map} of the Swiss roll dataset. The Swiss roll was created by using the \texttt{make\_swiss\_roll} methode from \cite{scikit-learn} with the parameters \texttt{n\_samples=2000, noise=0.01}. The \textit{Diffusion Map} parameters are $t=1$ and $\epsilon = 0.1$ by using all nearest neighbors.}
   \label{fig:swiss_roll_first_analysis}
\end{figure}

Now, let us consider a standard example of a curve that actually represents a two-dimensional manifold: the so-called Swiss roll. This three-dimensional dataset can be seen in figure \ref{fig:swiss_roll_original_color} and consists of a two-dimensional sheet which is rolled up like a spiral. It is used in other publications to demonstrate the results of dimensionality reduction methods (e.g. \cite{Tenebaum2000,roweis2000,DeSilva2002,Belkin2003,Nadler2007,Ferguson2011,Zhang2011}). \\

In our example, the length of this sheet is much longer than the width and should be therefore the most important natural parameter of the dataset. When we have a look at the \textit{Diffusion Map} of this dataset in figure \ref{fig:different_components_swissroll}, we can see that this most important natural parameter is represented by the first diffusion component $\Psi_1$. It is particularly interesting that in our results and in some literature \cite{Nadler2007,unwrapping_swiss_roll}, we observe the same behavior in the first diffusion components as shown in figure \ref{fig:one_dimensional_typical_shapes} for the one-dimensional spiral (dataset 3). 
Only in $\Psi_5(\Psi_1)$ the \textit{Diffusion Maps} of the one-dimensional and the two-dimensional curve begin to differ. Moreover, the diffusion components $\Psi_1$ to $\Psi_4$ contain almost no information about the width of the two-dimensional manifold, which is rolled up to the Swiss roll. It is only $\Psi_5$ that captures the width of this manifold. 
This is also mentioned in \cite{Nadler2007}. They conclude that sometimes higher diffusion components also provide important information about the dataset, even if previous components have already contributed redundant information.\\

Furthermore, from the spectrum of the Swiss roll (figure \ref{fig:spectrum_diffusion_swissroll}), it is not clear why $\Psi_5$ should be of interest or is the first component that decodes the width. At the same time, the diffusion components with redundant information ($\Psi_2,\Psi_3,\Psi_4$) have slightly higher eigenvalues. We can therefore say that eigenvalues cannot be used to sort the importance of the eigenvectors; in this case, this would even favor redundant information. Thus we must always be careful with the results of \textit{Diffusion Map}, especially when we see polynomial dependencies. The partially used approach to use such spectra to determine the used number of diffusion components \cite{Coifman2006,Lee2018} should therefore be reconsidered. \\

According to \cite{Nadler2007}, this behavior is different between linear and non-linear methods: in contrast to PCA, where all principal components are linearly uncorrelated and the data is projected onto the orthogonal directions of principal components, in the \textit{Diffusion Maps} \glqq several eigenvectors may encode for the same geometrical or spatial 'direction' of a manifold \grqq\cite{Nadler2007}. \\

\section{Finding the right parameters and data preprocessing}
\label{finding_parameter}

The results of a \textit{Diffusion Map} for a specific dataset depend on various parameters. In addition, normalizing and pre-processing the data beforehand can have an effect on the final result. These effects are represented in this section by using data which is lying on a Swiss roll as one non-linear example applying dimensionality reduction. Here we use the first two diffusion components $\Psi_1$ and $\Psi_2$ to illustrate the mentioned effects and how to identify the optimal parameters. This analysis will also guide us in determining the best approach for more complex datasets, such as those in social sciences.

\subsection{Width of the neighborhood $\epsilon$}
\label{finding_right_epsilon}

In section \ref{adding_locality}, we learned about the importance of measuring the similarities within local neighborhoods and the different methods to achieve this. The original approach proposed by \cite{lafon2004, Nadler2005, Coifman2005} for identifying neighborhoods involves introducing a Gaussian kernel, as defined in equation \ref{kernel_definition}. The parameter $\epsilon$ determines the width of the Gaussian neighborhood: A bigger $\epsilon$ creates a larger neighborhood. 

\paragraph*{Impact of different $\epsilon$ values on the Swiss roll dataset}~\\

\begin{figure}[H]
\centering
\subfigure[\textit{Diffusion Maps} with the first two diffusion components $\Psi_1$ and $\Psi_2$ for different $\epsilon$. $\epsilon = 0.1$ represents the expected result.]{
   \centering
   \label{fig:dmap_epsilon_swissroll}
   \includegraphics[width=0.9\linewidth]{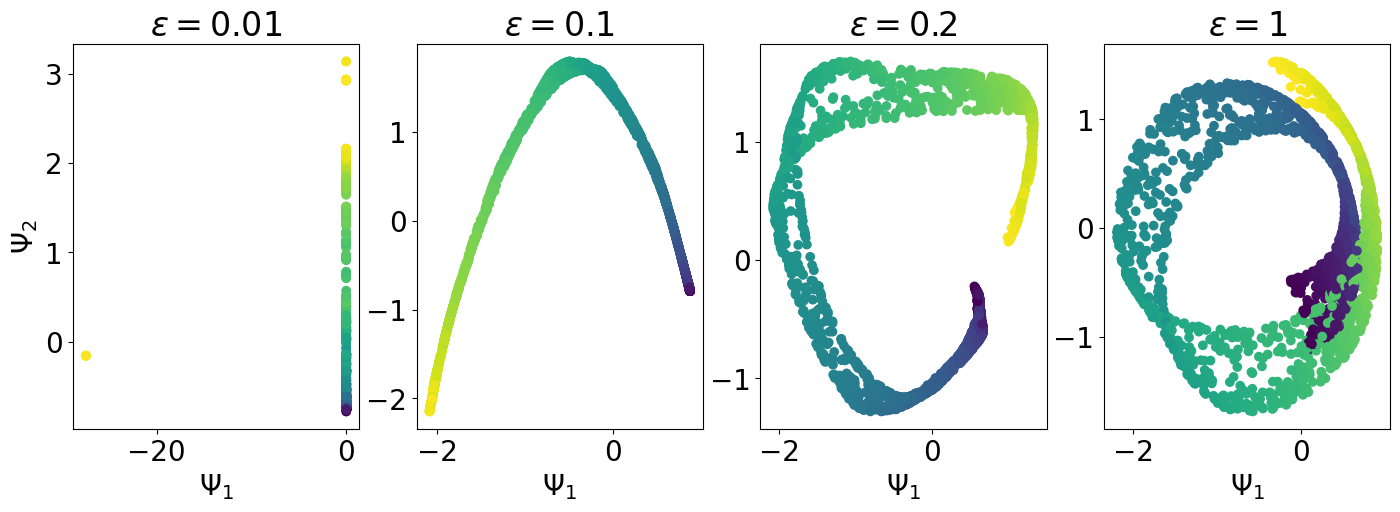}}\\
\subfigure[Illustrating the size of the $\epsilon$ neighborhood for $\epsilon=0.1$ and $\epsilon = 1$. The Swiss roll is color-coded to represent the magnitude of the entries of the symmetrized transition matrix $M_{s_{ij}}$.]{
   \centering
   \label{fig:swissroll_epsilon_Ms}
   \includegraphics[width=0.9\linewidth]{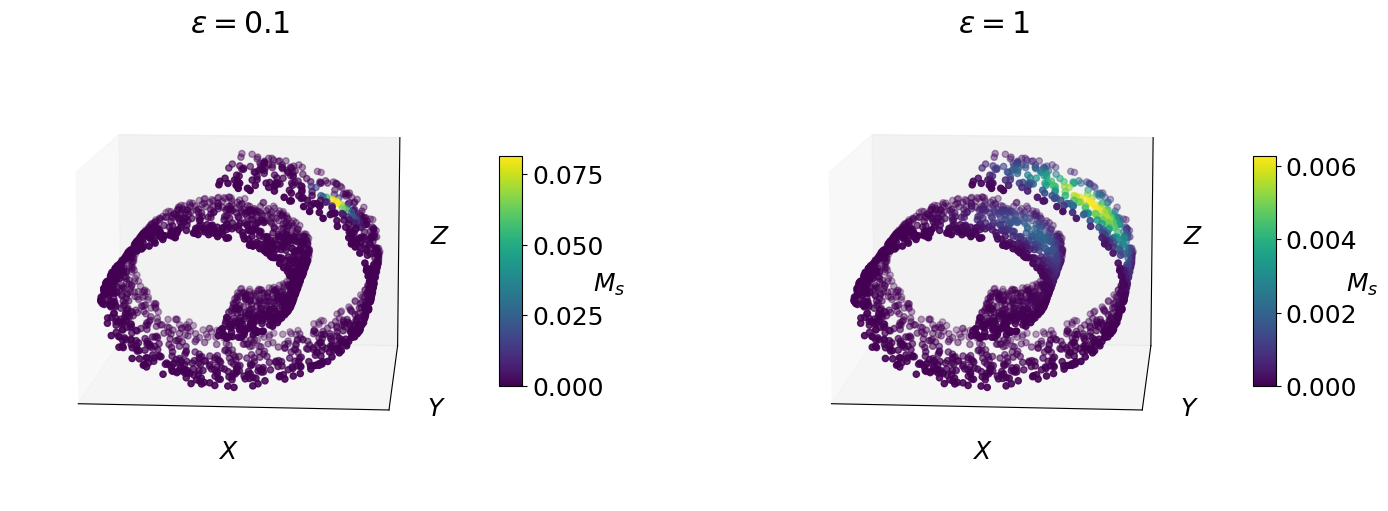}}\\
\subfigure[Illustrating the equivalence between PCA and the \textit{Diffusion Map} for a large parameter $\epsilon$. The \textit{Diffusion Map} shows the same result as the PCA for $\epsilon = 10000$.]{
   \centering
   \label{fig:PCA_diffmap_swissroll}
   \includegraphics[width=0.9\linewidth]{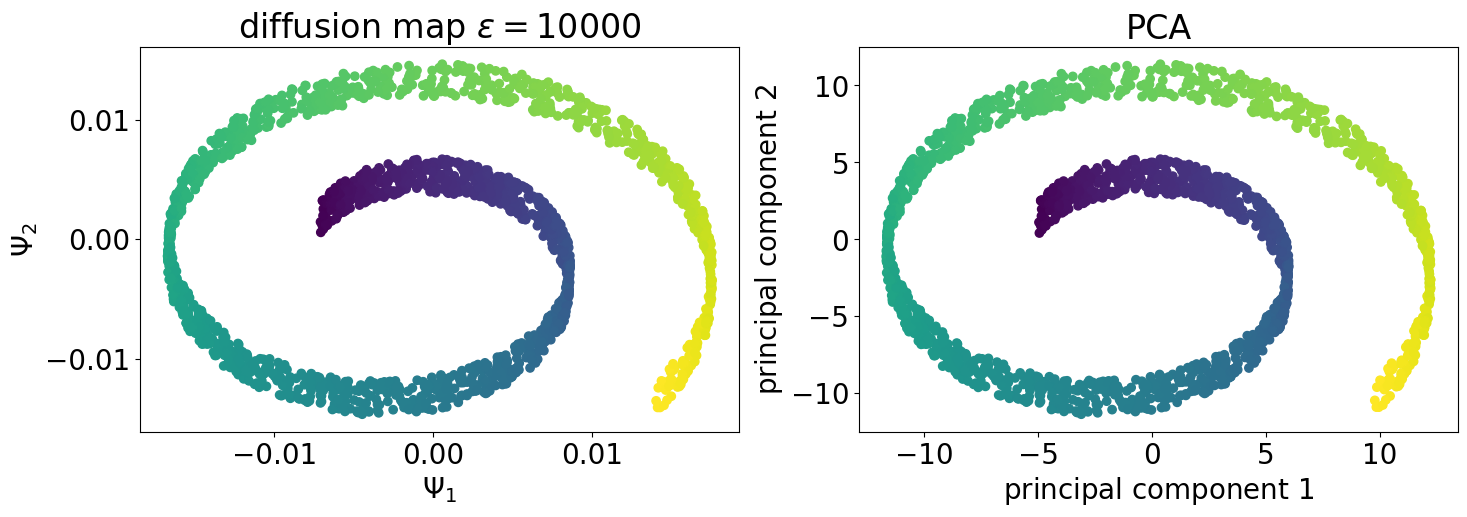}}
   \caption{\textbf{Visualizing the effect of the width parameter $\epsilon$} using the Swiss roll dataset shown in figure \ref{fig:swiss_roll_original_color}. The dataset was standardized beforehand. All nearest neighbors and $t=1$ were used.}
\end{figure}

We will now look at the Swiss roll dataset, which we already know from the analysis of figure \ref{fig:swiss_roll_first_analysis}.  We expect a parabolic shape, where the path coordinate along the length of the rolled-up sheet is indicated by the first diffusion component. 
This expected result is achieved for $\epsilon = 0.1$, see figure \ref{fig:dmap_epsilon_swissroll}. Varying $\epsilon$, while considering all nearest neighbors and keeping $t=1$ fixed, we move away from the expected result.

We see that if the width parameter is too small ($\epsilon=0.01$), the network defined by the kernel matrix starts to disconnect and no longer gives good results, as the original data is no longer resolved on a single manifold. If the width is too large ($\epsilon = 1$), the locality is lost, and the \textit{Diffusion Map} cannot find the intrinsic low-dimensional data. Additionally, shortcuts occur, causing the \textit{Diffusion Map} to be over-connected, losing the resolution of the intrinsic structure of the data.

In figure \ref{fig:swissroll_epsilon_Ms} we can see the size of the neighborhood for different values of $\epsilon$. We see that for $\epsilon=1$ the neighborhood is defined too large, as a jump across the boundaries of the manifold to another part of the manifold is possible. For the correct choice of $\epsilon = 0.1$, the neighborhood is restricted to the manifold and does not produce an over-connected network, leading to a non-representative manifold of the data.\\

Another interesting aspect is that for very large $\epsilon$, we find that the \textit{Diffusion Map} tends towards the embedding of the PCA (see figure \ref{fig:PCA_diffmap_swissroll}). This seems to be a general rule and can also apply to more complex datasets, as we will see in the sections \ref{vdem_PCA} and \ref{census_differentN_rotating}. However, this is not surprising, as an excessively large neighborhood width eliminates locality, the only significant difference between PCA and \textit{Diffusion Maps} as discussed in section \ref{equivalence_pca}.\\

In any case, an incorrect choice of parameters can have a significant impact on the result and lead to false conclusions, as shown, for example, in \cite{Zhang2011}, where it is falsely claimed the \textit{Diffusion Map} does not unroll the Swiss roll (which is the case for a wrong $\epsilon$).

\paragraph*{Various ideas for finding the right $\epsilon$}~\\

The correct value of $\epsilon$ depends on various factors, such as the number of data points, the dimension, the manifold volume or the manifold curvature \cite{Singer2006,Meila2024}. So the question arises: Are there methods to determine the optimal $\epsilon$ for a given dataset?\\

There are various ideas for finding a good $\epsilon$: 
A first approach is that the $\epsilon$ should be comparable to the square of the distance between the data points and their neighboring data points \cite{Cameron2021}. Therefore, it is suggested to take the minimum distances from the data points to other data points (with a different location) and compute their average \cite{lafon2004,Cameron2021}:
\begin{equation}
    \label{equation: find_optimal_epsilon}
    \epsilon = \frac{1}{n}\sum_{i=1}^n \underset{j:x_j\neq x_i}{min} ||x_i-x_j||^2
\end{equation}
However, this $\epsilon$ should be considered as a first guess and may need to be adjusted if necessary \cite{Cameron2021}. For the Swiss roll data described in figure \ref{fig:swiss_roll_first_analysis}, it resulted in a value of $0.015$ and had to be further adjusted, to get a useful result.\\

In \cite{Singer2009,Bah2008}, another method is described to find a correct $\epsilon$. The idea is that if $\epsilon$ is too large, then all the entries of the kernel matrix $K$ become too similar, since the entries tend towards $1$. On the other hand, if $\epsilon$ is too small, all entries tend towards $0$. 
Therefore, they suggest the following steps:
Constructing the kernel matrix $K$ for different choices of $\epsilon$. Then the sum 
\begin{equation}
    \label{finding_epsilon_singer}
    S(\epsilon) = \sum_{ij}K_{ij}(\epsilon)
\end{equation} 
is calculated. By plotting $S(\epsilon)$ we can analyze the asymptotes when $\epsilon \rightarrow 0$ and $\epsilon \rightarrow \infty$. The right $\epsilon$ can be chosen in between, where the log-log plot appears linear.

However, looking at the result of this method for the Swiss roll dataset, it does not seem to provide a better intuition, since the range of possible $\epsilon$ is very wide. It would accept values for $\epsilon$ such as $0.2$ or $1$ as good values, although these are unsuitable if we compare them with the results from figure \ref{fig:dmap_epsilon_swissroll}. 

\begin{figure}[h]
    \centering
    \includegraphics[width=0.5\linewidth]{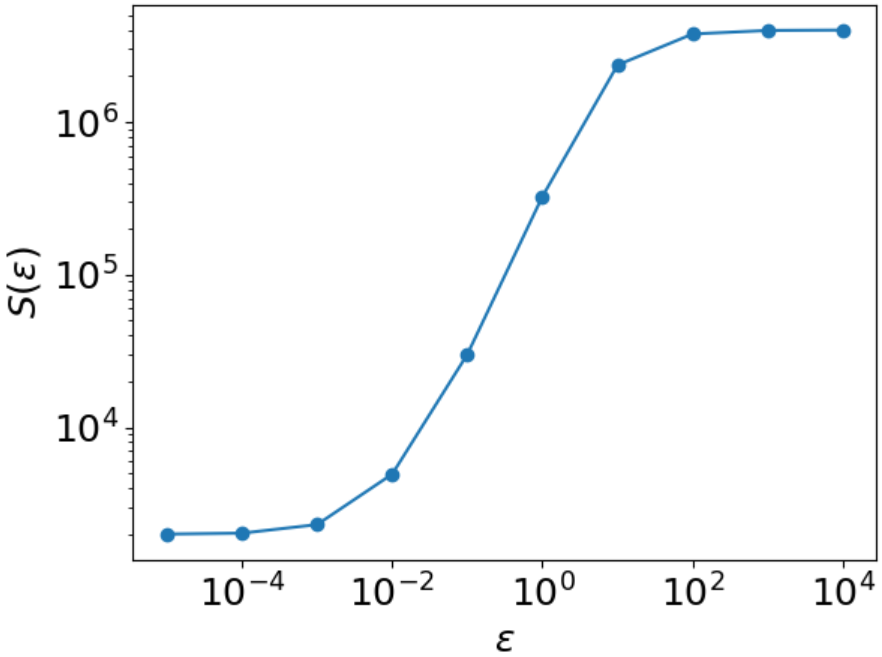}
    \caption{Method for finding the correct $\epsilon$ proposed by \cite{Singer2006} applied to the Swiss roll dataset.}
    \label{fig:finding_the_right_epsilon_singer}
\end{figure}

In this work, we will proceed as follows to find optimal values for $\epsilon$ for more complex datasets: We take the suggestion from \cite{lafon2004} as an initial guess and then vary $\epsilon$ around it. If the network becomes disconnected, the value of $\epsilon$ is too small. It is increased until a meaningful result is obtained. If the result approaches that of PCA, the value of $\epsilon$ is too large.

\subsection{Number of nearest neighbors $N$}
\label{finding_N}
\begin{figure}[h]
    \centering
    \includegraphics[width=\linewidth]{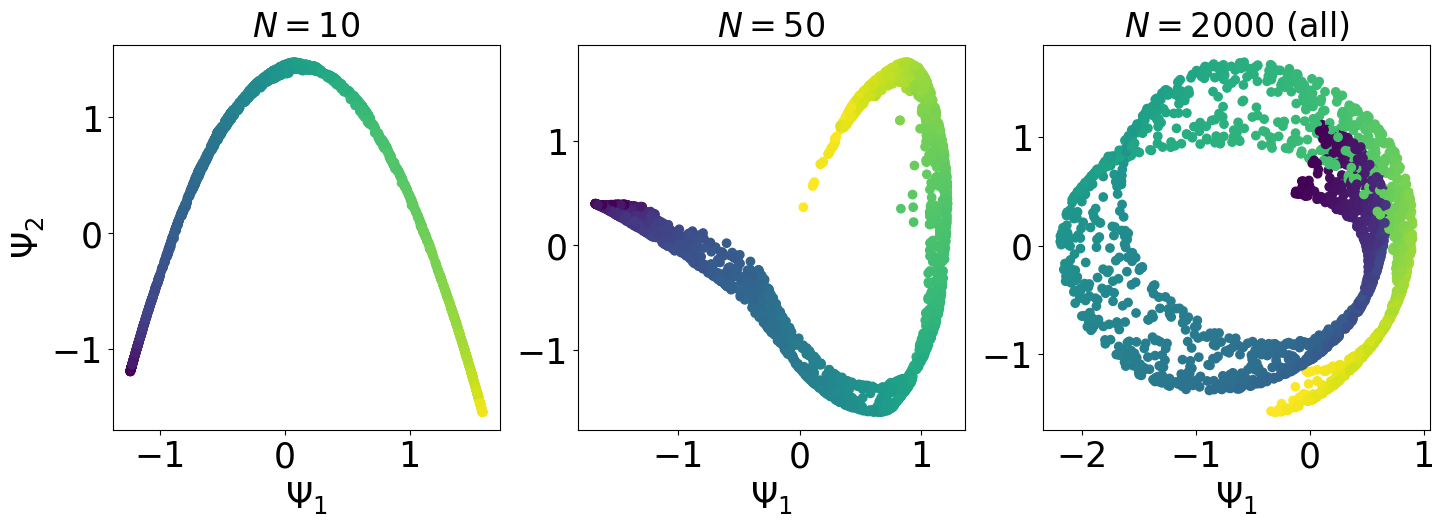}
    \caption{\textbf{Illustrating the effect of the number of considered nearest neighbors $N$} using the Swiss roll dataset shown in figure \ref{fig:swiss_roll_original_color}. $\epsilon =1$ and $t=1$ were used as \textit{Diffusion Map} parameters. The dataset was standardized beforehand.} 
    \label{fig:N_swissroll}
\end{figure}

Another parameter that can be used to establish locality is the number $N$ of considered nearest neighbors. As we can see in figure \ref{fig:N_swissroll}, this method can also be used to reduce and define the neighborhood and also to reduce calculation times, as discussed in section \ref{adding_locality}. When using both parameters, it should be noted that they can influence each other and do not only reduce the calculation time, but also change the structure of the \textit{Diffusion Map}. We can see this for our simple example in figure \ref{fig:N_swissroll}: if $\epsilon$ is chosen too large to properly unroll the manifold, a small $N$ (e.g. $N=10$) can still yield the expected result and therefore have a significant impact on the appearance of the \textit{Diffusion Map}. One could argue that in such cases, $N$ becomes the dominant neighborhood parameter.

\subsection{Time parameter $t$}
\label{time_parameter}

The parameter $t$ was introduced as a time parameter in fundamental works \cite{lafon2004,Nadler2005,Coifman2005}. It indicates that the transition matrix $M$ used for computing the \textit{Diffusion Map} embedding is raised to the power of $t$, thereby running forward the associated Markov process. $M^t$ corresponds to the time step $t$ of the Markov process associated with $M$.

In works such as \cite{Coifman2006,Porte2008}, it is stated that $t$ enables the exploration of the geometric structure of the original dataset at different scales and should therefore be considered a scale parameter. According to \cite{Coifman2006}, $t$, and thus the running forward of the Markov chain over time, represents one of the key ideas of the \textit{Diffusion Map} framework. This also constitutes a major difference from the Laplacian Eigenmap, and suggests that diffusion along the manifold plays a crucial role in achieving an appropriate embedding.
Moreover, $t$ is often attributed as the source of the method’s robustness to noise, as increasing $t$ effectively sums over all possible paths with $t$ timesteps, thereby smoothing out local perturbations \cite{lafon2004,Coifman2006}.\\

We will now present an opposing perspective, arguing that the parameter $t$ is often unnecessary and does not provide significant benefits for data analysis. Consequently, it is not required to run forward the Markov chain to cover the entire manifold. Instead, as described by \cite{Saul2003}, the local geometric information can be effectively linked through spectral decomposition to reveal the manifold structure, and this is always done, considering $t$ or not. The approach without using the additional parameter $t$ is also employed by many other spectral methods (see section \ref{related_methods_and_history}) and has been shown to yield meaningful and reliable results.\\

\begin{figure}[h]
    \centering
    \includegraphics[width=\linewidth]{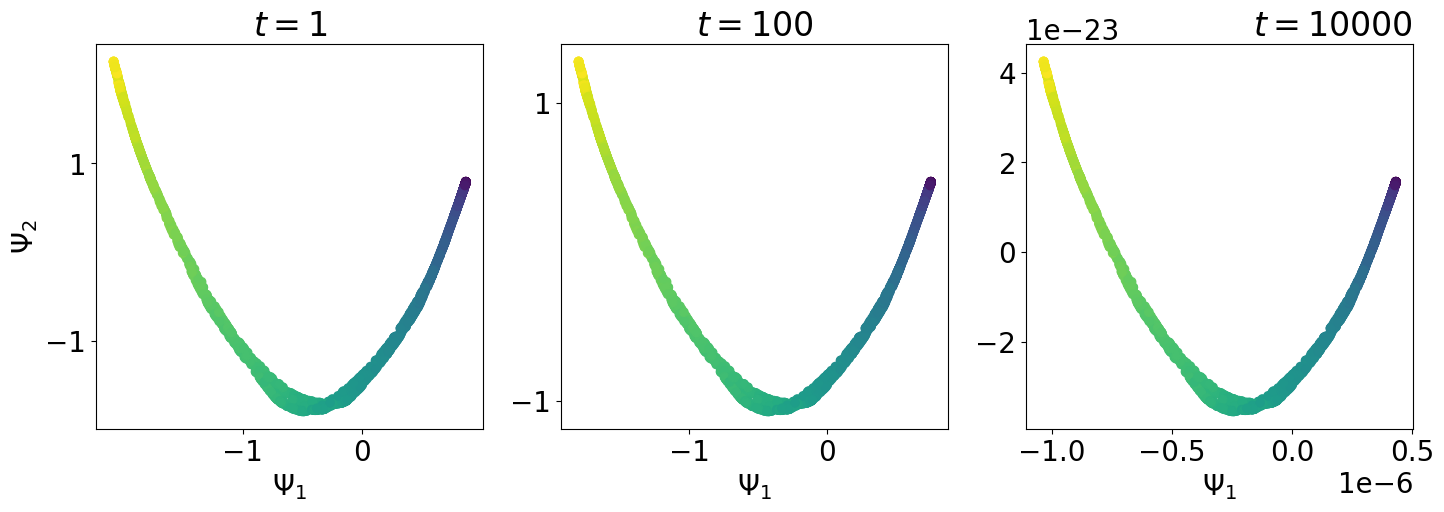}
    \caption{\textbf{Illustrating the effect of the time parameter $t$} using the Swiss roll dataset shown in figure \ref{fig:swiss_roll_original_color}. $\epsilon = 0.1$ and all nearest neighbors were used as parameters. The dataset was standardized beforehand.}
    \label{fig:influence_t_to_swissroll}
\end{figure}

To illustrate this, we consider an example of the application of $t$. In figure \ref{fig:influence_t_to_swissroll}, we can observe that the appearance of the \textit{Diffusion Map} does not change, when using different $t$. Only the axes, i.e. the diffusion components $\Psi_1$ and $\Psi_2$, are scaled to different extents. As $t$ increases, the \textit{Diffusion Map} becomes smaller, but otherwise, its structure remains unchanged.\\

This is expected considering the definition of the \textit{Diffusion Map} (equation \ref{equation:diffusion_map}). A dimension of the new embedding, i.e., the $i$th diffusion component, is given by $\Psi_i(x) = \lambda_i^t\psi_i(x)$.
There, $\psi_i$ and $\lambda_i$ are the eigenvectors and eigenvalues that result from the (symmetrized) transition matrix without considering the parameter $t$.\\

The only influence on the \textit{Diffusion Map} is thus the power to which the eigenvalue is raised. Since $\lambda_i < 1$, we can say that the eigenvectors are scaled down by $\lambda_i^t$. This means that a larger $t$ only causes the values of the diffusion components to decrease at different rates. Although this changes the distances within the \textit{Diffusion Map}, it does not change its structure. The effects of the time parameter could be important from a theoretical perspective, but it does not appear to be relevant for the qualitative analysis of the data. Hence it loses relevance when being used in the context of dimensionality reduction. Quantitative research involving more than one diffusion component seems to be difficult anyway, as the importance of the individual diffusion components in relation to each other is not clear (see discussion in section \ref{typical_shapes}). For the analysis of social science data, we conclude that the parameter $t$ is not necessary, and we use $t=1$.\\

To illustrate the effect of $t$, plots are sometimes created where the axes remain constant (e.g., \cite{Porte2008} figure 6). This visualization often leads to over-interpretation of the results, as it appears that the individual structures in the \textit{Diffusion Map} are merging into a single one. However, now we know that this is just an aspect of the perspective: if one would zoom in on this structure, the original structure would still be present.

\subsection{The influence of data normalization and scaling}
\label{normalization}

\begin{figure}[H]
\centering
\subfigure[Showing effect of no standardization and rescaling for the \textit{Diffusion Map} embedding. Comparing results with standardized results from figure \ref{fig:dmap_epsilon_swissroll} reveals huge differences in the optimal value of $\epsilon$.]{
   \centering
   \label{fig:swiss_roll_normalization}
   \includegraphics[width=0.75\linewidth]{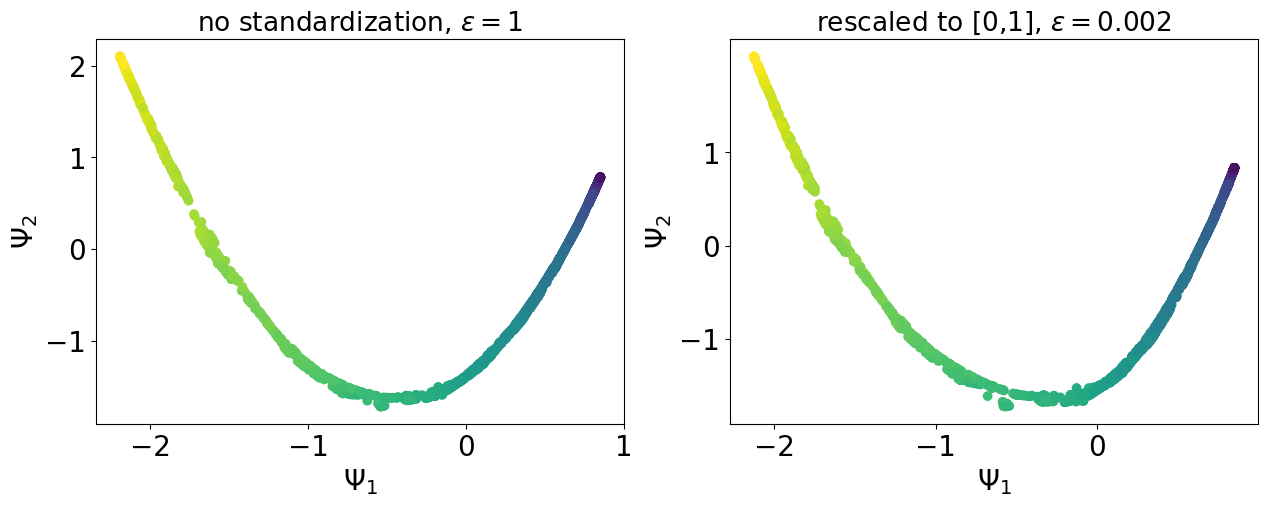}}\\
\subfigure[Showing the effect of rescaling the $Y$-axis and therefore the width of the Swiss roll. On the left the color code indicates the arc length along the Swiss roll. On the right it indicates the width of the Swiss roll in $Y$ direction. $\epsilon=0.1$ was used for all datasets.]{
   \centering
   \label{fig:swiss_roll_preprocessing_change_size_coordinate}
   \includegraphics[width=0.8\linewidth]{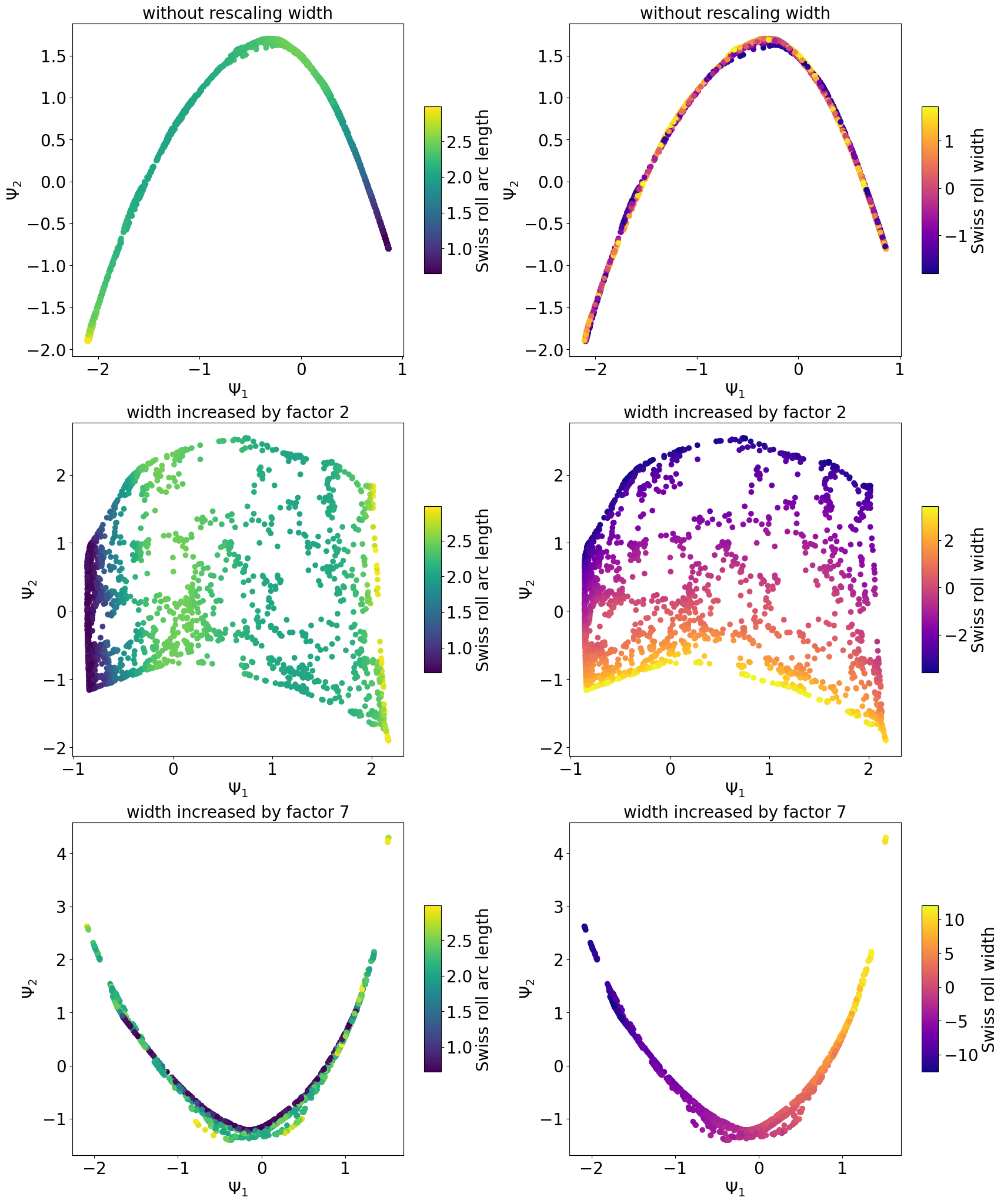}}
   \caption{\textbf{Illustrating the effect of preprocessing and normalization of the data} using the Swiss roll dataset shown in figure \ref{fig:swiss_roll_original_color}. All nearest neighbors and $t=1$ were used as parameters.}
   \label{fig:influence_normalization_to_swissroll}
\end{figure}

How the data is preprocessed has a significant influence on the appearance of the \textit{Diffusion Map}. We will now examine the impact of preprocessing steps, such as normalization, on the results of the algorithm. Comparing figure \ref{fig:swiss_roll_normalization} with figure \ref{fig:dmap_epsilon_swissroll}, we can already observe that different normalizations of the Swiss roll dataset lead to different optimal values of $\epsilon$. For instance, when standardizing the data, i.e. that each variable is adjusted to have a mean of zero and a standard deviation of one, (like in figure \ref{fig:dmap_epsilon_swissroll}) an optimal $\epsilon$ of around $0.1$ yields the expected parabolic shape. Without standardization and using the original scale from figure \ref{fig:swiss_roll_original_color}, the optimal value is around $\epsilon=1$, whereas rescaling all variables to the interval $[0,1]$ requires an $\epsilon$ of approximately $0.002$. This is not surprising, as every normalization alters the distances between data points. Since the \textit{Diffusion Map} algorithm fundamentally relies on these distances, the optimal neighborhood parameter $\epsilon$ changes accordingly.\\

However, such transformations can also lead to drastic changes in the range of some variables while leaving others unaffected. As a result, the importance of certain variables in the \textit{Diffusion Map} outcome may shift significantly. If one variable is disproportionately upscaled compared to others, it will have a greater influence on the pairwise distances between data points. We will investigate this effect using the Swiss roll dataset: If we increase the width of the Swiss roll, meaning we scale the $Y$ coordinate by a factor $>1$, the width becomes a more dominant feature and is no longer captured only in the fifth diffusion component, as discussed in section \ref{swiss_roll}.

For this purpose, we have shown the diffusion map for different width scalings in figure \ref{fig:swiss_roll_preprocessing_change_size_coordinate}. For the original scaling, we obtain the result shown in both plots at the top of the figure. Note that in this original case, the arc length is greater than the width and therefore more significant (as it can be seen with the color bar).

An interesting observation emerges when the width of the Swiss roll becomes comparable in scale to the arc length. In this case, the \textit{Diffusion Map} captures the two-dimensional structure of the manifold within the first two diffusion components $\Psi_1$ and $\Psi_2$ (see  figure \ref{fig:swiss_roll_preprocessing_change_size_coordinate}, middle) \cite{Nadler2007}. 
If the width is further increased to become significantly larger (the plots at the bottom of the figure) than the arc length, then $\Psi_1$  captures only the width parameter, while $\Psi_2$ again reflects the polynomial relationship. We can conclude: if one intrinsic dimension is significantly larger than the other, then the \textit{Diffusion Map} primarily represents this dominant dimension in the first two components and exhibits the typical polynomial behavior. \\

These findings highlight the need for caution when applying \textit{Diffusion Maps} to social science datasets. Adjusting the scaling of features can make certain dimensions of the dataset more or less influential for the resulting embedding. If a dataset contains features with particularly large ranges, one must consider whether rescaling is necessary or if the large range of the variable indicates that it carries substantial meaningful information. Additionally, it may be useful to reduce the range of less important variables to minimize their impact on the final results.

\subsection{Influence of redundant variables}
\label{swiss_roll_redundant_information}

In various real-world datasets, the same information can be stored in multiple variables. The presence of nearly redundant variables in a dataset depends on the data collection and selection process, as well as the degree to which these redundancies are identified and removed in advance. This issue is particularly relevant for datasets such as census data, which we will explore in a concrete example later in section \ref{census_britain_section}.\\

In this section, we aim to examine the potential effects of redundant variables. Understanding these effects will help us assess their impact on data analysis. Also, we present a possible method to remove linear redundancies. Therefore, we will focus on a specific case: the presence of equivalent variables.

For this, we use the Swiss roll dataset, as shown in figure \ref{fig:swiss_roll_first_analysis}, where we can also see the expected results of the \textit{Diffusion Map} for this dataset. Now, we will duplicate the width dimension of the Swiss roll (the Y-dimension) and add Gaussian noise. As a result, the new dataset will have a dimension of four for one redundant variable, a dimension of six for three redundant variables, and a dimension of 13 for ten redundant variables. \\

\begin{figure}[H]
\centering
   \includegraphics[width=1.05\linewidth]{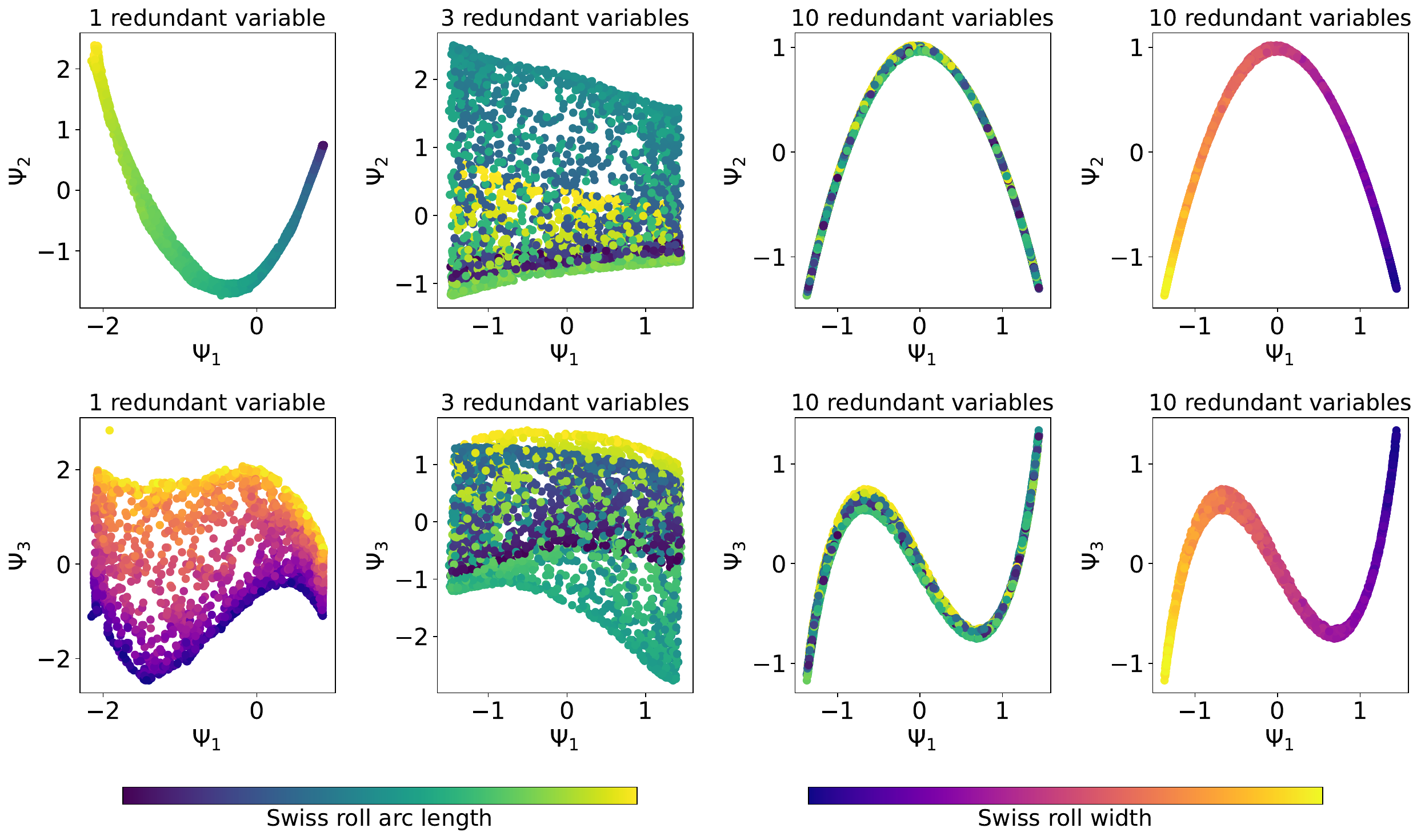}
   \caption{\textbf{Illustrating possible effects of redundant variables:} The original dataset is a Swiss roll, created by using the \texttt{make\_swiss\_roll} method from \cite{scikit-learn} with the parameters \texttt{n\_samples=3000, noise=0.01}, to which further redundant dimensions/variables have been added. These added dimensions are copies of the width dimension of the Swiss Roll (which is the Y-dimension comparing to figure \ref{fig:swiss_roll_first_analysis}) and Gaussian noise with a standard deviation of $1/10$ of the Swiss roll width was added. For the diffusion map parameters, all neighbors were considered, with $t=1$, and $\epsilon$ was set to twice the value obtained from equation \ref{equation: find_optimal_epsilon}. The duplicated (width-) dimension appears to have an increased influence on the diffusion map result. This effect becomes more pronounced as the number of redundant dimensions in the dataset increases (from 1 to 10).}
    \label{fig:swiss_roll_redundancies}
\end{figure}

The results of the \textit{Diffusion Map} for these new datasets can be seen in figure \ref{fig:swiss_roll_redundancies}. It shows that the width dimension of the Swiss roll becomes more important in the \textit{Diffusion Map}. This is because the repeated dimension increasingly influences the distance calculations between the data points. Structures that primarily lie in the width dimension become more significant in the context of the algorithm. The effect is comparable to the rescaling of individual dimensions, as demonstrated in figure \ref{fig:swiss_roll_preprocessing_change_size_coordinate}, where one dimension also gains significance. \\

We can conclude that when analyzing \textit{Diffusion Maps}, it is essential to carefully select the variables and to consider whether redundancies are intended or should be removed.

Another question is whether there is a way to remove these redundancies. In this simple example, this is possible by first eliminating linear redundancies using PCA. Since PCA combines linearly correlated variables into fewer, uncorrelated dimensions, it effectively reduces redundancy. If we then use the computed principal components as input for the \textit{Diffusion Map}, we can eliminate the effect of the redundancies and obtain a \textit{Diffusion Map} that no longer includes redundant variables.\\

\begin{figure}[H]
\centering
\subfigure[Eigenspectra of the PCA (crosses) and the \textit{Diffusion Map} (dots) of the test-dataset. For PCA, the explained variance is shown on a logarithmic scale.]{
   \label{swiss_roll_PCA_spectrum} 
   \centering
   \includegraphics[width=0.6\linewidth]{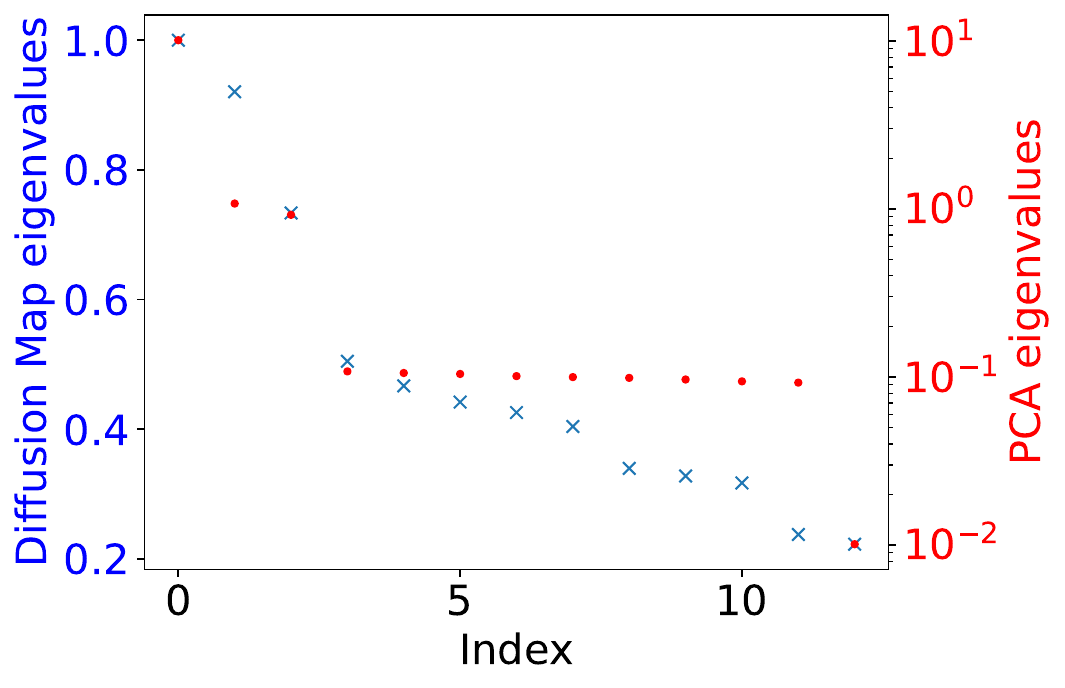}}\\
\subfigure[\textit{Diffusion Map} applied to the first three principal components obtained from the PCA of the test dataset.]{
   \centering
   \label{swiss_roll_PCA_restored} 
   \includegraphics[width=1\linewidth]{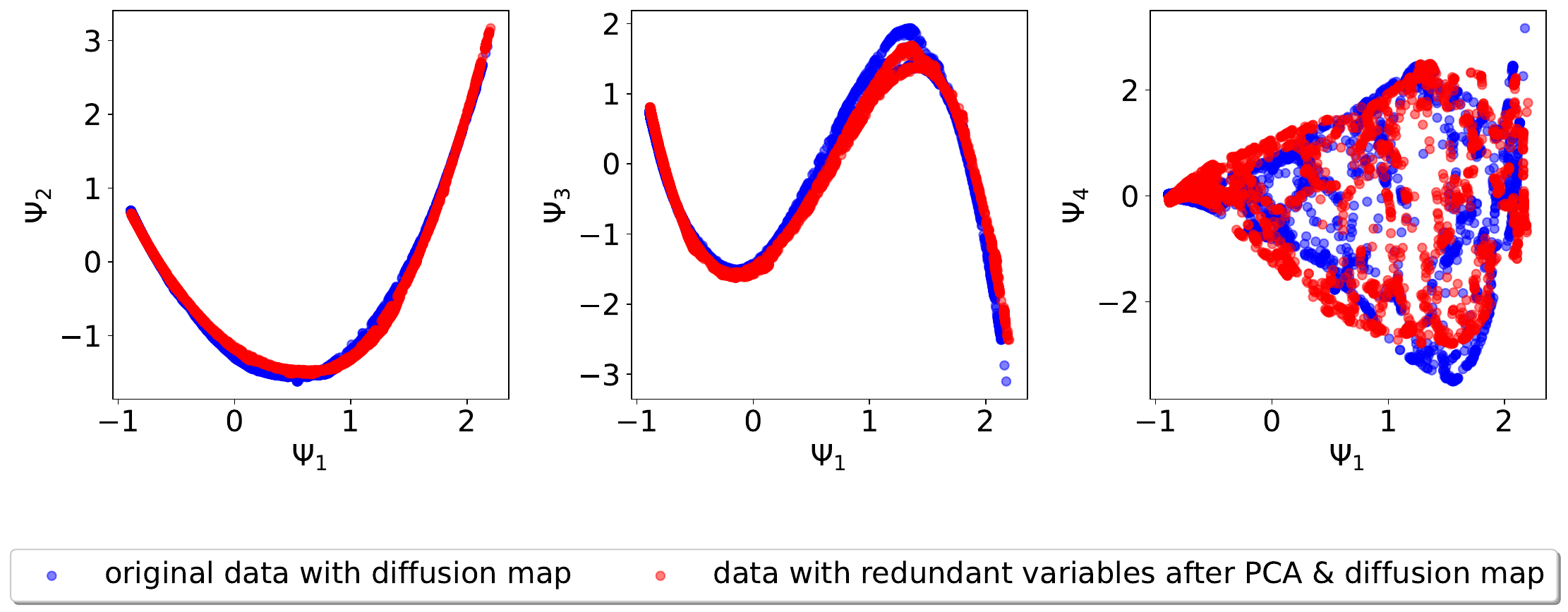}}
   \caption{\textbf{Using PCA to delete linear redundancies.} A 13-dimensional test-dataset containing the Swiss roll and ten copies of the width coordinate of the Swiss roll was used. Gaussian noise with a standard deviation of $1/10$ of the Swiss roll width was added to the ten redundant dimensions. PCA is able to eliminate the redundancies and the original \textit{Diffusion Map} result is restored.}
    \label{fig:swiss_roll_redundancies_PCA}
\end{figure}

To test this idea, we take our 13-dimensional dataset of the Swiss roll and 10 copies of the Y-coordinate. In this example, the size of the first eigenvalues of the variance matrix (figure \ref{swiss_roll_PCA_spectrum}) indicates that only three principal components are actually necessary to explain the variance of the dataset. (For illustrative purposes, the diffusion map spectrum of the same dataset is plotted to highlight the different underlying mechanisms. Note the different scales involved.)

If we now take the first three resulting principal components and compute the \textit{Diffusion Map} based on them, we obtain a result that closely resembles the outcome we achieved without redundant dimensions (see figure \ref{swiss_roll_PCA_restored}). The remaining noticeable deviations, especially in the third diffusion component, are a result of the Gaussian noise. This procedure also works for the other two original dimensions.\\

Nevertheless, it remains an open question whether this approach also performs well for more complex datasets.

\section{Sorting ability}
\label{sorting_ability}

A widely used example to illustrate the capabilities of dimensional reduction methods is the ability to organize images. The images are the data points, with each pixel of an image representing a single dimension of the feature space. An image with $l \times w$ pixels is therefore a data point in a $lw$ dimensional feature space. These images can then be displayed in a one- or two-dimensional space through dimensionality reduction. This has been demonstrated, for instance, in \cite{roweis2000, Tenebaum2000, lafon2004, Nadler2007}, as well as in various theses \cite{Bah2008, Roozemond2021}. In these works, various sets of images containing faces or objects are used and the respective spectral method is able to organize them according to their natural parameters such as rotation angle or facial expressions. \\

In this section, we aim to examine the underlying mechanisms that enable this sorting capability and identify the properties that the original data must possess to allow such a sorting. Through this analysis, we seek to determine potential limitations of the \textit{Diffusion Map} and its applicability to social science data. \\

\begin{figure}[H]
\centering
\subfigure[Dataset of $36$ data points/images with $300\times300$ dimensions. Clock hands are indicating the time. The dataset includes times ranging from 3:00 o'clock to 5:55 o'clock in 5-minute intervals.]{
   \centering
   \includegraphics[width=\linewidth]{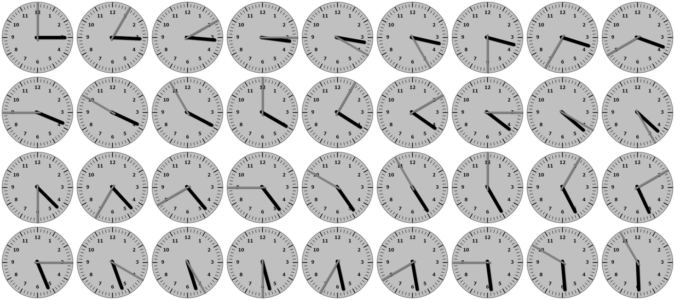}}\\
\subfigure[\textit{Diffusion Map} with parameters $\epsilon = 100$, using all nearest neighbors, $t=1$.]{
   \centering
   \includegraphics[width=0.6\linewidth]{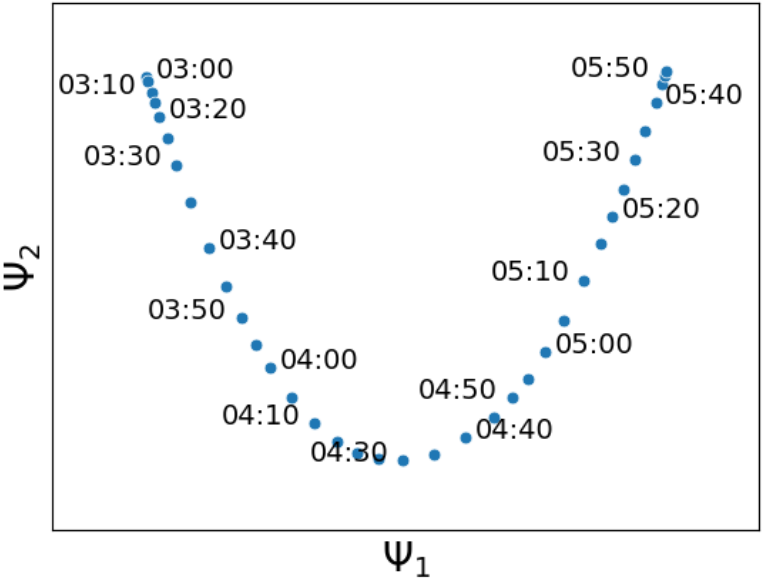}}
   \caption{\textbf{Illustrating sorting ability:} dataset consisting of images of clocks is organized according to the time using the first diffusion component $\Psi_1$.}
    \label{fig:normal_clock}
\end{figure}

For the analysis, we will use various images of analog clocks created using \cite{clock}. These are images with a dimension of $300\times300$ pixels.
First, we use a dataset (figure \ref{fig:normal_clock} (a)) consisting of $36$ images of clock faces displaying times from 3:00 o'clock to 5:55 o'clock. The \textit{Diffusion Map} is capable of organizing the images according to their natural parameter, namely the time or, equivalently, the angles of the clock hands. 

The second diffusion component shows the typical parabolic behavior (see figure \ref{fig:normal_clock}(b)), which we already know from section \ref{typical_shapes}. This suggests that the original dataset is likely a one-dimensional, non-intersecting curve embedded in $300\times300$ dimensions.

\begin{figure}[H]
\centering
\subfigure[Dataset of $32$ data points/images with $300\times300$ dimensions. For the hours 2,5,7 and 10 there are 8 data points with minutes ranging from 0 to 21.]{
   \centering
   \label{fig:clock_gaps_dataset}
   \includegraphics[width=0.9\linewidth]{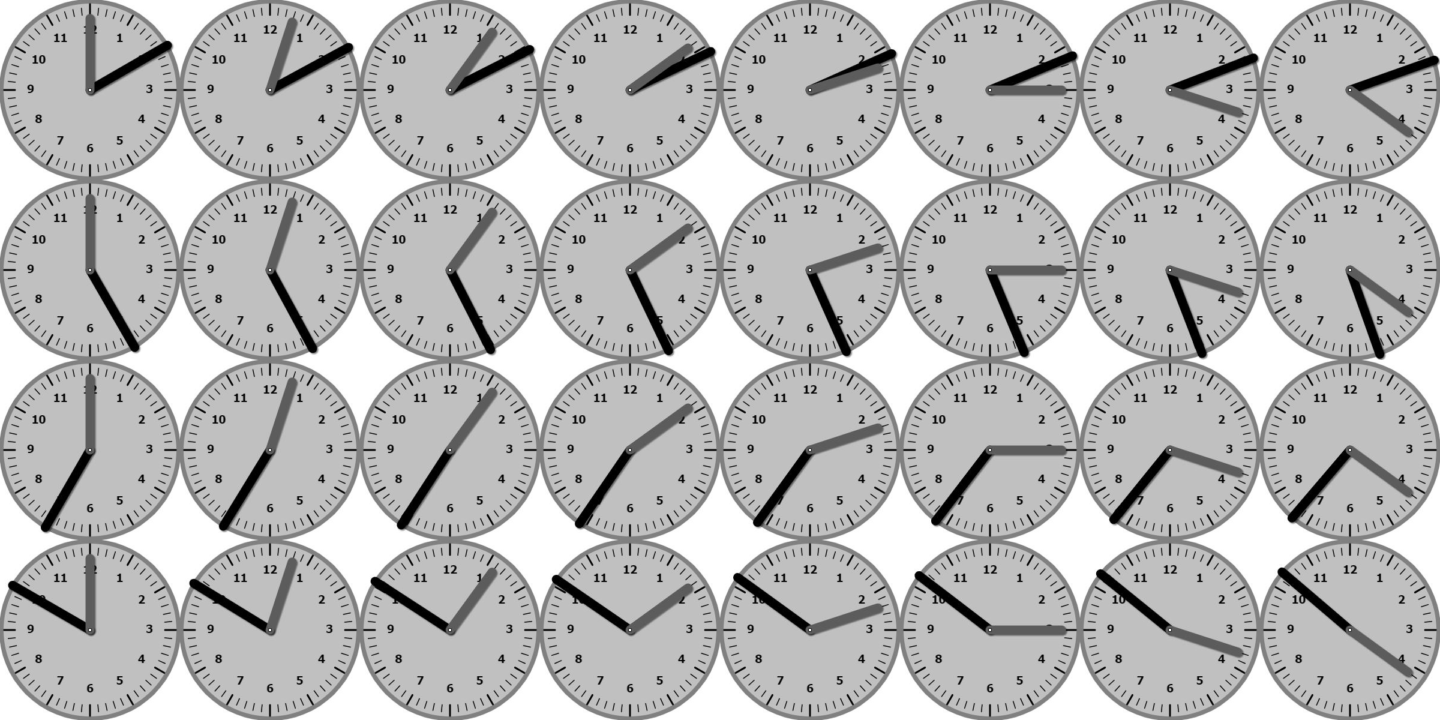}}\\
\subfigure[\textit{Diffusion Map} with parameters $\epsilon = 100$, using all nearest neighbors, $t=1$]{
   \centering
   \label{fig:clock_gaps_diffmap}
   \includegraphics[width=0.6\linewidth]{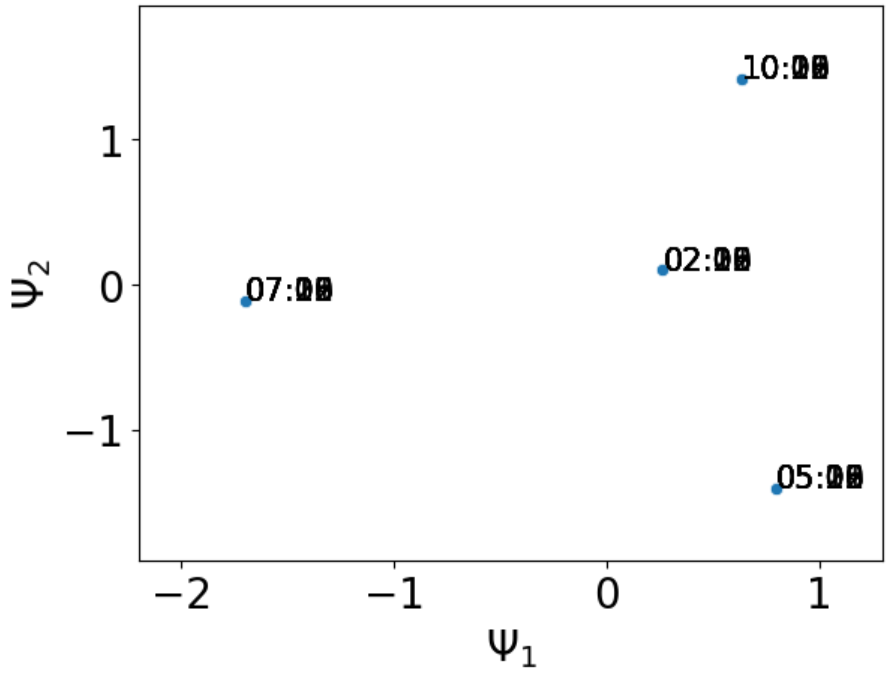}}~
\subfigure[Zooming into one cluster of the \textit{Diffusion Map} shown in \ref{fig:clock_gaps_diffmap}]{
   \centering
   \label{fig:clock_gaps_diffmap_zoom}
   \includegraphics[width=0.35\linewidth]{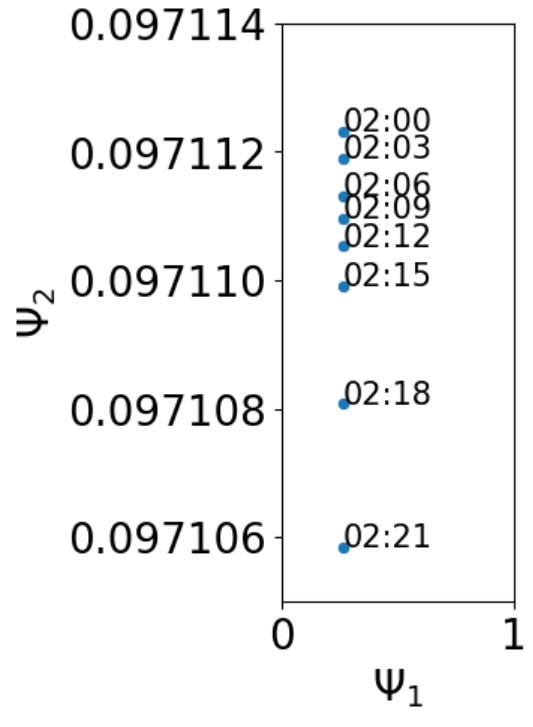}}
   \caption{\textbf{Visualizing clustering and sorting ability inside of clusters:} the most similar points that belong to one hour are organized into clusters. Within the clusters, there is a meaningful sorting of the data points. Clusters do not appear to be meaningfully sorted against each other.}
   \label{fig:clock_gaps}
\end{figure}

But why is the \textit{Diffusion Map} able to organize the images? To explore this, we examine a second dataset that contains eight images for each of the hours 2, 5, 7, and 10 o'clock, covering minutes 0 to 21 (see figure \ref{fig:clock_gaps_dataset}). In this case, we have large time gaps between the hours and smaller gaps within each hour.
For this dataset, we observe that the \textit{Diffusion Map} is no longer able to correctly sort the hours (see figure \ref{fig:clock_gaps_diffmap}). However, the \textit{Diffusion Map} demonstrates its clustering capability, discussed in section \ref{clustering}, grouping the data points corresponding to each hour into distinct clusters. The clusters seem to be arranged randomly. 
However, on the smaller scale, if we zoom into one cluster, the \textit{Diffusion Map} still appears to organize the times relatively well (see figure \ref{fig:clock_gaps_diffmap_zoom}).\\

The sorting capability does not arise from the \textit{Diffusion Maps} ability to infer the angle of the hour hand, but rather requires that data points have to be very similar to each other.

It is helpful to examine what we know how the \textit{Diffusion Map} assigns data points to each other. \cite{lafon2004} says that the euclidean distance can only relate pictures which are very similar, which is the reason we use a local kernel. This leads to the conclusion, that the parameter (in this example the angle of the clock hand) has to be sufficiently sampled, so that the agglomerated local information of the euclidean distances reveals the global structure of the dataset \cite{lafon2004}.\\

Within an hour, the local information appears to be sufficiently sampled, but between the hours, it is not. The question that remains is why the data is sometimes sufficiently sampled and at other times not.\\

\begin{figure}[H]
\centering
\subfigure[Dataset of $13$ data points/images with $300\times300$ dimensions/pixels. Squares of the size $12 \times 12$ dimensions/pixels are indicating the time. These squares overlap with the ones of similar times]{
   \centering
   \label{fig:connected_data}
   \includegraphics[width=0.39\linewidth]{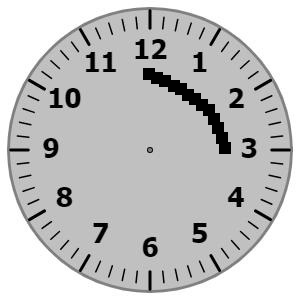}}~~
\subfigure[\textit{Diffusion Map} with parameters $\epsilon = 5$, using all nearest neighbors, $t=1$ from \ref{fig:connected_data}. $\Psi_1$ sorts the data points meaningfully according to their time.]{
   \centering
   \label{fig:connected_dmap}
   \includegraphics[width=0.55\linewidth]{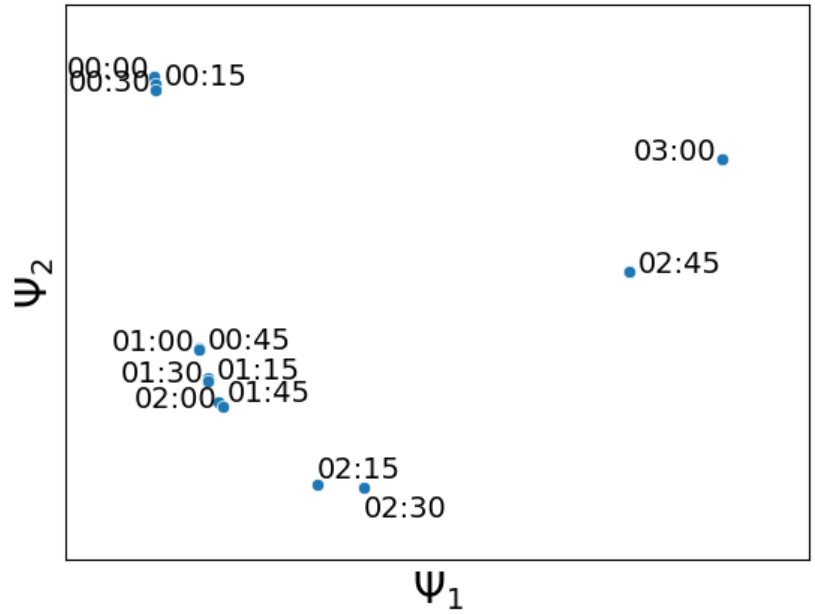}}\\
\subfigure[Dataset of $7$ data points/images. Squares of the same size as in (a) have no overlap with data points with similar times.]{
   \centering
   \label{fig:not_connected_data}
   \includegraphics[width=0.39\linewidth]{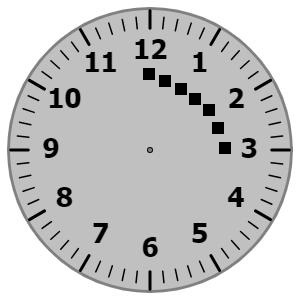}}~~
\subfigure[\textit{Diffusion Map} with parameters $\epsilon = 5$, using all nearest neighbors, $t=1$ from \ref{fig:not_connected_data}. The results are similar for different epsilons tested. The data points are not sorted by time in a meaningful way.]{
   \centering
   \label{fig:not_connected_dmap}
   \includegraphics[width=0.55\linewidth]{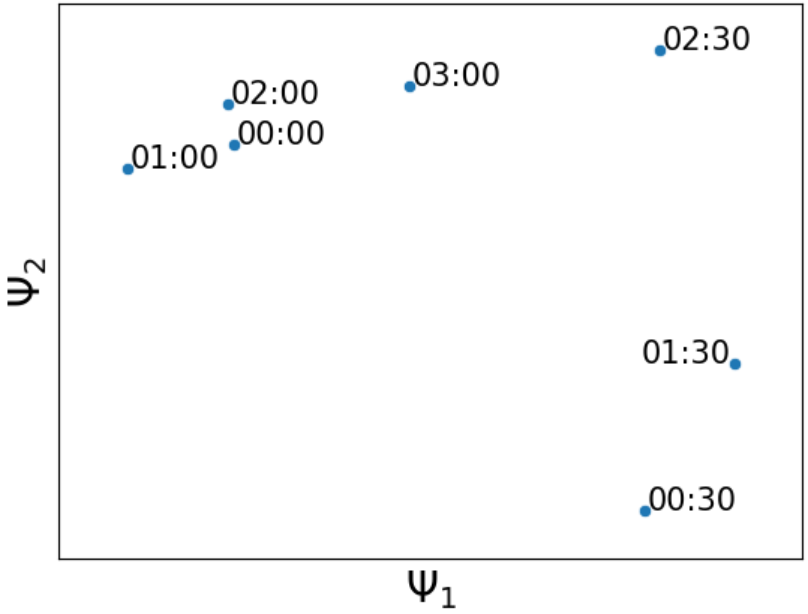}}
   \caption{\textbf{Revealing limits of the sorting ability:}   if the clock hands overlap (see (a)), the data points from the \textit{Diffusion Map} are arranged meaningfully as shown in (b). If the data points no longer have any similarity because the clock hands (in form of the $12 \times 12$ squares) do not overlap (see (c)), then the \textit{Diffusion Map} can no longer sort the data points/images correctly according to the natural parameter time as shown in (d).}
\end{figure}

Examining the next dataset will provide further clarity. In figures \ref{fig:connected_data} and \ref{fig:connected_dmap}, we present a dataset consisting of 13 data points, where each time is now represented by a single square-hand of size $12 \times 12$ pixels. In figure \ref{fig:connected_data}, we overlay the squares from the individual images. We observe that these overlap, meaning that several pixels remain the same across neighboring points, so they are similar to each other.
For this dataset, we see in figure \ref{fig:connected_dmap} that the \textit{Diffusion Map} is still able to sort the times correctly, as the first diffusion component effectively arranges the data points.

However, if we remove some of the data points from the dataset so that the point-hands no longer overlap (see figure \ref{fig:not_connected_data}), the \textit{Diffusion Map} loses its ability to sort this modified dataset as well (see figure \ref{fig:not_connected_dmap}).

In conclusion, we have learned that similar images or data points are positioned next to each other in the \textit{Diffusion Map}. However, the \textit{Diffusion Map} does not magically recognize orientation or other natural parameters of the dataset. 

If two points are very dissimilar and there are no similar points between them, then it is no longer possible to sort these points in a meaningful way (comparable with figure \ref{fig:not_connected_dmap}). Thus, it seems that there must be a sufficient number of similar points between the two dissimilar points to form a "similarity chain" from one to the other, enabling a meaningful organization of these dissimilar points within the embedding. 

Based on the discussed example of the clocks and in agreement with examples of moving and rotating objects from the literature, we can conclude that probably these examples only work because the considered objects undergo slight movements and possess a certain width, such that two images/data points are sufficiently similar. This width leads to overlaps of various pixels that retain the same values across two images and create similarity. The sufficiently small movements then establish the similarity chain.

The same also applies to the distance and location of different clusters to each other. They must be considered carefully, since these do not seem to be arranged meaningfully in terms of natural parameters (see figure \ref{fig:clock_gaps_diffmap} - data points of different hours are not sorted in a meaningful way). 

We therefore conclude that when analyzing social science datasets, we have to be careful when we see areas where there are too few similar data points. Here, we have to critically determine whether the organization of the data in the diffusion space is useful for the analysis and whether there are not enough data points for a 'similarity chain'.

\section{Clustering ability}
\label{clustering}

\paragraph*{Origin of the ability}~\\
Clustering is the task of grouping data points into clusters based on their similarity. The goal is to ensure that data points within a cluster are more similar to each other than to other data points.
There are publications stating that spectral dimensionality reduction methods emphasize the clustering of data \cite{Belkin2003, Meila2024}. In the case of the \textit{Diffusion Map}, this property can be explained by the concept of the diffusion distance (equation \ref{equation:diffusion_distance}), which ensures that highly connected (very similar) data points are mapped close to each other in one cluster, while barely or not connected points are placed further apart in different clusters.\\

Indeed, when embedding the data, one can observe that a smooth embedding occurs in regions with many neighbors on the manifold, while clustering emerges when accumulations of data points are not or only slightly connected \cite{Meila2024}. This effect, where smooth embedding occurs simultaneously with clustering, can also be observed in our clock example from figure \ref{fig:clock_gaps}. The data points corresponding to different hours are not connected, resulting in the formation of four distinct clusters. However, within each cluster, the data points are well connected, leading to a smooth embedding.

\paragraph*{Spectral clustering}~\\
The clustering capability is also utilized by more advanced methods: Thus, spectral clustering is a method that combines spectral embedding with k-means clustering \cite{Ng2001}. In section \ref{related_methods_and_history}, we have already discussed that the \textit{Diffusion Map} provides a probabilistic interpretation of spectral embedding \cite{Nadler2007} and that applying k-means clustering in the diffusion space is equivalent to spectral clustering. Spectral clustering has been shown to outperform traditional clustering methods, such as applying k-means directly to the original data \cite{Luxburg2007}. 

\begin{figure}[H]
\centering
\subfigure[Original dataset of three rings.]{
   \centering
   \label{fig:rings_data}
   \includegraphics[width=0.3\linewidth]{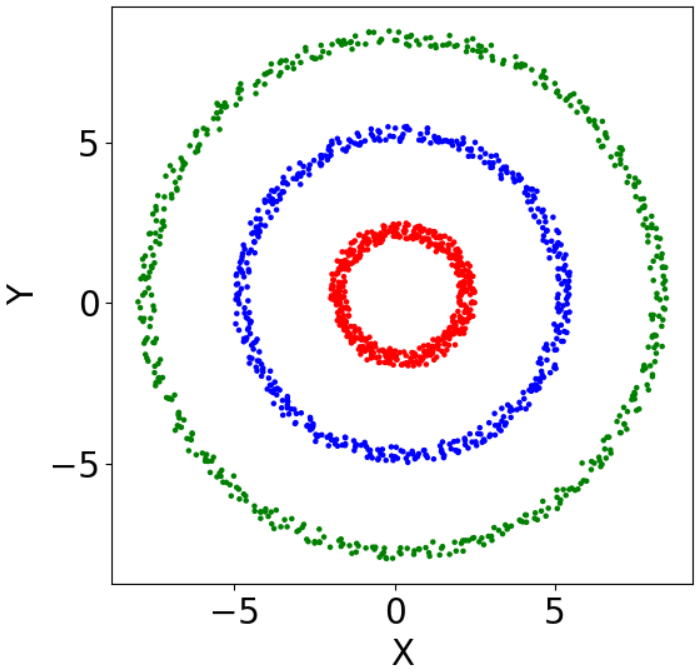}}~
\subfigure[Histogram of the data according to the first diffusion component $\Psi_1$. The \textit{Diffusion Map} parameters are $\epsilon = 0.01$, using all nearest neighbors and $t=1$.]{
   \centering
   \label{fig:rings_diffmap}
   \includegraphics[width=0.39\linewidth]{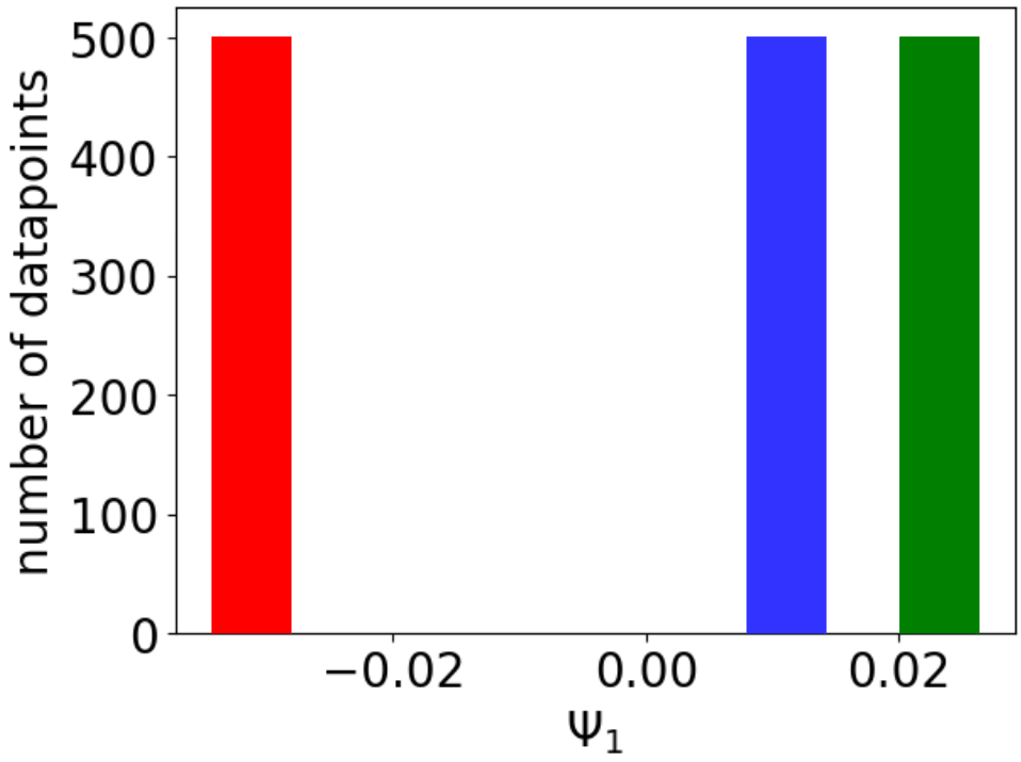}}\\
\subfigure[Transition matrix for $t=1$. There is only a transition probability to neighboring points.]{
   \centering
   \label{fig:rings_t1}
   \includegraphics[width=0.31\linewidth]{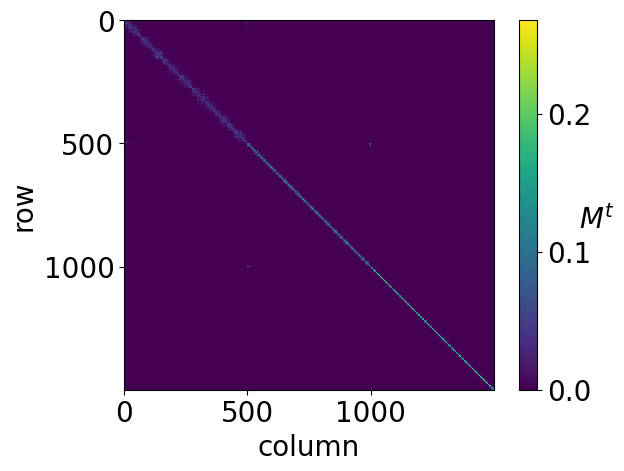}}~~
\subfigure[Transition matrix for $t=1000$. It can be observed that for the smallest ring, the transition probability becomes uniform more quickly across the entire ring, whereas for the larger rings, it takes longer to get from one side of the ring to the other.]{
   \centering
   \label{fig:rings_t1000}
   \includegraphics[width=0.31\linewidth]{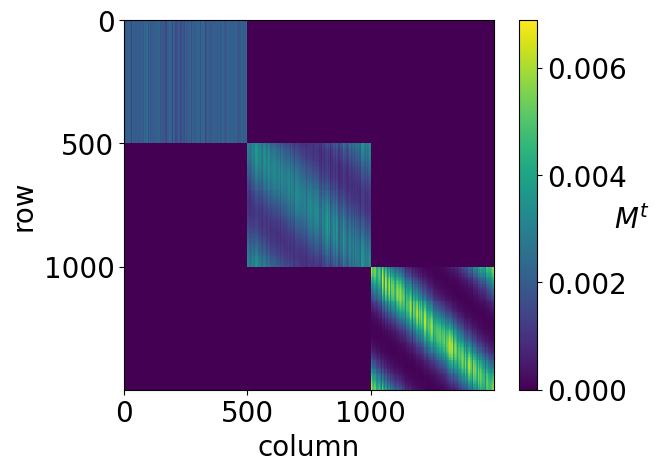}}~~
\subfigure[Transition matrix for $t=10000$. The transition probability becomes uniform.]{
   \centering
   \label{fig:rings_t10000}
   \includegraphics[width=0.31\linewidth]{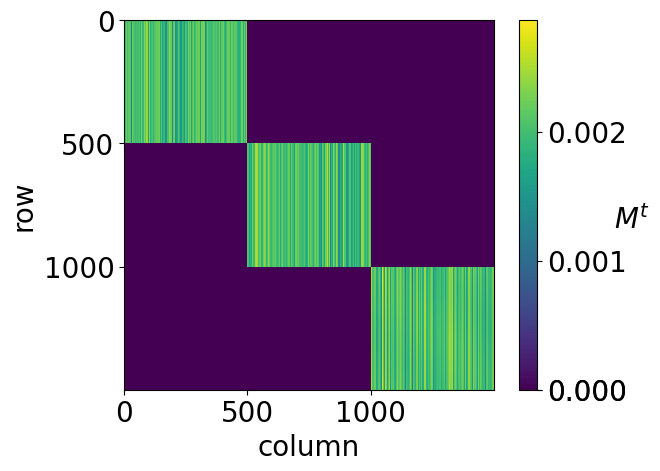}}
   \caption{\textbf{Example for the clustering ability of the \textit{Diffusion Map}:} The Data consist of $3$ rings with $500$ data points each. The \textit{Diffusion Map} is able to cluster the data points of each ring by the first principal component $\Psi_1$. This is independent from the chosen parameter $t$.} 
\end{figure}

\paragraph*{Example and influence of $t$}~\\
Let us consider one example: in figure \ref{fig:rings_data} we see a non-linear dataset with $3$ rings that cannot be properly clustered using k-means alone \cite{comparison_scikit}. However, when we examine the first diffusion component $\Psi_1$ in figure \ref{fig:rings_diffmap}, we can already distinguish the individual rings as separate clusters, because all data points of one ring get the same value in $\Psi_1$.\\

Referring to the effect of the time parameter $t$ discussed in section \ref{time_parameter}, figures \ref{fig:rings_t1}, \ref{fig:rings_t1000} and \ref{fig:rings_t10000} display the transition matrices $M^t$ for different values of $t$ for the ring dataset. We observe that for larger $t$, the probability of jumping from one point to another within the same cluster increases, which leads to the formation of three distinct blocks of higher jumping probability in the matrix, each corresponding to one ring.
This further illustrates that the time parameter $t$ is not necessary to obtain the correct \textit{Diffusion Map} result, as the embedding was computed with $t = 1$. Moreover, it is not required that, as in the case of $t = 10000$, all data points within a cluster exhibit high transition probabilities among themselves. It suffices, as seen for $t = 1$, that the similarity is preserved only between directly neighboring data points. For $t = 1000$ and $t = 10000$, the effect on $\Psi_1$ is only a rescaling, which also contradicts the descriptions for example in \cite{Porte2008,Roozemond2021}, where the time parameter $t$ is assigned a significant role also in clustering.

%% file: Chapter/41_vdem.tex
\section{V-Dem democracy dataset: electoral democracies an one-dimensional concept?}
\label{vdem_section}

The first social science dataset is describing democracies. Understanding democracies and the processes through which individuals and different interest groups interact within this complex system to shape society is becoming increasingly important. Issues such as democratic backsliding, populism, polarization, and the potential collapse of democratic systems are relevant questions in academia and politics.

We aim to investigate whether the \textit{Diffusion Map} can reveal an intrinsic low-dimensional representation of democracy. To do so, we first examine the nature of the dataset in sections \ref{vdem_the_dataset} and \ref{vdem_motivation_of_the_analysis}, and identify the appropriate parameters in sections \ref{vdem_PCA}. In section \ref{vdem_interpretation}, we attempt to understand the embedding of the democracy data and explain various distinct structures using the different democracy variables and their particular characteristics. Finally, in section \ref{vdem_one_dimensional_curve}, we explore whether the data can be described by a one-dimensional manifold, as seen in section \ref{typical_shapes}.\\

In the joint publication \cite{Pirker-Diaz2024}, the \textit{Diffusion Map} of the same dataset was analyzed, but with a different research focus: examining the diffusive behavior of countries inside the diffusion space.

\subsection{The dataset}
\label{vdem_the_dataset}

The used data on the quality of democracy in various countries is taken from the "Varieties of Democracies" (V-Dem) project \cite{VDemV12}. This dataset quantifies different attributes of countries worldwide with respect to democratic standards. \\

These attributes are called indicators, which are created from surveys by experts. For each year, country, and indicator, around five independent researchers are consulted. Details on the aggregation method can be found in \cite{Pemstein2022}.

There are also low-, mid- and high-level indices, which are composed of various indices or indicators and describe increasingly abstract characteristics to democracy. Certain aggregation rules are applied to create higher indices and can be found in \cite{Coppedge2022_codebook}. Many of these low-level indices are determined using point estimates from a Bayesian factor analysis \cite{Pemstein2022}.

The high-level indices describe various key principles and definitions of democracy, such as the idea of liberal, participatory, deliberative or egalitarian democracies. The most important high-level index of this dataset is the electoral democracy index (EDI), because it enters all other high-level indices \cite{Coppedge2022_codebook}. The EDI answers the question of whether those in government are responsive to their citizens and whether there is a fair electoral contest in which those in power can be held to account and voted out of power by the sovereign people. It is based on the work of \cite{Dahl1971}, in which the quality levels and characteristics of ‘democratic’ systems is examined and in which real-world and thus incomplete examples of democracy are described as polyarchy . Since the EDI uses this definitional basis, it is also often referred to as a polyarchy index. \\   

For our analysis we want to use all subordinate indicators of the EDI as our dataset. Since the subordinate indicators of the low-level indices \texttt{v2x\_elecoff} are logical (taking values 0 or 1) and not numerical and \texttt{v2x\_elecoff} should not be interpreted as an important element of democracy but is particularly useful to aggregate higher indices after \cite{Coppedge2022_codebook}, we used the 24 indicators of the remaining low-level indices without  \texttt{v2x\_elecoff}. So the results are also comparable to the analysis of the same 24 variables with the PCA method in \cite{Wiesner2024a}. The 24 used variables can be found in table \ref{table:vdem_variables}. 

This gives us one 24-dimensional data point per country and year. We use the time period from 1900-2022 and therefore consider 12296 data points.

\begin{table}
\footnotesize
\centering
\begin{tabular}{r||l||l}
\textbf{Index name} & \textbf{Description} & \textbf{Associated low level index}                 \\
\hline 
\hline 
%
%
\texttt{v2cldiscm}              & Freedom of discussion for men & Freedom of expression and       \\
\texttt{v2cldiscw}              & Freedom of discussion for women &  alternative sources of information,    \\
\texttt{v2clacfree}             & Freedom of academic and cultural expression & v2x\_freexp\_altinf \\
v2mecenefm             & Government censorship effort – Media &    \\
\texttt{v2mecrit}               & Print/broadcast media critical &     \\
\texttt{v2merange}              & Print/broadcast media perspectives &   \\
\texttt{v2meharjrn}             & Harassment of journalists &    \\
\texttt{v2meslfcen}             & Media self-censorship &   \\
\texttt{v2mebias}               & Media bias &        \\
\hline 
\texttt{v2psbars}               & Barriers to parties & Freedom of association,                            \\
v2psparban             & Party ban &   v2x\_frassoc\_thick                                   \\
\texttt{v2psoppaut}             & Opposition parties autonomy &                     \\
\texttt{v2cseeorgs}             & Civil society entry and exit &                   \\
\texttt{v2csreprss}             & Civil society repression &                      \\
\texttt{v2elmulpar}             & Elections multiparty &                            \\
\hline 
\texttt{v2x\_suffr}             & Share of population with suffrage & Suffrage, \texttt{v2x\_suffr} \\
\hline 
\texttt{v2elembaut}             & Election management body autonomy & Clean elections, v2xel\_frefair    \\
\texttt{v2elintim}              & Election government intimidation &                \\
\texttt{v2elfrfair}             & Election free and fair &   \\
\texttt{v2elpeace}              & Election other electoral violence &               \\
\texttt{v2elembcap}             & Election management body capacity &               \\
\texttt{v2elrgstry}             & Election voter registry &                        \\
\texttt{v2elvotbuy}             & Election vote buying &                            \\
\texttt{v2elirreg}              & Election other voting irregularities &      \\
      
\end{tabular}
\caption{List of the 24 indicators used to calculate the EDI and analysed in this section. The descriptions come from \cite{Coppedge2022_codebook}.}
\label{table:vdem_variables}
\end{table}


\subsection{Motivating the analysis: exploring the nature of electoral democracies}
\label{vdem_motivation_of_the_analysis}

An important question when creating democracy indices such as the EDI is how the collected data is aggregated into one or more indices. There are various approaches that can be used, depending on the way the indices depend on the individual indicators. According to \cite{Munck2002}, these aggregation rules could be: 

\begin{itemize}
    \item Adding indicators if the indicators have the same importance and weight. 
    \item Multiplying indicators if the indicators describe necessary features.
    \item Taking the highest indicator if all indicators are sufficient. 
\end{itemize}
So far, it has been necessary for the researchers to develop an explicit theory of how the indices are related in order to adjust the rules. 

The V-Dem project also proposes a different idea of aggregation. The EDI is calculated from the average of the multiplication and addition of lower indices:
{
\scriptsize
\begin{align*}
\text{EDI} = \frac{1}{2} \cdot (\texttt{v2x\_elecoff} \cdot \texttt{v2x\_el\_frefair} \cdot \texttt{v2x\_frassoc\_thick} \cdot \texttt{v2x\_suffr} \cdot \texttt{v2x\_freexp\_altinf}) + \\
\frac{1}{2} \cdot \left( 
    \frac{1}{8} \cdot \texttt{v2x\_elecoff} + 
    \frac{1}{4} \cdot \texttt{v2x\_el\_frefair} + 
    \frac{1}{4} \cdot \texttt{v2x\_frassoc\_thick} + 
    \frac{1}{8} \cdot \texttt{v2x\_suffr} + 
    \frac{1}{4} \cdot \texttt{v2x\_freexp\_altinf} 
\right)
\end{align*}
}
The project stated that this rule is a compromise between adding and multiplying: The addition term enables a compensation across the indicators and the multiplication term gives penalties for particularly weaknesses in one single index \cite{Teorell2016,Coppedge2022_methodology}. The lower weights of the indices \texttt{v2x\_elecoff} and \texttt{v2x\_suffr} are also rough approximations. Other rules are conceivable, and overall it would be desirable if the result could be achieved without artificial aggregation rules.\\

Therefore, another idea is pursuited: If explicit relationships between the indices cannot be established, \cite{Munck2002} suggests the employing of alternative data analysis techniques, such as PCA. For this democracy dataset, this was done in \cite{Wiesner2024a}. It was shown here that principal component 1 is highly correlated with the EDI, but that principal component 2 indicates that the EDI does not sufficiently represent election capability and civil liberties. Thus, the question arises whether the EDI describes the dataset in the best possible way. \\

Since PCA mainly works well for cases in which the data have linear relationships, it makes sense to use non-linear methods like the \textit{Diffusion Map} to see whether the data lies on a non-linear manifold and whether this manifold can be used to develop alternative democracy indices with the help of the natural parameters of this manifold. Another question is: Are more than one dimension needed to represent the dataset and the idea of an electoral democracy, like it is stated in \cite{Dahl1971,Coppedge2008}? Is a single one-dimensional index, such as the EDI, sufficient? We want to address these questions with the help of the \textit{Diffusion Map}.

\subsection{Finding the right parameters}
\label{vdem_PCA}
In order to analyze the dataset using the \textit{Diffusion Map} method, we first have to find the correct parameters for the method. First of all, we have to ask ourselves which locality parameter (width of the neighborhood $\epsilon$, number of considered nearest neighbors $N$) to use and how large the neighborhood should be. In section \ref{time_parameter} we have already established that the use of the time parameter $t$ seems to have no purpose in the qualitative analysis of the \textit{Diffusion Map}. Since previous ideas for finding a suitable size of neighborhood did not work accurately (see section \ref{finding_right_epsilon}) and only give us a first good guess, we instead investigate the realizations of the \textit{Diffusion Map} with different parameters and find an appropriate size. The goal is to find a realization that tells us as much as possible about the nature of the data.   \\

\begin{figure}[H]
  \centering
   \subfigure[For different widths of the Gaussian kernels $\epsilon$ using all neighbors.]{
   \centering
   \label{fig:parameter_vdem1}
   \includegraphics[width = 0.8\textwidth]{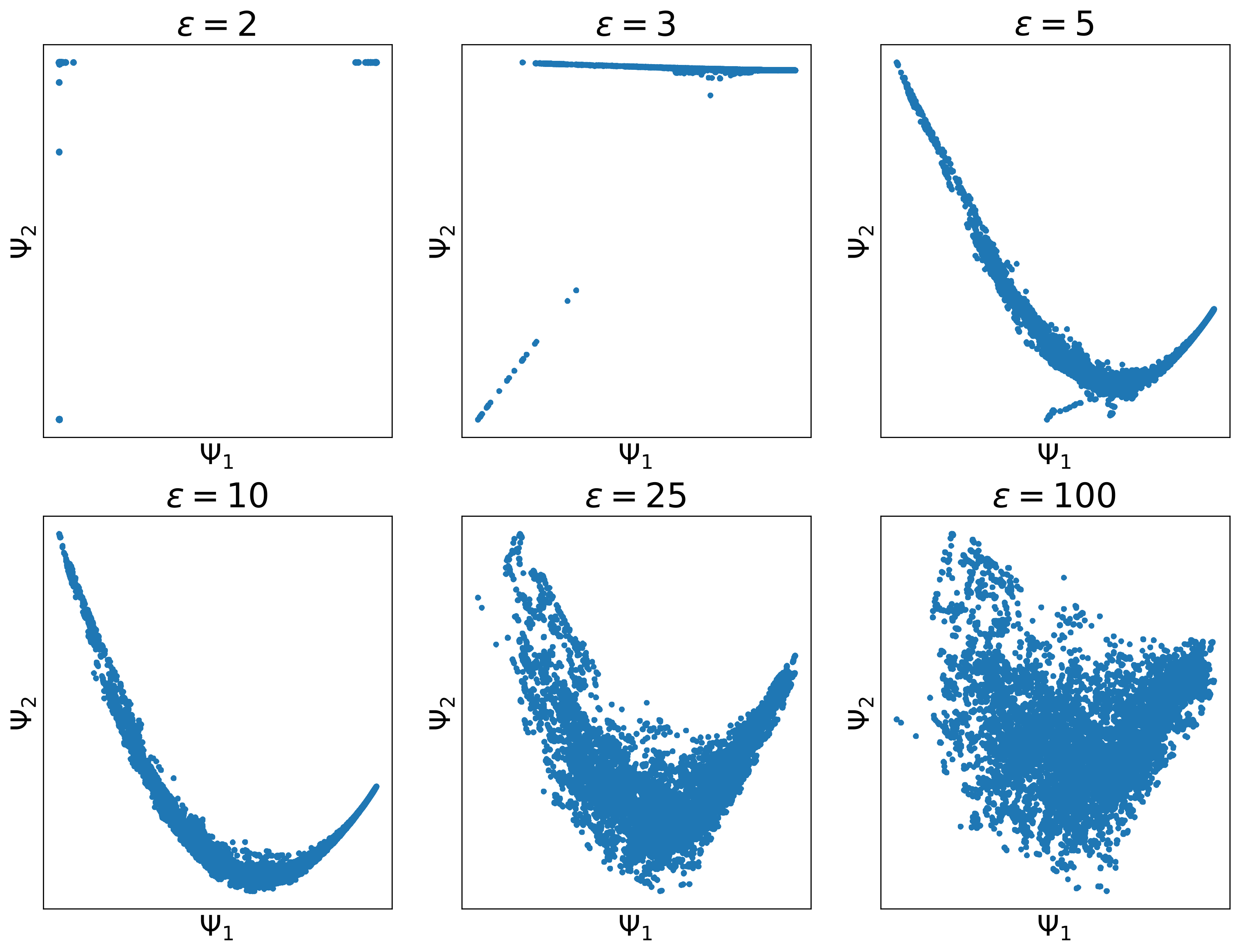}}\\
   \subfigure[For $\epsilon = 10$ fixed and varying numbers of nearest neighbors $N$.]{
   \centering
   \label{fig:parameter_vdem2}
   \includegraphics[width = 1\textwidth]{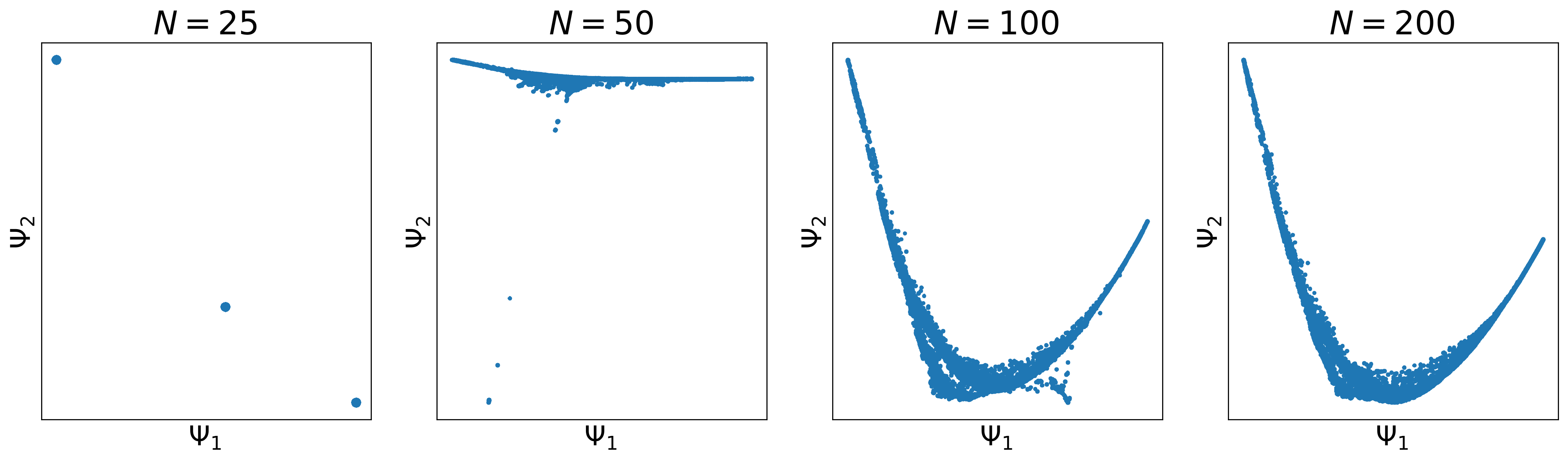}}\\
   \subfigure[For varying $\epsilon$ and considering only the $N=100$ nearest neighbors.]{
   \centering
   \label{fig:parameter_vdem3}
   \includegraphics[width = 1\textwidth]{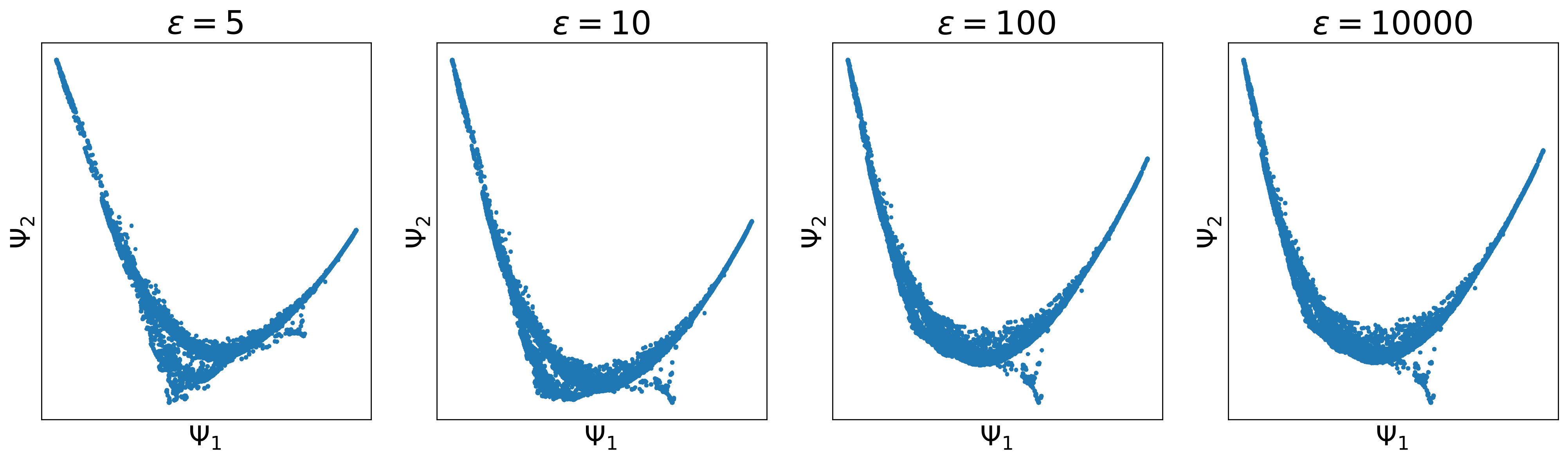}}\\
  \caption{\textbf{Find the right neighborhood parameters:} figures are showing the \textit{Diffusion Map} representations of the standardized V-Dem data (variables listed in Table \ref{table:vdem_variables}) using different values for the parameters $N$ and $\epsilon$. Using $t=1$.}
  \label{fig:parameter_vdem}
\end{figure}

%
\paragraph*{Investigating the effects of the neighborhood parameter}~\\
In order to compare the individual dimensions/variables, we first of all standardized the data, so that the data for each of the 24 indicators has a mean of zero and a variance of one. The method proposed by \cite{lafon2004} for determining the optimal $\epsilon$ (see Equation \ref{equation: find_optimal_epsilon}) provides an initial estimate of $\epsilon = 0.46$. Also the method of \cite{Singer2009} gives us a wide range of possible $\epsilon$ from approximate $\epsilon = 0.1$ to $\epsilon = 100$. 
Thus, we will now plot the \textit{Diffusion Maps} within this range of $\epsilon$. However, for $\epsilon < 2$, the eigendecomposition could not be performed, and the results in figure \ref{fig:parameter_vdem1} reveal the reason why:
figure \ref{fig:parameter_vdem1} illustrates that the \textit{Diffusion Map} becomes disconnected when $\epsilon$ is too small (e.g., $\epsilon = 2$, $\epsilon = 3$). For values of $\epsilon$ that are too large (e.g. $\epsilon = 25$ ,$\epsilon=100$), the clear structure disappears, and the \textit{Diffusion Map} becomes wider and wider. Since at $\epsilon = 5$ we still observe artifacts from the increasingly disconnected \textit{Diffusion Map} at $\epsilon = 3$ (where the data points outside the clear one-dimensional structure belong exclusively to a single country (Zambia) suggesting that this additional structure does not convey much meaningful information of electoral democracies in general) we consider $\epsilon = 10$ as the optimal value.\\

We want also to consider the effect of using only $N$ nearest neighbors. Figure \ref{fig:parameter_vdem2} shows the interaction between a fixed $\epsilon = 10$ and varying values of $N$. When $N$ is too small (for example, $N = 25$ or $N=50$), the \textit{Diffusion Map} also becomes disconnected. We can also see that for $N=100$ different additional structures in the form of a gap and a peak occur, which we want to preserve and examine in section \ref{vdem_interpretation}, as the goal is to learn as much as possible from the \textit{Diffusion Map} representation, so it makes sense to examine structures that would disappear again for too large a neighborhood. These features disappear again for larger $N$. Nevertheless, it can be stated that the parameter $N$ has a significant impact on the appearance of the \textit{Diffusion Map} and not merely on accelerating the computations.\\

In figure \ref{fig:parameter_vdem3} we pursuit the results for a fixed $N=100$ and different $\epsilon$. We see here that the peak is more pronounced for big $\epsilon$ and the gap is more pronounced for small $\epsilon$. That is why we should take an intermediate value of $\epsilon$ in order to have the possibility to analyze both.\\

However, we observe that the overall structure remains relatively stable for a wide range of parameter pairs around $\epsilon = 10$ and $\epsilon = 100$. In the following sections, we will examine this \textit{Diffusion Map} representation.

\begin{figure}[H]
    \centering
    \includegraphics[width=1\linewidth]{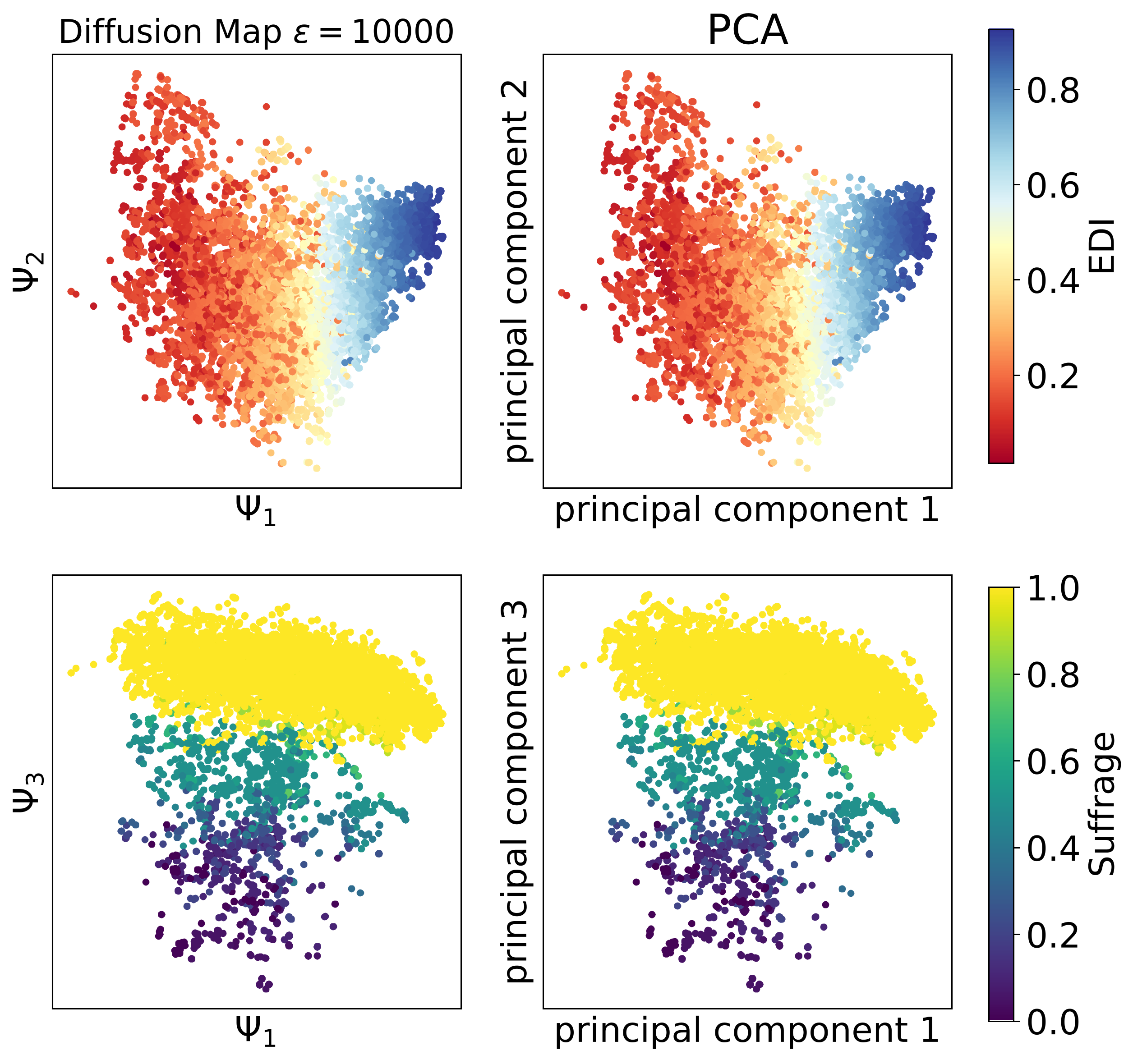}
    \caption{\textbf{Illustrating the equivalence of \textit{Diffusion Map} and PCA for large $\epsilon$.} Representation of the first three diffusion components and principal components of the V-Dem data, color-coded with the values for EDI and suffrage.}
    \label{fig:vdem_PCA}
\end{figure}

\paragraph*{Comparison with the PCA} ~\\
As already mentioned in section \ref{equivalence_pca} and shown for the Swiss roll in section \ref{finding_parameter}, it appears that the \textit{Diffusion Map} converges to the PCA representation for a large neighborhood. This observation also holds for this complex dataset, as can be seen in figure \ref{fig:vdem_PCA}. From \cite{Wiesner2024a} we know that principal component 1 is strongly correlated with the EDI and principal component 3 with the suffrage index. The figures have therefore been color-coded for better comparison.

\subsection{Exploring the resulting \textit{Diffusion Map}}
\label{vdem_interpretation}

In this section, we want to check whether the \textit{Diffusion Map} of the V-Dem data is meaningful in any way, what information we can get from the \textit{Diffusion Map} and if we can learn more about the nature of electoral democracies. 

\subsubsection{Consideration of specific countries}
\label{vdem:consideration_specific_countries}
\begin{figure}[H]
  \centering
   \subfigure[]{
   \centering
   \includegraphics[width = 0.89\textwidth]{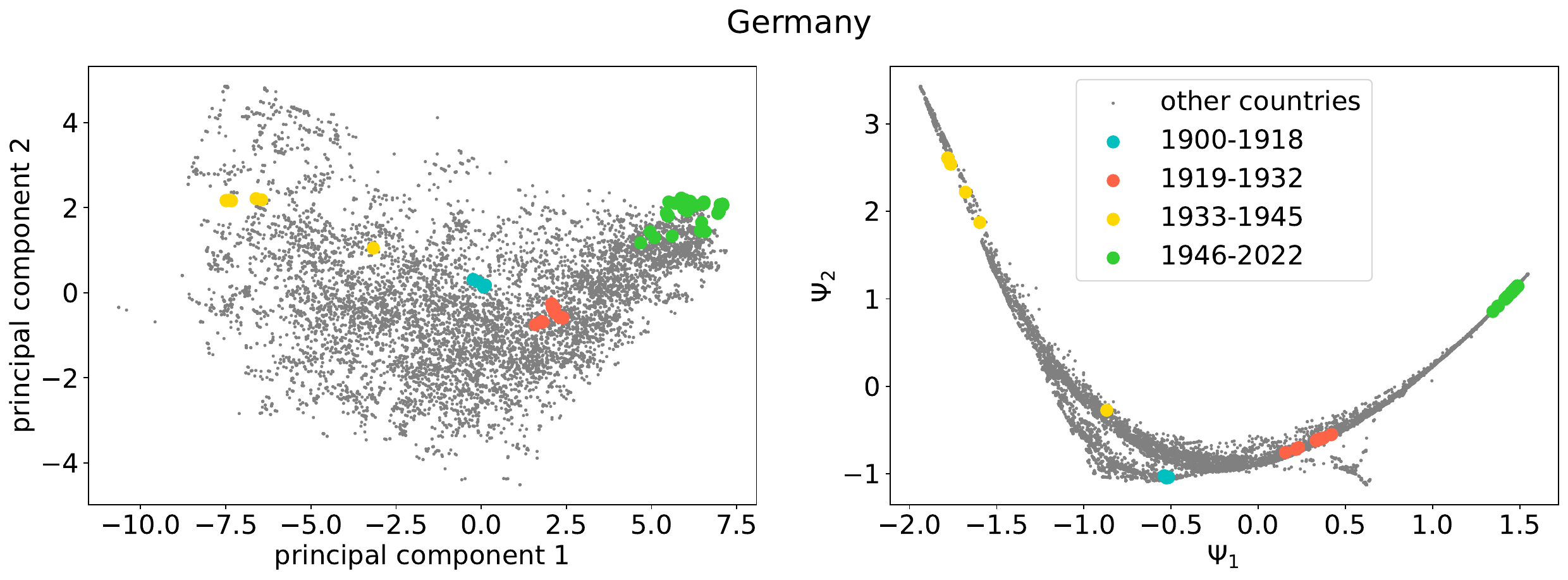}}\\
   \subfigure[]{
   \centering
   \includegraphics[width = 0.89\textwidth]{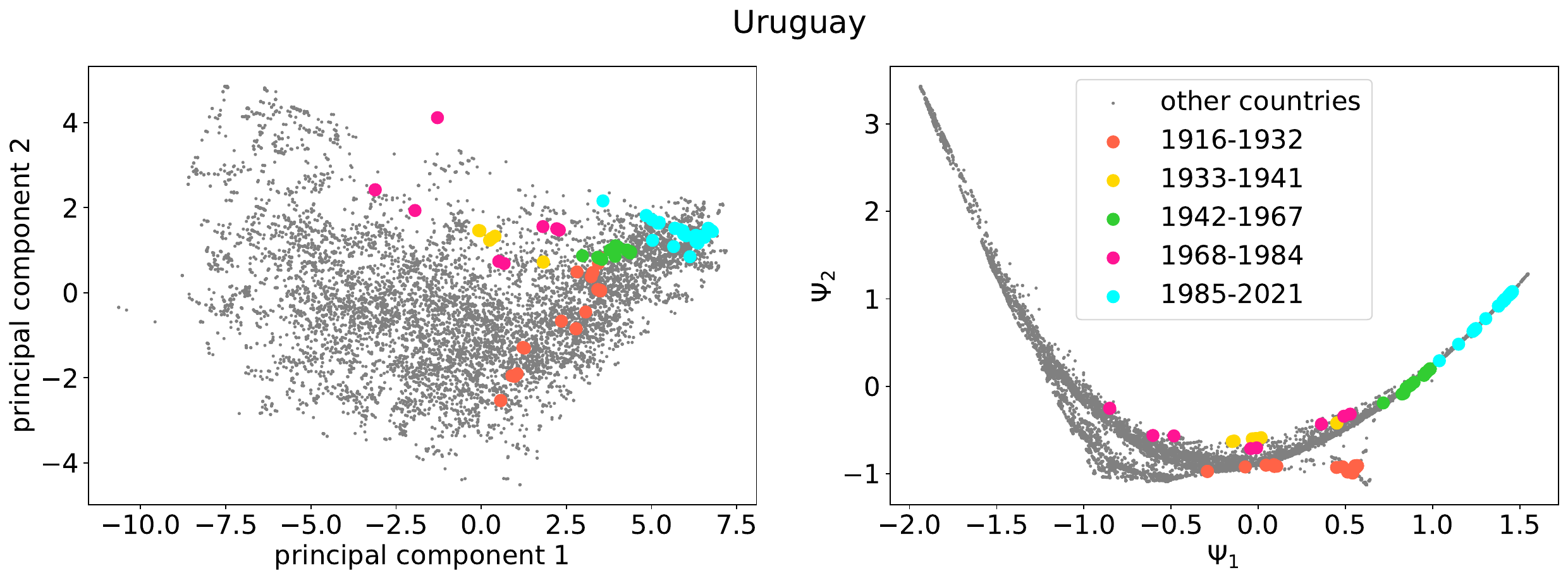}}\\
   \subfigure[]{
   \centering
   \includegraphics[width = 0.89\textwidth]{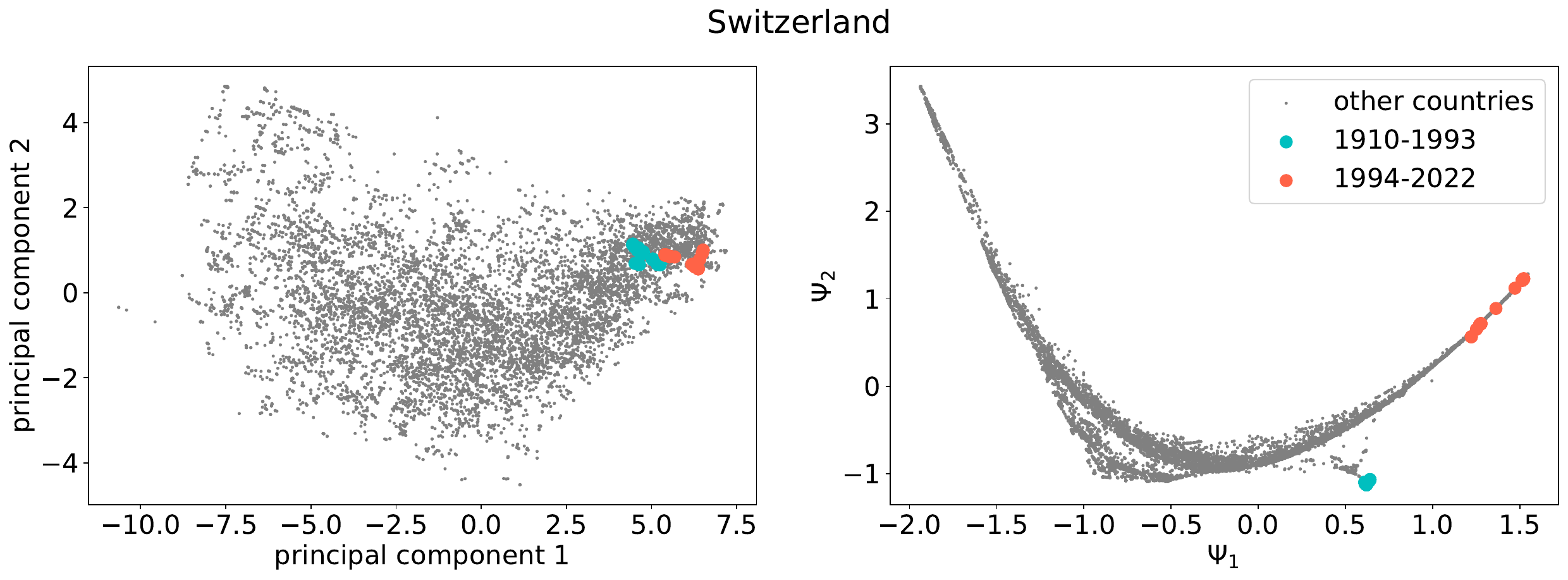}} \\
   \subfigure[]{
   \centering
   \includegraphics[width = 0.89\textwidth]{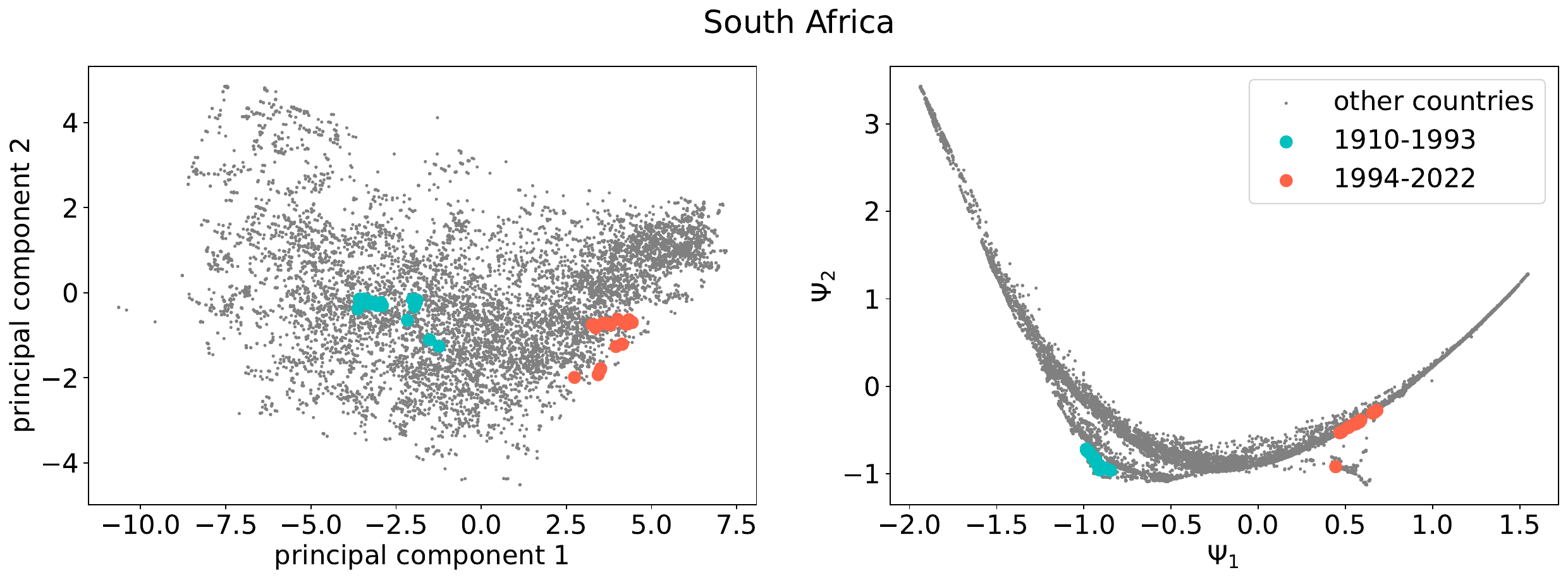}}
  \caption{The first both components of the PCA (left) and the \textit{Diffusion Map} (right) representations of the V-Dem dataset. Different time periods of different countries are marked in different colors. The PCA data comes from \cite{Wiesner2024a}. The \textit{Diffusion Map} was created with the parameters $\epsilon = 10$, $N=100$.}
  \label{fig:meaningfullness}
\end{figure}

~\\
First, we check whether the \textit{Diffusion Map} of V-Dem data can depict the history of individual countries accurately. To do this, we will look at the four examples of Germany, Uruguay, South Africa and Switzerland. In figure \ref{fig:meaningfullness} we compare the result of the principal component analysis of the dataset (which was done, in \cite{Wiesner2024a}) with that of the \textit{Diffusion Map}. The trajectory of various other countries in the diffusion space can be viewed in \cite{Pirker-Diaz2024}. \\ ~\\

\underline{\textbf{1. Germany}}\\
The example of Germany (figure \ref{fig:meaningfullness}(a)) shows the rough nature of the \textit{Diffusion Map}: During the democratic period, Germany is on the right side of the \textit{Diffusion Map}. The data points of Germany during the dictatorship and monarchy are on the left side. \\

We can see in combination with the figures \ref{fig:vdem_edi} that this left-right split is also generally true. On the left of the \textit{Diffusion Map} are countries with low EDI (autocracies). On the right are countries with a high EDI (democracies).\\

\underline{Historical information:}
\begin{description}
    \item[\footnotesize 1900-1918] German empire: End 1918 with the end of World War I
    \item[\footnotesize 1919-1932] Weimar Republic: End with the  beginning of 1933 with the appointment of Adolf Hitler as chancellor
    \item[\footnotesize 1933-1945] Nazi Germany: End with the end of World War II
    \item[\footnotesize 1946-2022] Federal republic of Germany (in this plot East Germany is not considered)
\end{description}
~\\

\underline{\textbf{2. Uruguay}}\\
Uruguay (figure \ref{fig:meaningfullness}(b)) is a very interesting example because of its many democratic and autocratic phases. In the 20th century, we have 3 democratic and 2 autocratic regimes, during which the country moves in one direction or another in the \textit{Diffusion Map} embedding. \\

\underline{Historical information:}
\begin{description}
    \item[\footnotesize 1900-1932] Presidential democracy. First "social" democracy of the South-Africa: Implementation of welfare, compulsory education, freedom of assembly and of the press. Implementation of Women suffrage at the end of the phase 1932.
    \item[\footnotesize 1933-1942] Takeover and dictatorship 1933
    \item[\footnotesize 1943-1967] Restoration of democracy in 1942. For the first time Uruguay is a democracy with women's suffrage.
    \item[\footnotesize 1968-1984] Re-introduction of the presidential system, declaration of a state of emergency, suspension of civil rights, state of war, several military coups. 
    \item[\footnotesize 1985-2021] Mass protests and a return to a democracy.
\end{description}
~\\

\underline{\textbf{3. Switzerland}}\\
With Switzerland (figure \ref{fig:meaningfullness}(c)), we have a country that was democratic throughout the entire observation period. It is interesting to observe that there is a long phase at the beginning of the period in which the country is on the democratic right side in the recognizable peak. We can easily identify what is in this peak using figure \ref{fig:vdem_suffrage} or the history of Switzerland. When leaving the peak, in 1994, women were given full voting rights for the first time. Hence, we can identify this peak with democracies that do not have full voting rights. \\

\underline{Historical information:}
\begin{description}
    \item[\footnotesize 1910-1993] Democracy without or only partial women's suffrage.
    \item[\footnotesize 1994-2022] Introduction of women's suffrage in all cantons.
\end{description}

~\\

\underline{\textbf{4. South Africa}}\\
But how can we explain the other gap for autocracies at the left side? It is interesting to see that South Africa (figure \ref{fig:meaningfullness}(d)) is below the gap during the autocratic phase of apartheid, whereas Germany and Uruguay are above the gap during their autocratic phase. In combination with figure \ref{fig:vdem_suffrage} we can see that this has something to do with the suffrage. 
Because in Uruguay and in Germany there were nevertheless elections during these autocratic phases, which were held with universal suffrage. In contrast, many people were excluded from political participation in South Africa. As a result, the suffrage here is very low. Above the gap are autocracies with full voting rights. Below are autocracies with only partial voting rights.\\

\underline{Historical information:}
\begin{description}
    \item[\footnotesize 1910-1993] Colony and apartheid. 
    \item[\footnotesize 1994-2022] First democratic elections in 1994.
\end{description}

From these examples, we find that two-dimensional representation of the \textit{Diffusion Map} reveals greater structure than the PCA and provides insights into a country's level of democracy and suffrage.

\subsubsection{\textit{Diffusion Map} is ordering the data from autocracy to democracy}
\label{vdem_autocracy_to_democracy}

\begin{figure}[H]
  \centering
   \subfigure[V-Dem data in diffusion space is color coded by the EDI.]{
   \label{fig:vdem_edi1}
   \centering
   \includegraphics[width = 0.6\textwidth]{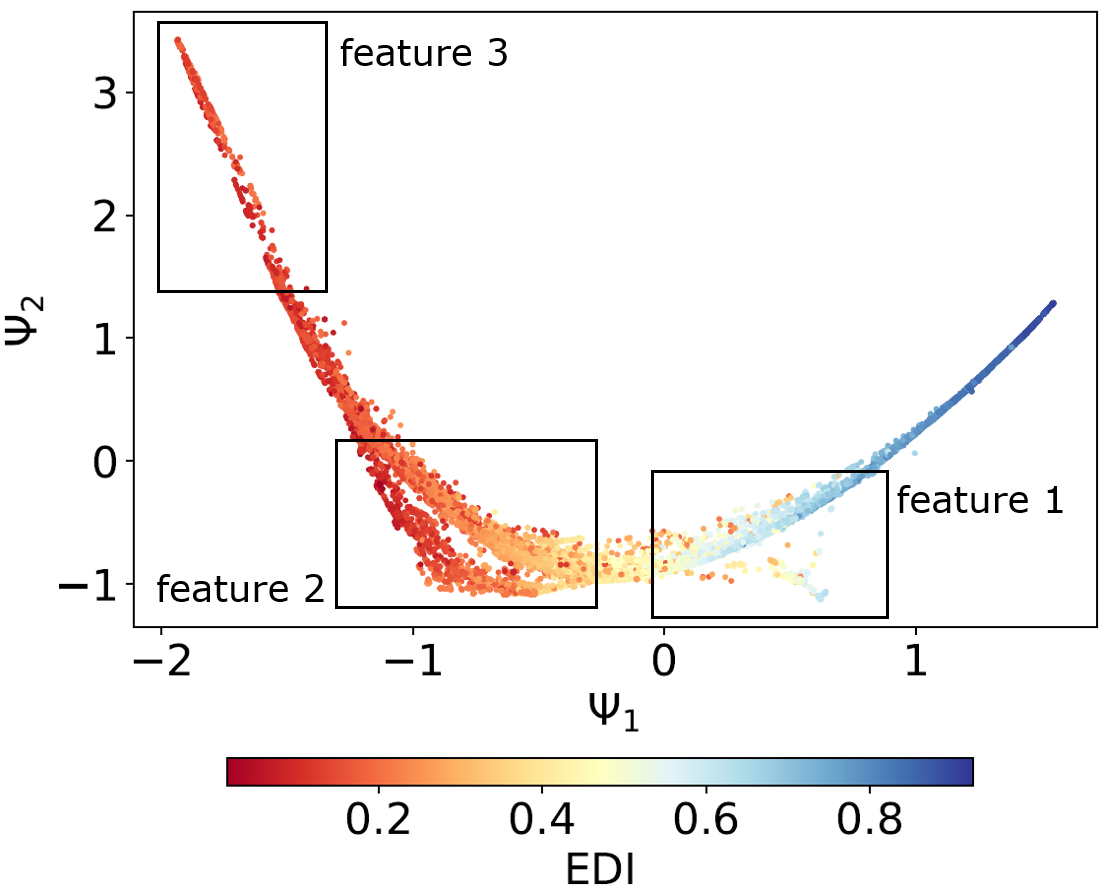}}\\
   \subfigure[Showing the correlation of $\Psi_1$ with the EDI.]{
   \label{fig:vdem_edi2}
   \centering
   \includegraphics[width = 0.6\textwidth]{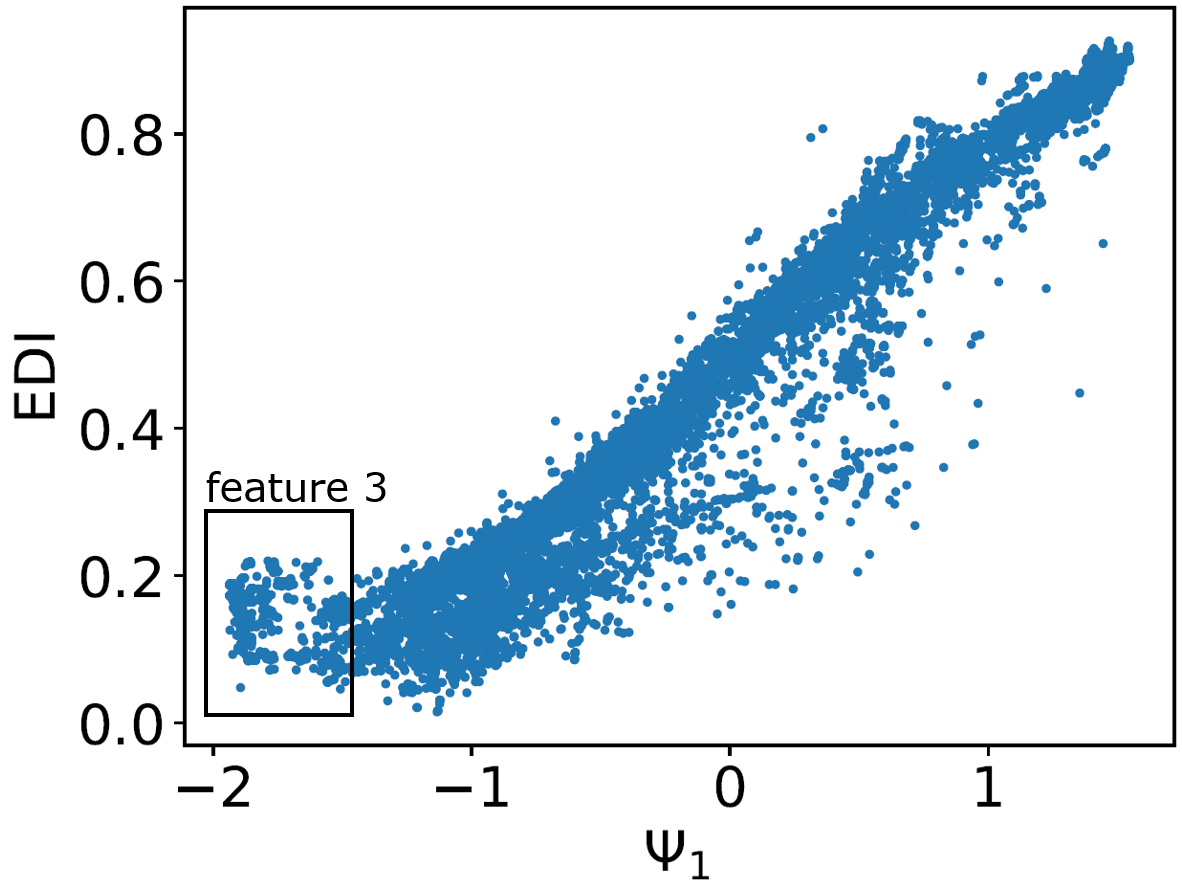}}\\ 
    \caption{\textbf{Meaning of the first diffusion component $\Psi_1$.} Authoritarian states are on the left for low EDI scores. Democratic states are on the right for high EDI scores. Three features are marked in the illustrations. The origin of the features are explained in sections \ref{vdem_suffrage} and \ref{vdem_quality_of_election} in detail. The \textit{Diffusion Map} parameters $\epsilon = 10$, $N=100$, $t=1$ are used.}
    \label{fig:vdem_edi}
\end{figure}

As we have seen in the examples of trajectories of the individual countries in section \ref{vdem:consideration_specific_countries}, the diffusion is rich in information, which allows for the accurate tracing of historical events. In figure \ref{fig:vdem_edi1}, the color coding reveals that $\Psi_1$ sorts the data according to the EDI. Additionally, figure \ref{fig:vdem_edi2} shows a strong linear correlation between $\Psi_1$ and the EDI, with a Pearson correlation coefficient of $0.96$. Also, the correct order of data points is crucial for creating and comparing indices, which is why we are also considering the Spearman rank correlation, revealing a strong monotonic relationship with a correlation of $0.96$.\\

Overall, the \textit{Diffusion Map} shows that this must be a roughly one-dimensional data structure. However, there are three features that we have to consider more closely. In figure \ref{fig:vdem_edi1}, these are marked and resemble a peak (feature 1), a gap/loop inside the manifold (feature 2), and slightly separated data points at one end of the curve (feature 3). In the next sections, we will examine the origin of these characteristics, which arise from the peculiarities of individual variables (such as the suffrage index) in the V-Dem dataset.

In figure \ref{fig:vdem_edi2}, we can observe that the same data points from feature 3 also cause a significant discrepancy between $\Psi_1$ and the EDI. Therefore, it is crucial to identify which of the two variables accurately classifies the data points regarding their democracy. 

\subsubsection{Suffrage - Influence of different variable types}
\label{vdem_suffrage}

With the trajectories of Switzerland and South Africa from section \ref{vdem:consideration_specific_countries}, we have already got a first guess that features 1 and 2 from figure \ref{fig:vdem_edi1} depend on the suffrage, i.e. the proportion of people who are allowed to vote. Feature 1 may show the 'democratic' states without universal suffrage (mostly without women's suffrage) and feature 2 may show the split of authoritarian states into authoritarian states with universal suffrage (e.g. Nazi Germany) and authoritarian states where certain groups of the population have no right to vote (South Africa in times of apartheid).\\

\begin{figure}[H]
    \centering
    \includegraphics[width=1\linewidth]{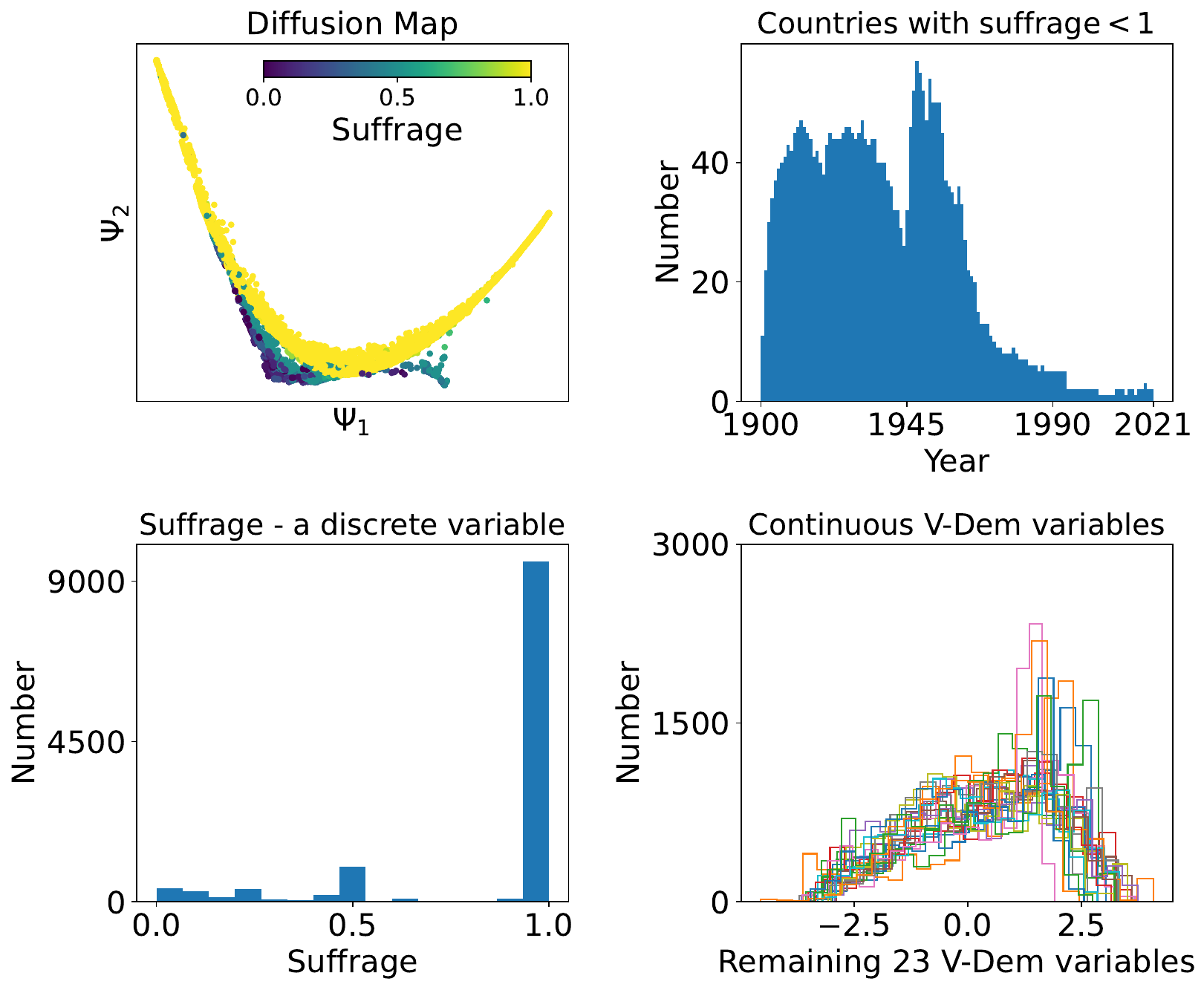}
    \caption{\textbf{Illustrating the effect of the discrete suffrage variable} on the \textit{Diffusion Map}. On the top-left is the \textit{Diffusion Map} color-coded with the suffrage index. In the top-right there is a histogram revealing the number of countries with a suffrage index smaller then $1$. At the bottom-left is a histogram with all values of the suffrage index, showing that the suffrage index is a discrete variable with mainly 3 values nearly $0$(exclusion of population groups), $0.5$(no women suffrage), $1$(universal suffrage). At the bottom right there is a plot with the histograms of all other variables of the dataset (from table \ref{table:vdem_variables}) indicating that these are continuous.}
    \label{fig:vdem_suffrage}
\end{figure}

\paragraph*{Suffrage - a discrete variable}~\\
This hypothesis is further supported by figure \ref{fig:vdem_suffrage} (left): The color code clearly indicates that these structures correspond to different suffrage values. Additionally, figure \ref{fig:vdem_suffrage} (middle) demonstrates that the effect of varying suffrage values is primarily a phenomenon of the first half of the twentieth century and has almost completely disappeared today. In the present day, nearly every state with elections in the dataset has a suffrage value of 1. The only exceptions in 2021 are Somalia (which previously had a clan-based voting system, now being replaced by a universal voting system \cite{suffrage_somalia}) and Thailand (which has a suffrage value of $0.994$, as a small group of monks is not allowed to vote \cite{suffrage_thailand}). \\

The suffrage index having such a profound impact on the appearance of the \textit{Diffusion Map} can be explained by the fact that it is the only index which is discrete. This effect can be seen if one compares both histograms in figure \ref{vdem_suffrage} bottom left (showing the suffrage index) and bottom right(showing the remaining 23 variables). The suffrage index is almost always $1$ (universal suffrage), $0.5$ (no women's suffrage) or very small (if large population groups are excluded e.g. because of racism and apartheid). So the relative difference between different suffrage values is very large, leading to larger distances between data points. All other used low-level indices are continuous and therefore do not have such a huge impact on the calculation of distances and the resulting \textit{Diffusion Map}. In figure \ref{fig:vdem_higher_components_without_suffrage} we can indeed see that if we remove suffrage as an index when calculating the \textit{Diffusion Map}, the feature $1$ and $2$ disappear from the representation in the diffusion space.

\paragraph*{Comparison to other examples}~\\
If we compare feature 1 and feature 2 with the example datasets in figure \ref{fig:complex_typical_shapes}, we can further explain the appearance of the V-Dem \textit{Diffusion Map}. The patterns of the \textit{Diffusion Map} embedding of datasets 4 and 5 closely resemble features 1 and 2. This suggests that such structures can be easily reproduced artificially. 

We can conclude that whenever we observe such loops or peaks in \textit{Diffusion Maps}, it indicates that similar loops and peaks exist in the original data. This is because the \textit{Diffusion Map} preserves the topology of the manifold.

However, it should be critically examined whether variables like the suffrage parameter, so discrete variables (or more precisely variables with large gaps between their possible values) can be handled well by the \textit{Diffusion Map}. This observation rather suggests that working with different types of variables (continuous and discrete) might be challenging, because the discrete suffrage variable gains increased importance primarily due to the structure of the data. The key question here is whether an alternative rescaling of discrete variables would be beneficial, in which the distances between the individual possible values are reduced. Also, an argument could be made to omit discrete variables and discuss them separately from the other variables (as we do it in the next section of this thesis).

This issue regarding the handling of different types of variables with the \textit{Diffusion Map} is not addressed in the theoretical literature, as classical datasets are assumed where the dimensions of the feature space are of the same type. For example, when considering geometric structures like the Swiss roll, the coordinate axes $x$, $y$, $z$ have the same nature of spatial coordinates. The same holds true for images, where all pixels are assumed to have comparable values.

\subsubsection{Quality of election - Defining two kinds of autocracies}
\label{vdem_quality_of_election}

Now, the question remains whether feature 3 of the \textit{Diffusion Map} (see figure \ref{fig:vdem_edi}) can also be explained by the influence of specific variables and if there is a valid reason why these data points are separated from the rest of the curve by $\Psi_1$ or positioned at the very edge of the autocratic side, while they are not particularly emphasized in the EDI?\\

\paragraph*{Important variables for the feature}~\\
To visualize which variables differ significantly in the feature 3 compared to the rest of the curve, a simple approach is used: 
We standardize the data and divide it into two groups: the data points within the feature (i.e., those with $\Psi_1<-1.7$) and the adjacent data points in the rest of the curve (i.e., those with $-1.7<\Psi_1<-1.2$).
We then calculate the means of the 24 variables for both groups, denoted as $m^{variable}_{(\Psi_1<-1.7)}$ and $m^{variable}_{(-1.7<\Psi_1<-1.2)}$. Then we are calculating the difference between this means given by 
\begin{equation}
    \label{difference_variables_feature3}
    \Delta^{variable} = |m^{variable}_{(\Psi_1<-1.7)} - m^{variable}_{(-1.7<\Psi_1<-1.2)}|
\end{equation}
and can be seen in table \ref{tab:differences_feature3}.\\

\begin{figure}[H]
    \centering
    \includegraphics[width=1\linewidth]{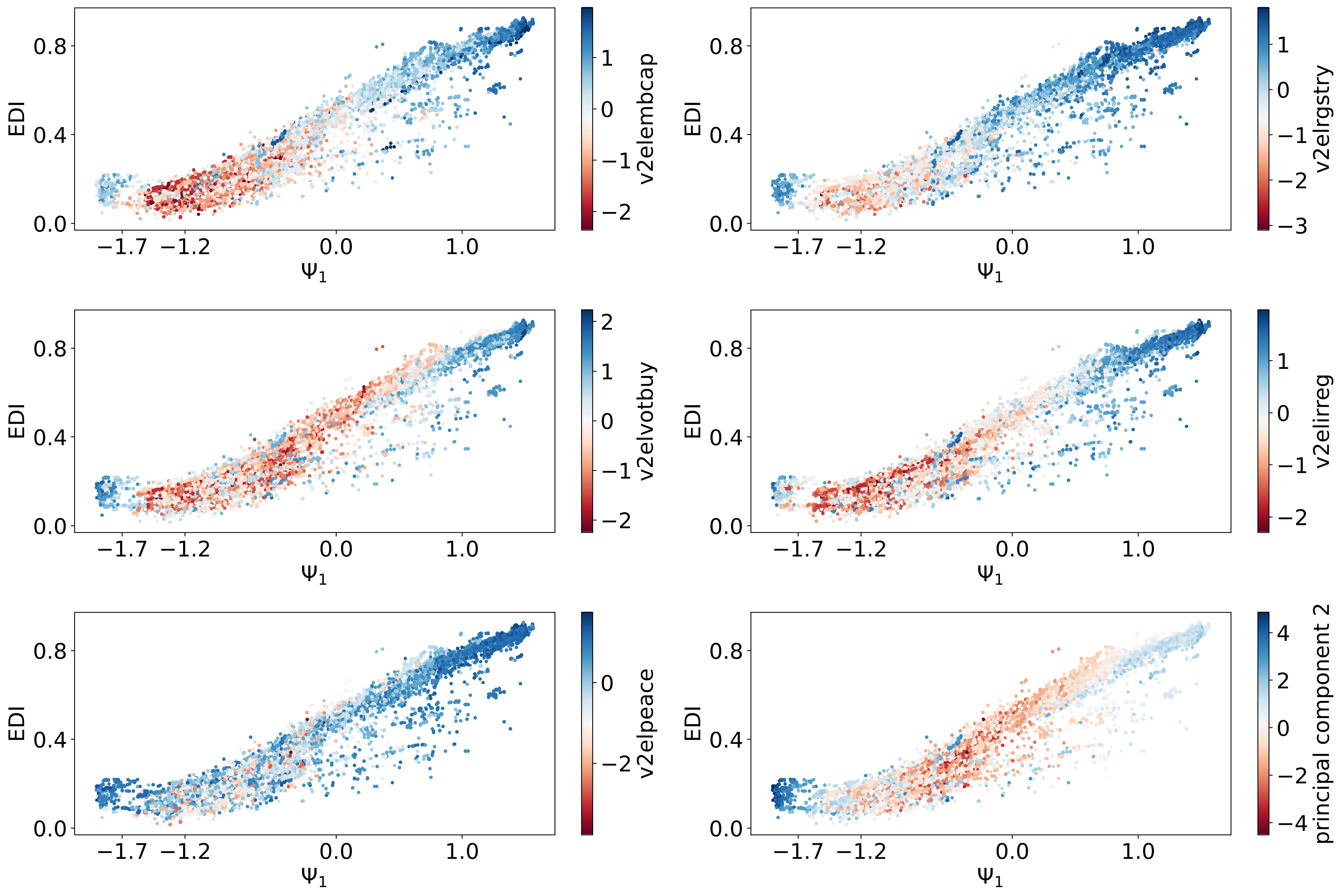}
    \caption{\textbf{Influence of the quality of election} on the \textit{Diffusion Map} embedding. The color-code is indicating the values of the variables \texttt{v2elembcap}, \texttt{v2elrgstry}, \texttt{v2elvotbuy}, \texttt{v2elirreg}, \texttt{v2elpeace} and the second principal component of the V-Dem data, which are indicating the quality of the elections and their organization (for the meaning of the principal component see \cite{Wiesner2024a}). The \textit{Diffusion Map} was calculated without the suffrage variable to avoid the effects discussed in section \ref{vdem_suffrage} (see figure \ref{fig:vdem_edi2} for the correlation plotted including the suffrage variable). $\epsilon = 10$, $N=100$ were used.}
    \label{fig:quality_of_election}
\end{figure}

The largest differences between the means inside and outside feature 3 are given for the variables \texttt{v2elrgstry}, \texttt{v2elembcap}, \texttt{v2elvotbuy} and to a smaller amount also \texttt{v2elirreg} and \texttt{v2elpeace}. These five variables are shown as a color code in figure \ref{fig:quality_of_election} and stand for an accurate voter registry, the amount of vote/turnout buying, whether the election management body has sufficient resources for an election, if there are other voting irregularities and if there is violence during the election period, respectively. We see that a clear and abrupt transition from the separated data points and the rest of the curve can be seen, especially for the first three indices (\texttt{v2elrgstry}, \texttt{v2elembcap}, \texttt{v2elvotbuy}). The separated data points have significantly higher index values.
Thus, the differences are particularly pronounced in variables that, according to \cite{Wiesner2024a}, have a high positive loading on the second principal component of the data, and intend to stand for the election capability of specific countries. Therefore, the countries in the third feature exhibit a better organized electoral process compared to other autocratic regimes.\\

\paragraph*{Comparison to similar similar structures}~\\
We know that such \textit{Diffusion Map} structures as feature 3 in figure \ref{fig:vdem_edi} can also emerge when there are not enough similar data points connecting two strands of data points, as observed in the example dataset 6 from figure \ref{fig:complex_typical_shapes}. So it seems possible that this feature exists because there are probably not enough similar data points to form a meaningful curve (see also the limits of the sorting ability discussed in section \ref{sorting_ability}) between both of the two different types of autocracies: those with high and those with low electoral organization. 

The reason for such gaps in this particular dataset may also be that not every possible state in the democracy feature space has been realized over the past 100 years, leaving gaps in the space that can complicate an accurate description using the \textit{Diffusion Map}. This does not mean that such states could not exist. However, some democratic states are difficult to conceive (for instance, the coexistence of free media with otherwise entirely autocratic characteristics). This also explains why the data may lie on a low-dimensional manifold within the 24-dimensional feature space.

\paragraph*{Connection to election capabilities}~\\
So we have learned that if we would interpret the diffusion component $\Psi_1$ as a democracy index, it would rank autocratic states with strong election organization more autocratic in comparison to the EDI. This seems to be an unwanted side effect. However, it might also make sense to arrange these autocracies with a high election capability in a more autocratic region, since a strong organizational ability could indicate that the autocratic system is less contested, more consolidated, and stable. 
There are various discussions on the circumstances under which elections in autocracies have a stabilizing effect or not \cite{Seeberg2014,Knutsen2017}. However, high election capabilities may indicate a high level of state capacity. Therefore, high state capabilities \cite{Seeberg2014} and resulting high election capabilities \cite{Wiesner2024a} may be a sign of stable autocratic regimes.

The results support the statement of \cite{Wiesner2024a}, that these two types of autocracies (with and without enough election capabilities) cannot be sufficiently separated and described by the EDI, but they also do not seem to be properly represented by $\Psi_1$ and can easily be missed when the curve is not analyzed carefully and without considering the points raised in the previous sections \ref{typical_shapes} and \ref{sorting_ability}, because the results cannot be interpreted intuitively. Nevertheless, one can deduce from $\Psi_1$ that one should consider both kinds of autocracies as different phenomena, which is not possible relying only on the EDI.

\begin{table}[H]
\footnotesize
\centering
\begin{tabular}{r||l}
\textbf{Variable} & $\Delta^{variable}$ \\
\hline
\hline
\texttt{v2elrgstry} & 1.7 \\
\texttt{v2elembcap} & 1.6 \\
\texttt{v2elvotbuy} & 1.6 \\
\texttt{v2elirreg} & 1.0 \\
\texttt{v2elpeace} & 1.0 \\
\hline
\texttt{v2mecrit} & 0.6 \\
\texttt{v2x\_suffr} & 0.6\\
\texttt{v2psbaran} & 0.5 \\
\texttt{v2cseeorgs} & 0.4 \\
\texttt{v2csreprss} & 0.4 \\
\texttt{v2meslfcen} & 0.4 \\
\texttt{v2mebias} & 0.4 \\
\texttt{v2merange} & 0.4 \\
\texttt{v2elfrfair} & 0.4 \\
\texttt{v2psoppaut} & 0.4 \\
\texttt{v2elmulpar} & 0.3 \\
\texttt{v2clacfree} & 0.2 \\
\texttt{v2cldiscw} & 0.2 \\
\texttt{v2elintim} & 0.2 \\
\texttt{v2psbars} & 0.1 \\
\texttt{v2cldiscm} & 0.1 \\
\texttt{v2elembaut} & 0.1 \\
\texttt{v2meharjrn} & 0.0 \\
\end{tabular}
\caption{$\Delta^{variable}$ (equation \ref{difference_variables_feature3}) sorted from highest to lowest for different variables.}
\label{tab:differences_feature3}
\end{table}

\subsection{Dimensionality of the dataset}
\label{vdem_one_dimensional_curve}

Now, we want to summarize and examine whether electoral democracy — or at least the available data from the V-Dem dataset — is best represented by a one-dimensional index, i.e. the data lies on a 1-D manifold. Specifically, we ask whether the first diffusion component $\Psi_1$ and also the EDI  provides a good description of all manifestations of electoral democracy and in which cases it fails to do so.

\begin{figure}[H]
  \centering
   \subfigure[With all 24 indices mentioned in table \ref{table:vdem_variables}.]{
   \centering
   \label{fig:vdem_higher_components_with_suffrage}
   \includegraphics[width = 1\textwidth]{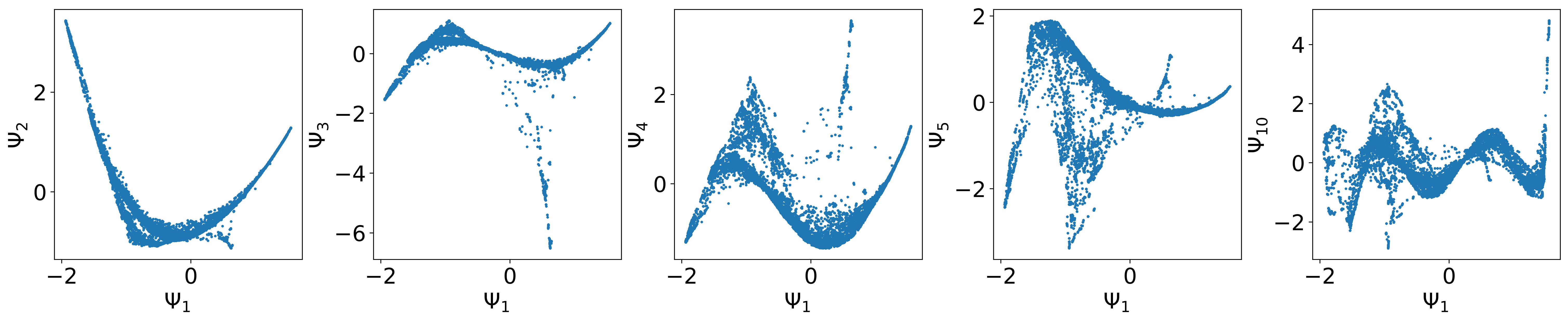}}\\
   \subfigure[Without the \texttt{v2x\_suffr} index.]{
   \label{fig:vdem_higher_components_without_suffrage}
   \centering
   \includegraphics[width = 1\textwidth]{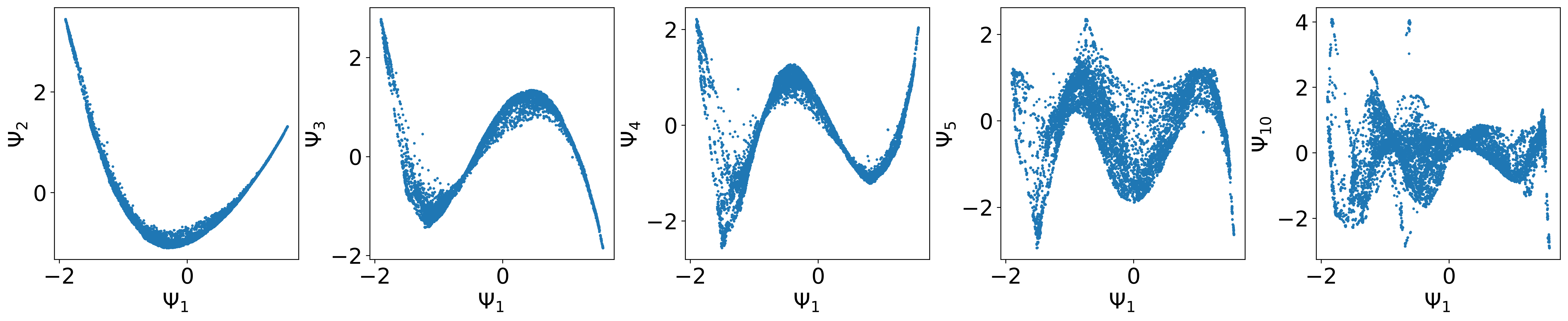}}\\
\subfigure[Without the \texttt{v2x\_suffr}, \texttt{v2elvotbuy}, \texttt{v2elrgstry}, \texttt{v2elembcap}, \texttt{v2elirreg} and \texttt{v2elpeace} indices. 18 variables are remaining.]{
   \label{fig:vdem_higher_components_without_6}
   \centering
   \includegraphics[width = 1\textwidth]{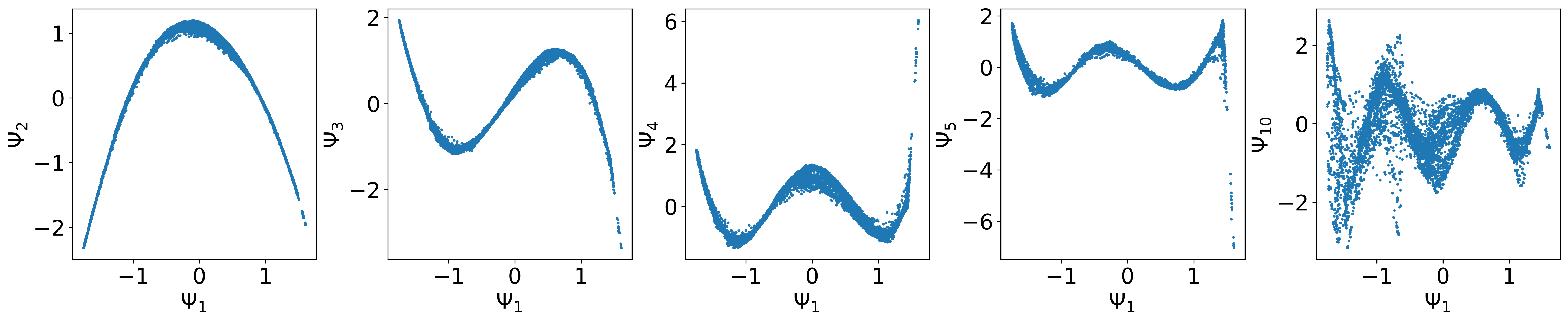}}\\
  \caption{\textit{Diffusion maps} of the V-Dem dataset with and without consideration of the suffrage index in the calculation. The different columns show the dependence of the various diffusion components to each other. The first diffusion component $\Psi_1$ is shown in relation to $\Psi_2$, $\Psi_3$ and $\Psi_{10}$. The \textit{Diffusion Maps} were created with the parameters $\epsilon = 10, N=100$.}
  \label{fig:vdem_higher_components}
\end{figure}

In order to investigate the question of one-dimensionality of the data, we reconsider the higher diffusion components. From \cite{Nadler2007} and section \ref{typical_shapes}, we know that if the data is truly one-dimensional, we should observe characteristic polynomial dependencies. These polynomial dependencies serve as an indicator and a necessary condition for the data to lie on a one-dimensional manifold. In figure \ref{fig:vdem_higher_components_with_suffrage}, we recognize these characteristic forms, although we also observe disturbances that lead to stronger deviations in higher diffusion components.\\

We have identified the reasons for these deviations. One deviation from one-dimensionality (feature 1, 2 marked in figure \ref{fig:vdem_edi1}) is associated with the suffrage index and its discrete nature. In both the original feature space and the diffusion space, the data is effectively split into two parts: countries with and without universal suffrage. This effect is particularly pronounced in the \textit{Diffusion Map} because suffrage is the only discrete variable among otherwise continuous variables (see section \ref{vdem_suffrage}).

A second deviation (feature 3 marked in figure \ref{fig:vdem_edi1}) arises from variables, which indicate stronger election capabilities. These variables distinguish autocracies with well-developed election systems from those without. In the \textit{Diffusion Map}, this results in autocracies with strong election capabilities potentially being classified with a lower $\Psi_1$ as more autocratic. However, this effect might also stem from an insufficient number of data points bridging these two groups, preventing the \textit{Diffusion Map} from correctly capturing the underlying manifold (see section \ref{vdem_quality_of_election}).\\

When we account for both effects and exclude these variables from the \textit{Diffusion Map} calculation, the distortions decrease (see figures \ref{fig:vdem_higher_components_without_suffrage} and \ref{fig:vdem_higher_components_without_6}).

This result provides strong indications that the remaining 18 variables could lie on a one-dimensional manifold. In this case, $\Psi_1$ serves as a meaningful descriptor of the intrinsic one-dimensional data which is embedded in 18 dimensions and its natural parameter (after \cite{Nadler2007}). We can therefore interpret $\Psi_1$ as a purely data-driven democracy index, which should further be complemented by electoral capacities and suffrage to cover all aspects of electoral democracies.\\

Overall the results also confirm on one hand the legitimacy of the EDI due to its high correlation with the data driven created diffusion component $\Psi_1$: therefore the correlation between the EDI and $\Psi_1$ is $0.96$ (Pearson) and $0.96$ (Spearman) when using all 24 variables (and it is $0.94$ (Pearson) and $0.95$ (Spearman) using 18 variables). 

On the other hand, it also indicates one weakness if we have a closer look at the data: as also previously noted by \cite{Wiesner2024a}, the EDI (as also $\Psi_1$ when considering all 24 variables) does not reveal the difference between autocracies with robust electoral processes and those with weak or non-existent electoral frameworks. This could be a reason to consider electoral democracy as a two- or multi-dimensional phenomenon, which has to be described by more than one index.
Nevertheless, it should be noted that the one-dimensional EDI still works very well in capturing the data, if we take the result of the data-driven $\Psi_1$ as a benchmark. The identified improvement would affect only a small number of data points, meaning that, in most cases, the EDI provides an accurate and meaningful representation of the underlying data.

%% file: Chapter/43_census_britain.tex
\section{Census data of Bristol}
\label{census_britain_section}

In this section, we will examine a census dataset in which citizens of a country are surveyed on various aspects of their living conditions, with the goal of using this information to inform political decisions. 

With this dataset, we will learn how to determine the meaning of individual diffusion components using a correlation analysis and identify the most exploratory variables and compare these results to previous literature results, but also address difficulties in the application (section \ref{census_exploring_correlation}). 
Furthermore, we will examine the extent to which the feasibility and outcome of this correlation analysis depend on the choice of the neighborhood parameter $N$, and how comparable the results of this dataset are to those obtained using PCA (section \ref{census_differentN_rotating}).
In addition, we will assess to what extent the method proposed in section \ref{census_redundant_information} for handling linearly redundant variables can be applied. The main question at the end is whether the application of the \textit{Diffusion Map} is appropriate for this dataset (section \ref{census_conclusion}).

\subsection{The dataset and motivation for the analysis}
\label{census_motivation}

We will analyze the 2011 census data from England and Wales, which is publicly available at \cite{census_great_britain}. This dataset was also studied using the \textit{Diffusion Map} method in \cite{Barter2019, Xiu2023}. The article by \cite{Barter2019} is particularly interesting because it identifies the density of students and poverty as the most important variables describing the dataset. The authors identify the locations of university campuses using the first diffusion component and show that low values of the second diffusion component coincide with locations of social housing.

This result about the first diffusion component is interesting because at first sight it is not obvious why the number of students is particularly important for describing a country's population. That is why we want to take a closer look at the results presented in \cite{Barter2019}, reproduce them, and address the open questions regarding why we obtain such results in order to gain a deeper understanding of the data and the \textit{Diffusion Map} method. \\

This dataset comprises several thousand different variables, with a detailed description provided in the guidance document available from \cite{census_great_britain}. The most important indices for the analysis in this chapter are listed in the table \ref{tab:census_description} with a description. Instead of listing the responses of individual persons or households, the data is aggregated into geographic units named output areas. Each output area contains information from approximately 100 households. Notably, output areas in rural regions tend to be larger than those in urban areas. In total, there are $175434$ output areas across England and Wales.

As \cite{Barter2019} we will focus also on the city of Bristol and its surrounding areas, which together comprise 3634 output areas. A complete list of the used output areas can be found at \url{https://github.com/SoenBeier/diffusion_map}. Output areas can be visualized and found by using visualization tools from the Office for National Statistics \cite{OA_visualization}.\\

All available variables were used for the analysis. The data types "count", “sum” and “average” were divided by the number of inhabitants to obtain a ratio, so that output areas with different numbers of inhabitants are not treated differently. Existing variables of the type “percentage” or “ratio” were neglected to avoid duplicates. Also, duplicate variables with exactly the same entries and variables containing only zeros are discarded. In total, we have $1770$ variables in our dataset. 

\begin{table}[]
    \centering
    \footnotesize
    \begin{tabular}{r|l}
         \textbf{Index} & \textbf{Description} \\
         \hline
         \hline
         \texttt{CT00100002} & Number of inhabitants who identify as British (including all regions of Great Britain) \\
         \hline
         \texttt{KS202EW0007} & Number of inhabitants who identify as Welsh \\
         \hline
         \texttt{KS404EW0007} & Number of cars in the specific output area \\
         \hline
         \texttt{KS501EW0007} & Number of inhabitants with the qualification level 4 out of 4 \\
         \hline
         \texttt{KS501EW0010} & Number of inhabitants who are schoolchildren and students aged over 18\\
         \hline
         \texttt{KS608EW0003} & Number of inhabitants with a professional occupation\\
         \hline
         \texttt{KS609EW0003} & Number of male inhabitants with a professional occupation\\
         \hline
         \texttt{KS611EW0002} & Number of inhabitants with higher managerial,\\
         &administrative and professional occupations \\
         \hline
         \texttt{KS611EW0004} & Number of inhabitants with higher professional occupations\\
         \hline
         \texttt{QS106EW0002} & Number of inhabitants without a second address \\
         \hline
         \texttt{QS408EW0002} & Number of households with more then 2 rooms \\ 
         \hline
         \texttt{QS502EW0005} & Number of inhabitants with a higher diploma (GCSE Grades A$^*$-C)\\
         \hline
         \texttt{QS502EW0008} & Number of inhabitants with higher school certificate (A-levels)\\
         \hline
         \texttt{QS502EW0010} & Number of inhabitants with a higher scientific degree (MA, PhD)\\
         \hline
         \texttt{QS611EW0005} & Number of inhabitants with lowest social grade D or E; index refer to semi-/\\
         &  unskilled manual and lowest grade occupations and unemployed \cite{social_grade_reference}
    \end{tabular}
    \caption{Selected census variables referred in this paper. Descriptions are according to the guidance document available on \cite{census_great_britain}. All these variables were divided by the number of the inhabitants of the according output area. In the following analysis, the indices are therefore referenced as a ratios.}
    \label{tab:census_description}
\end{table}

\subsection{Exploring the \textit{Diffusion Map} by correlation analysis}
\label{census_exploring_correlation}

Now, we will attempt to reproduce the results of \cite{Barter2019}. It should be noted that \cite{Barter2019} uses a different kernel function, $K_{ij}=1/\mathcal{D}_{ij}$. However, we will use our Gaussian kernel to ensure comparability with the other results examined in this work. Furthermore, the exact output areas from \cite{Barter2019} are not provided, so the dataset is not identical; nevertheless, both datasets include Bristol with neighboring output areas.\\

First, we will investigate which \textit{Diffusion Map} parameters are optimal for reproducing the results. In \cite{Barter2019}, the number of nearest neighbors is specified as $N=10$. As in \cite{Barter2019}, we first standardize the dataset. Since we have already demonstrated that $t$ has little impact on the results—and given that \cite{Barter2019} does not use this parameter—we set $t=1$. Consequently, we now wish to determine the appropriate $\epsilon$ for this dataset. According to formula \ref{equation: find_optimal_epsilon}, we obtain an initial guess of $\epsilon = 3135$. We will now vary this value and examine whether the \textit{Diffusion Map} remains stable in this range.

\begin{figure}[h]
    \centering
    \includegraphics[width=\linewidth]{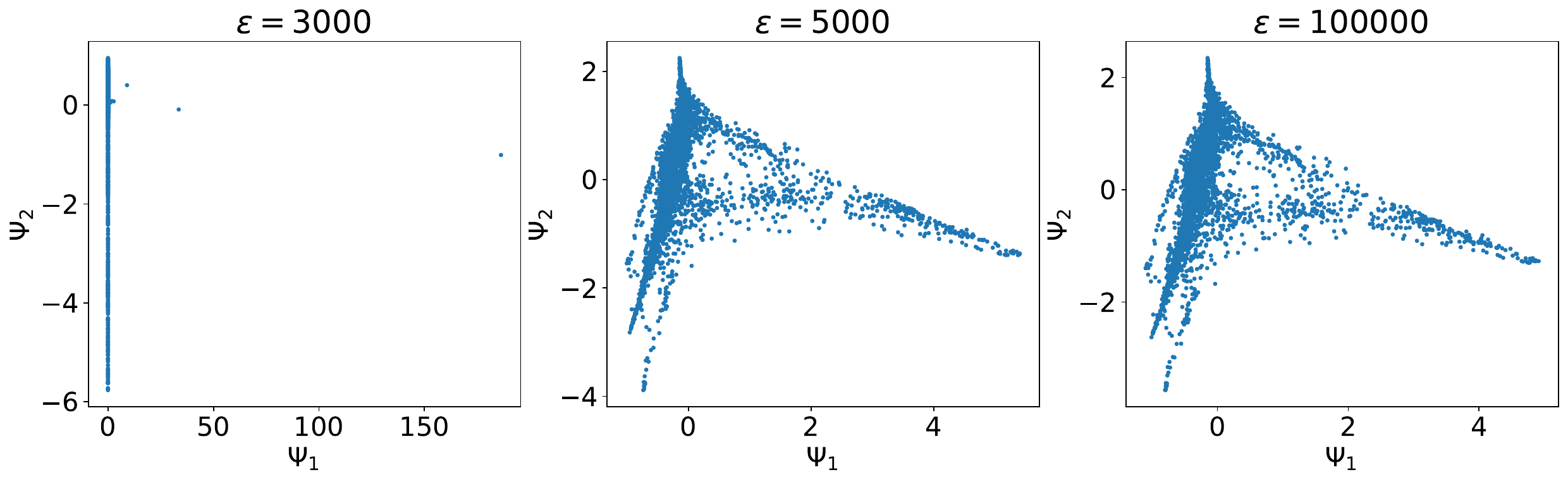}
    \caption{Consideration of the effect of different $\epsilon$ for the \textit{Diffusion Map} of the census dataset, when using $N=10$.}
    \label{fig:census_finding_epsilon_N10}
\end{figure}

The result can be seen in figure \ref{fig:census_finding_epsilon_N10}. We observe that the \textit{Diffusion Map} yields stable results for $\epsilon > 5000$. We also note that once $\epsilon$ is sufficiently large to prevent the data from disconnecting, it no longer influences the \textit{Diffusion Map}. Thus, $N$ is the dominant neighborhood parameter, which is also an argument that the results should remain the same despite the different kernel function, since the parameter $N=10$ only allows a very small neighborhood.

Now, the question is which meaning the \textit{Diffusion Map} structure hold and what interpretations can be assigned to the individual diffusion components to enable a comparison with the results from \cite{Barter2019}. For this purpose, we will examine the correlations between individual indices and the diffusion components.

\subsubsection{First diffusion component $\Psi_1$ - indicating school education}

We begin by considering $\Psi_1$ and examining the highest correlations. In this case, $\Psi_1$ exhibits high Spearman rank correlations $\rho_{s}$ with indices that indicate a high educational level among the inhabitants. These include, for example, \texttt{QS502EW0005}, \texttt{QS502EW0008}, \texttt{QS502EW0010}, and \texttt{KS501EW0007}, each with $\rho_{s} > 0.88$, which correspond to high school and academic qualifications. Additionally, several indices that indicate a higher level of professional occupations among the inhabitants, such as \texttt{KS608EW0003}, \texttt{KS609EW0003}, \texttt{KS611EW0002}, and \texttt{KS611EW0004}, all show Spearman correlations of $\rho_{s} > 0.84$. This collectively suggests a high number of scientific personnel residing in the output areas. In contrast, the number of students shows a relatively low Spearman correlation of $\rho_{s} = 0.31$. \\

Now, the question arises: where does the discrepancy with the results of \cite{Barter2019} originate, given that they represent the number of students through $\Psi_1$ and consider this index to be the most significant explanatory variable for the dataset?\\

This could possibly be explained on the one hand by examining the linear Pearson correlation $\rho_{p}$ instead. Here, we find the variables that indicate student numbers.  Among these are the number of students and schoolchildren \texttt{KS501EW0010} with $\rho_{p} = 0.86$, the negative correlation with the number of inhabitants without a secondary address (as students typically have a second home address) \texttt{QS106EW0002} $\rho_{p} = -0.87$, and, as seen previously with the Spearman correlations: inhabitants with a higher school certificate \texttt{QS502EW0008} $\rho_{p} = 0.86$. This significant difference between both correlation measures may also arise from the fact that the data are not Gaussian distributed, which is a condition for the meaningful application of the Pearson correlation.\\

\begin{figure}[H]
\centering
\subfigure[]{
   \centering
   \label{fig:census7}
   \includegraphics[height = 4.9cm]{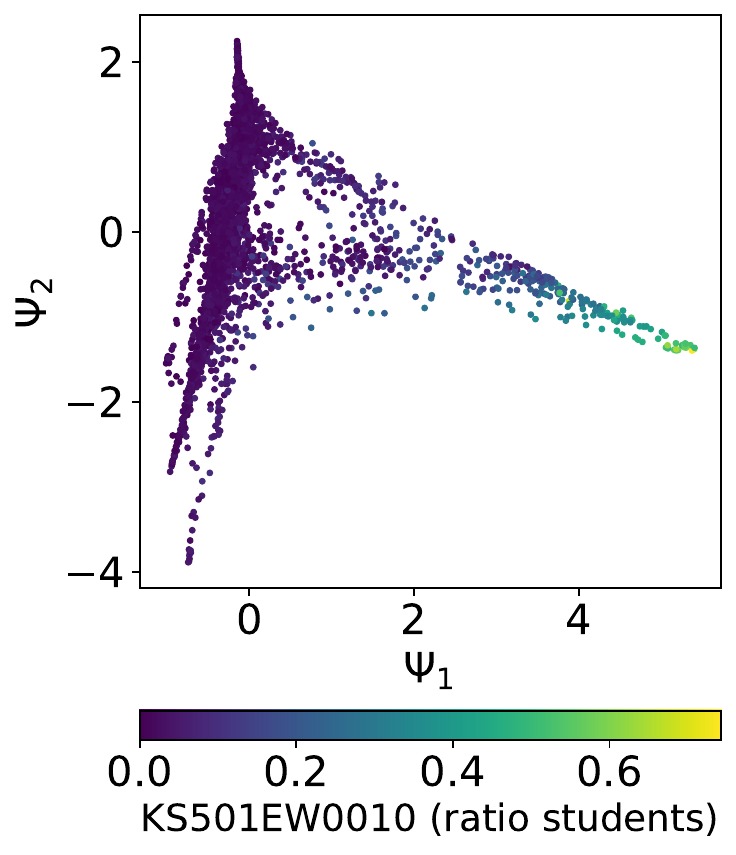}}~
\subfigure[]{
   \centering
   \label{fig:census8}
   \includegraphics[height = 4.9cm]{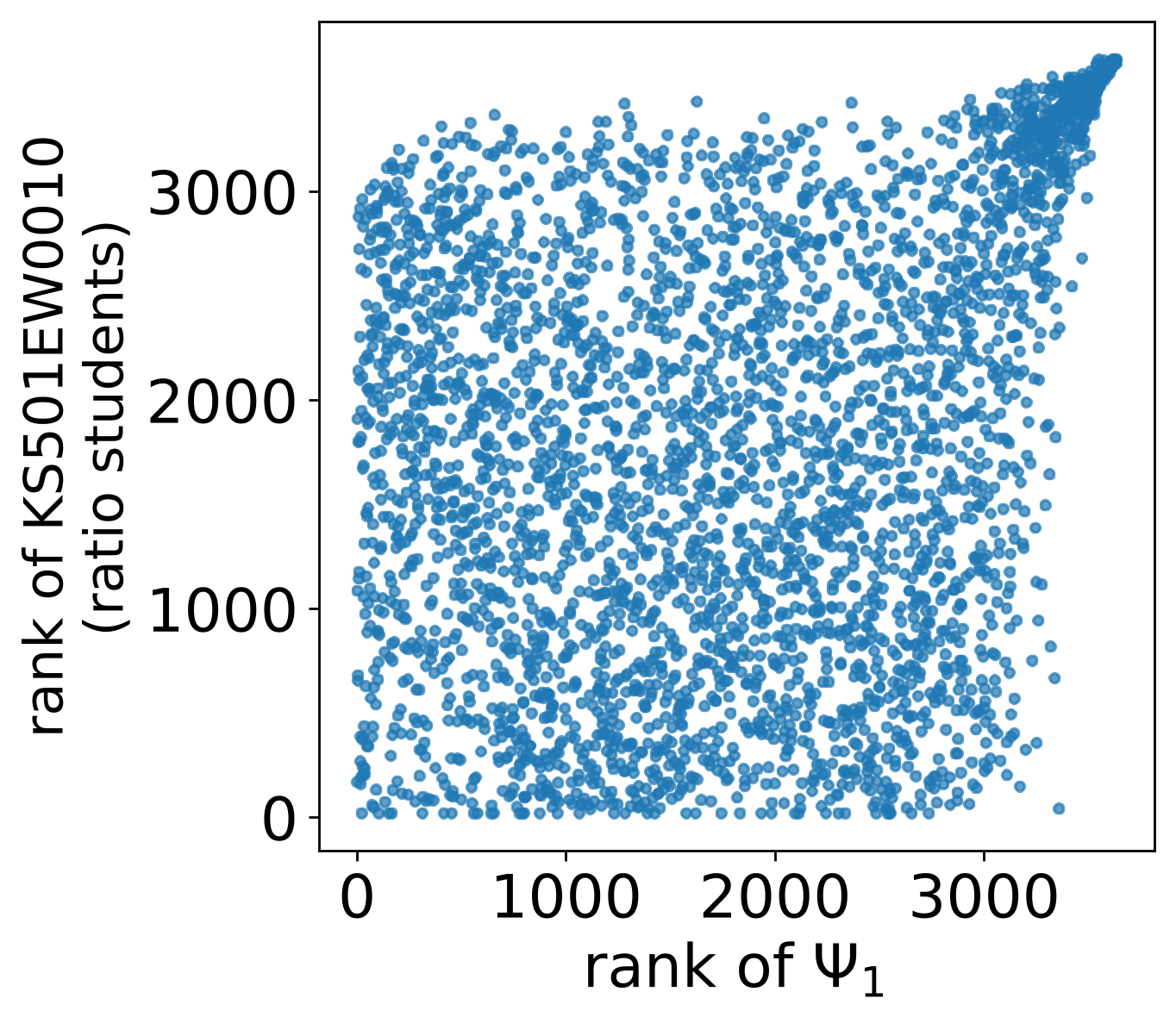}}~
\subfigure[]{
   \centering
   \label{fig:census9}
   \includegraphics[height = 4.9cm]{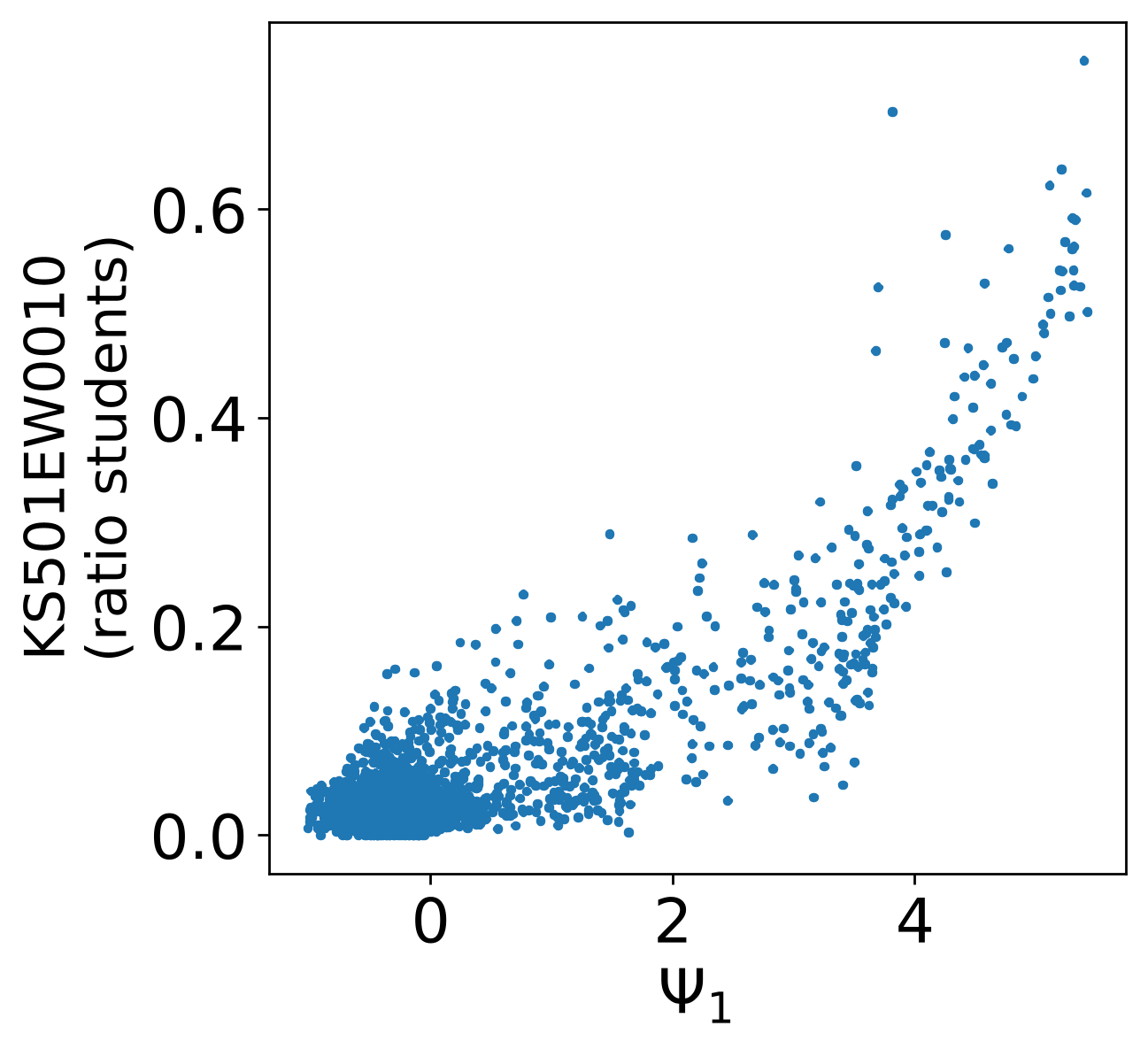}}\\
\subfigure[]{
   \centering
   \label{fig:census4}
   \includegraphics[height = 4.9cm]{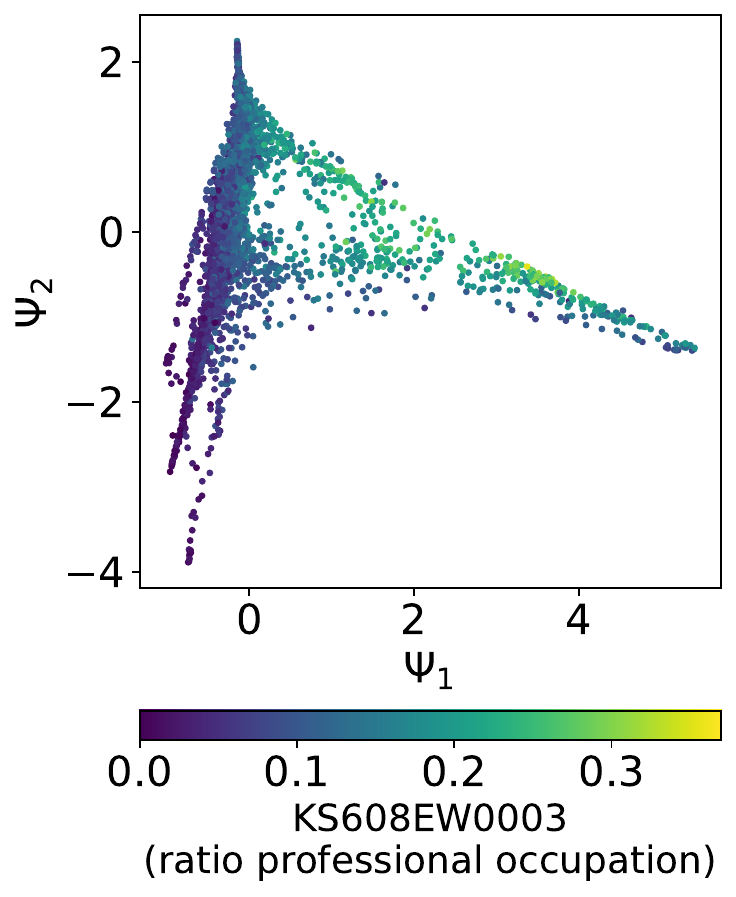}}~
\subfigure[]{
   \centering
   \label{fig:census5}
   \includegraphics[height = 4.9cm]{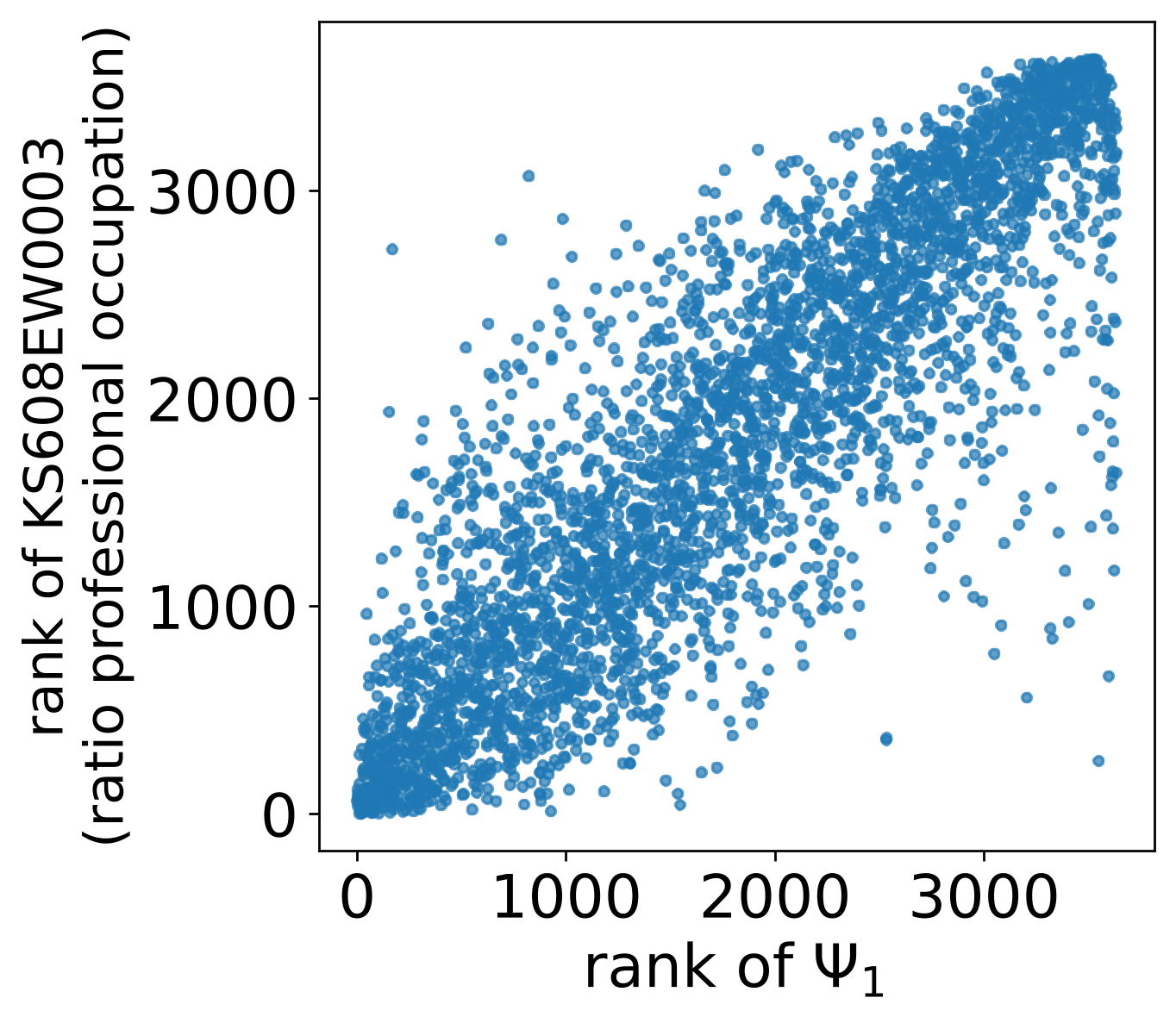}}~
\subfigure[]{
   \centering
   \label{fig:census6}
   \includegraphics[height = 4.9cm]{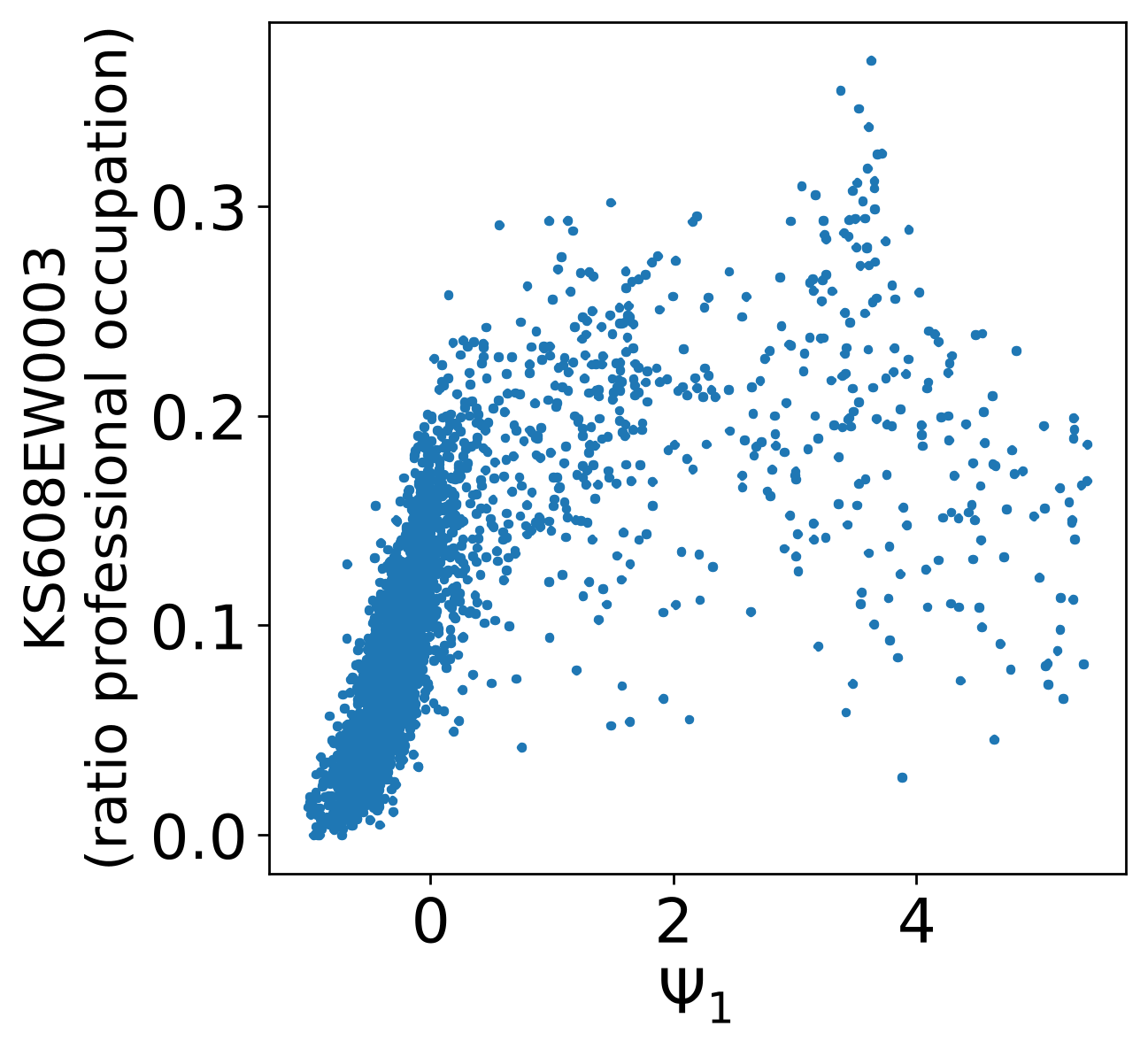}}\\
\subfigure[]{
   \centering
   \label{fig:census_1}
   \includegraphics[height = 4.9cm]{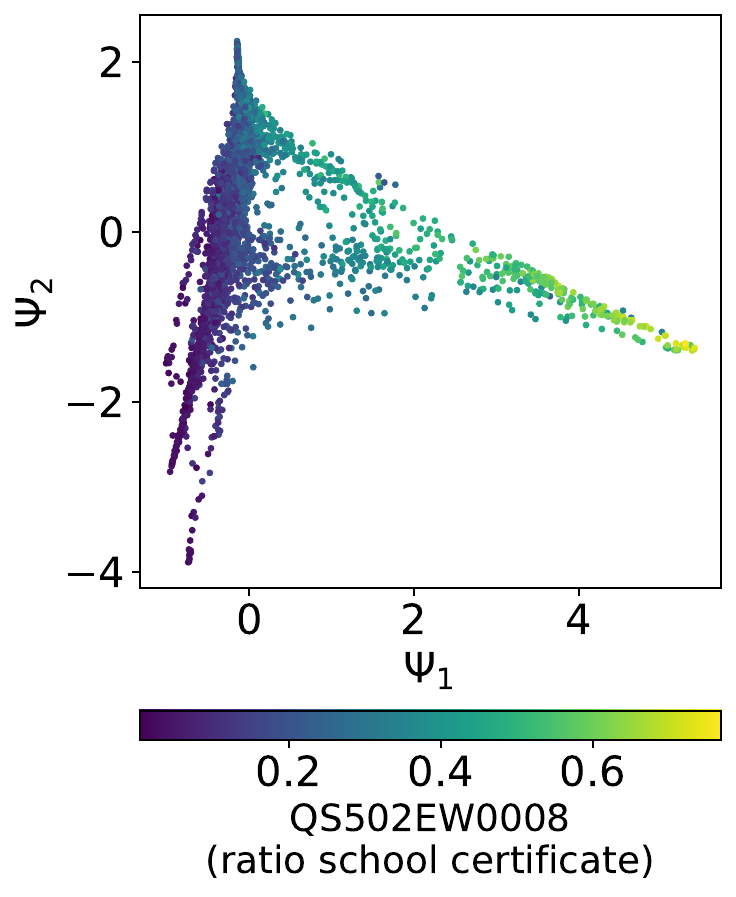}}~
\subfigure[]{
   \centering
   \label{fig:census2}
   \includegraphics[height = 4.9cm]{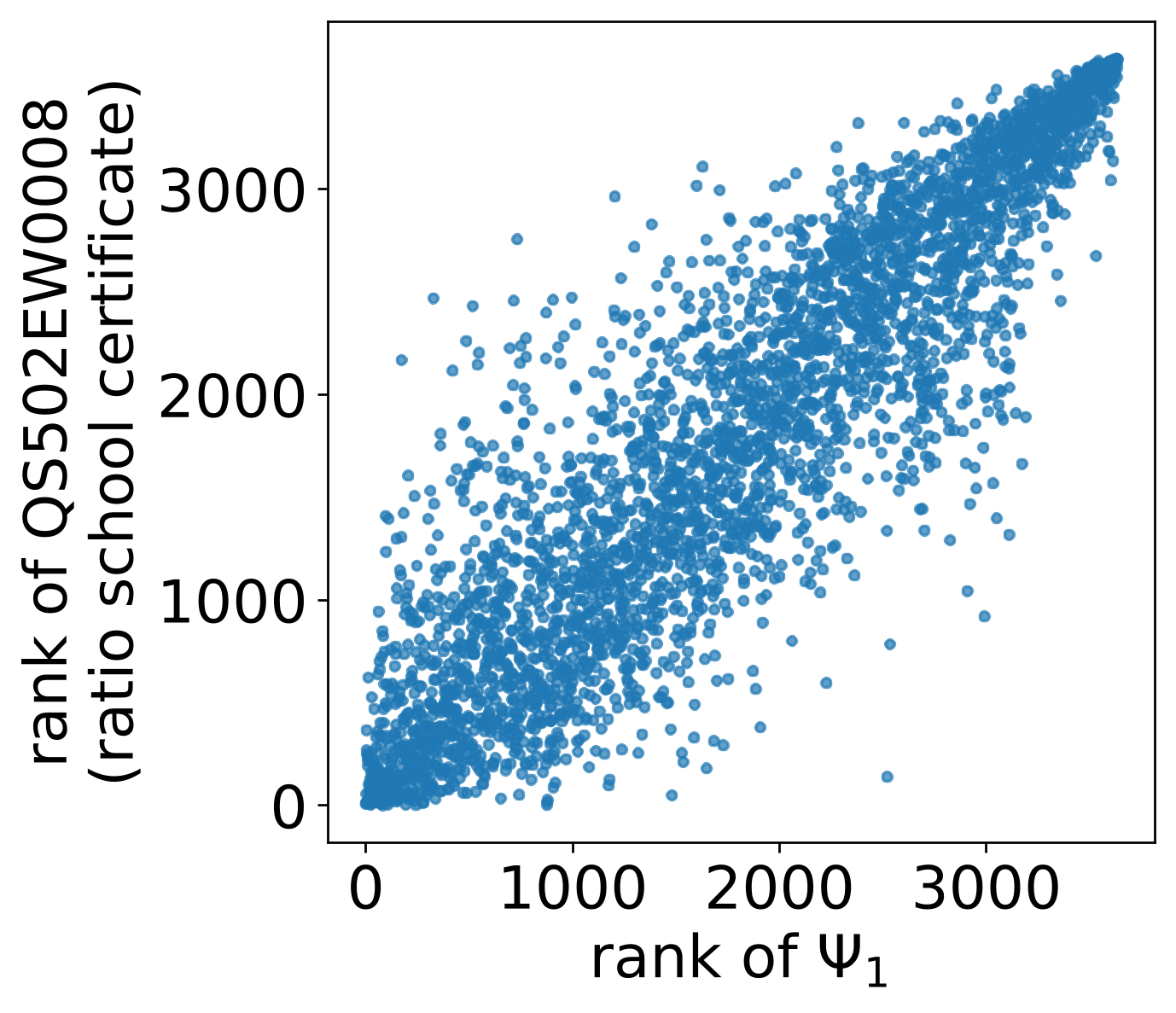}}~
\subfigure[]{
   \centering
   \label{fig:census3}
   \includegraphics[height = 4.9cm]{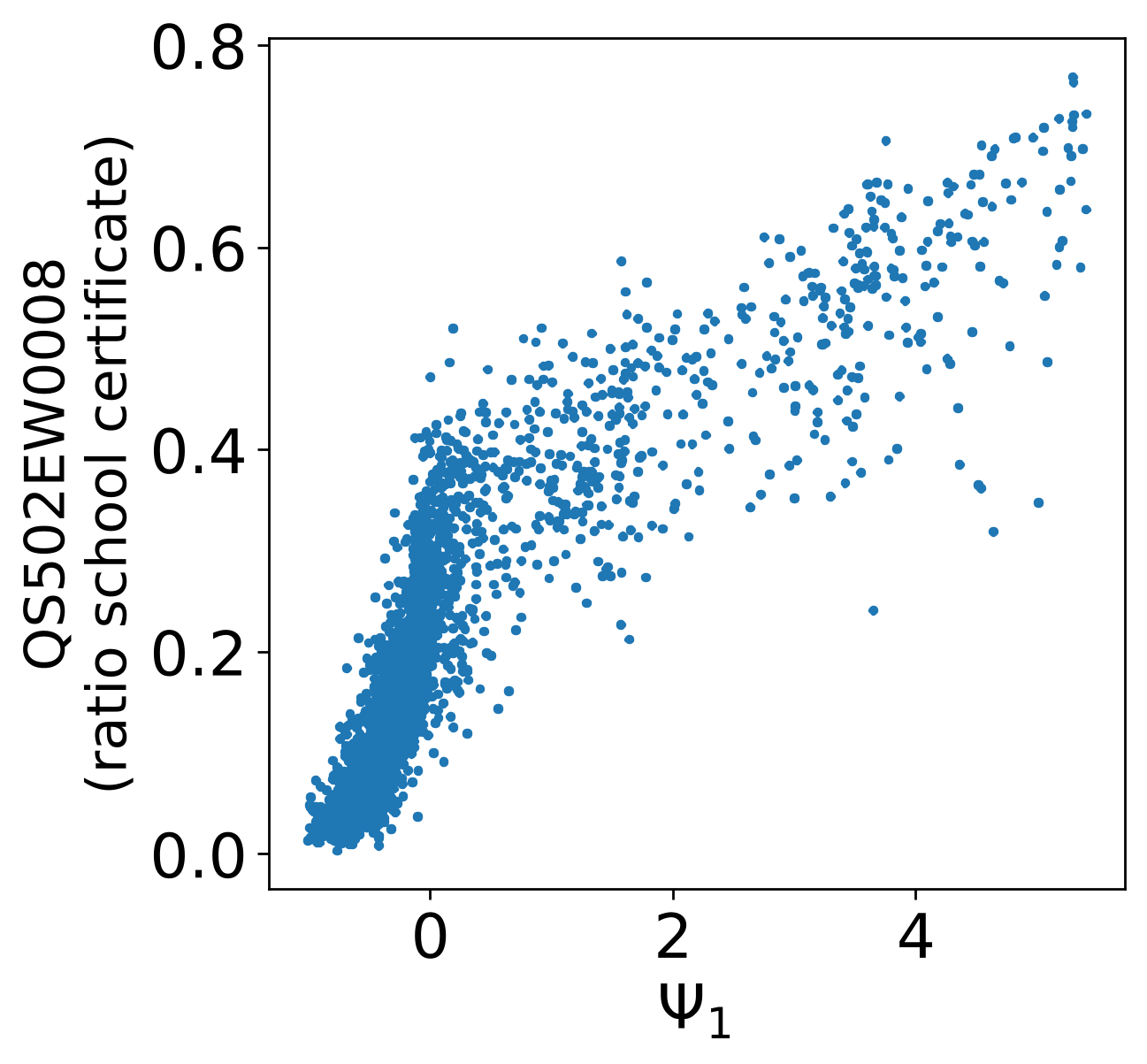}}\\
   \caption{\textbf{Finding interpretation of the first diffusion component $\Psi_1$:} The columns illustrate the relationships with the individual indices \texttt{QS502EW0008} (ratio school certificate), \texttt{KS608EW0003} (ratio professional occupation), and \texttt{KS501EW0010} (ratio students), see table \ref{tab:census_description} for their exact meanings. Each row, respectively, shows the \textit{Diffusion Map} with the first two diffusion components—with a color scale corresponding to the respective index—the rank–rank distribution of the index versus $\Psi_1$ (which can be used to illustrate the Spearman rank correlation), and the index plotted against $\Psi_1$.\\~\\}
   \label{fig:census_meaning_psi1}
\end{figure}

On the other hand, \cite{Barter2019} derives their statement from the observation that the highest values of $\Psi_1$ occur in areas where student dormitories are located. This statement is correct and can be explained by examining figure \ref{fig:census_meaning_psi1}, which illustrates selected variables from the dataset and their correlations. As seen in the two-dimensional \textit{Diffusion Map}, which is color-coded by the student index (see figure \ref{fig:census7}) locations with a high concentration of students are indeed positioned at the top of the \textit{Diffusion Map}, corresponding to large values of $\Psi_1$. Furthermore, figures \ref{fig:census8} and \ref{fig:census9} show a strong Spearman and Pearson correlation in regions with high $\Psi_1$. However, figure \ref{fig:census8} particularly highlights that for small $\Psi_1$, this correlation no longer holds, indicating that the number of students is no longer associated with $\Psi_1$ in this area.\\

When examining the plots and correlations of the ratio of professional occupations (figures \ref{fig:census4}, \ref{fig:census5}, and \ref{fig:census4}), we can observe that, conversely to the index of student numbers, this index is not correlated with data points exhibiting large $\Psi_1$. In figure \ref{fig:census4}, the peak for high $\Psi_1$ appears darker, indicating a lower proportion of professional occupations, although both should be positively correlated. In the rank distribution shown in figure \ref{fig:census4}, a wide distribution of data points is observed for large values, and in figure \ref{fig:census5}, the data points are not correlated for approx. $\Psi_1 > 1$.
Thus, the number of professional occupations alone is not sufficient to fully explain the observed patterns.\\

However, we can observe that a single index can unify both effects: As previously described, the index for the ratio of school certificates exhibits the strongest Spearman correlation. Moreover, in figures \ref{fig:census_1}, \ref{fig:census2}, and \ref{fig:census3}, we see that this index remains well correlated with $\Psi_1$ across its entire range.

This is a logical finding, as this index combines the effects of both the student index and the professional occupations index. In both cases, one requires high school certificates to enroll in university, complete the studies, and subsequently enter professional occupations.

Thus, contrary to the claim made by \cite{Barter2019}, it is not the number of students per se which has explanatory power for $\Psi_1$ and the dataset, but rather the level of educational qualifications, which appears to be one of the key explanatory indices. This interpretation seems also more plausible as these qualifications should have a major influence on the living and housing situations surveyed by the census, although the exact causal relationships are not clear \cite{Alexiu2010}.

\subsubsection{Second diffusion component $\Psi_2$ - indicating financial status}

Now, we examine the second diffusion component, and we observe that it exhibits strong correlations with multiple variables that indicate the financial situation of the inhabitants.\\

\begin{figure}[H]
\centering
\subfigure[]{
   \centering
   \label{fig:census10}
   \includegraphics[height = 4.9cm]{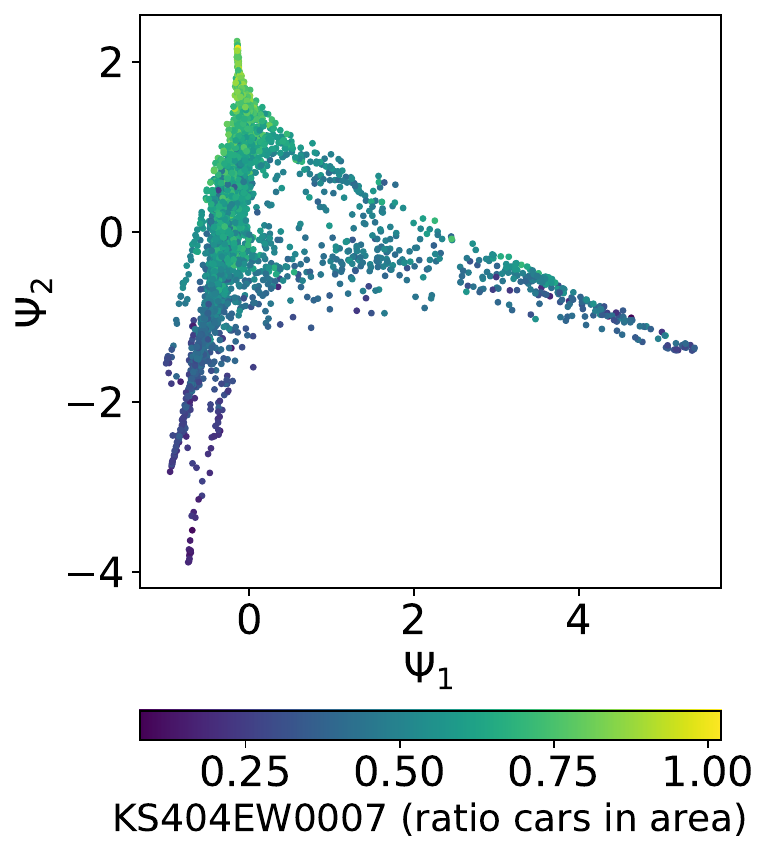}}~
\subfigure[]{
   \centering
   \label{fig:census11}
   \includegraphics[height = 4.9cm]{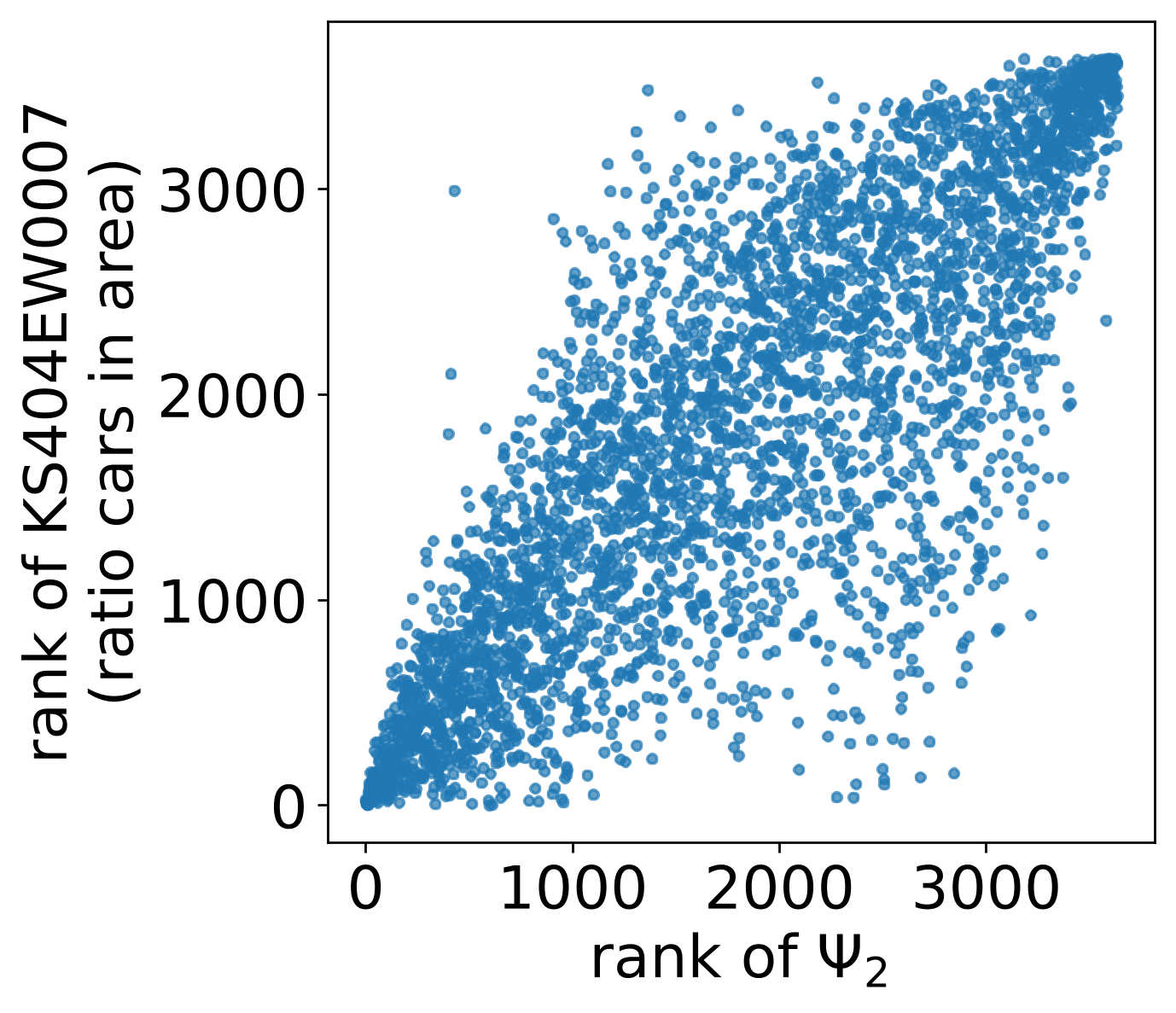}}~
\subfigure[]{
   \centering
   \label{fig:census12}
   \includegraphics[height = 4.9cm]{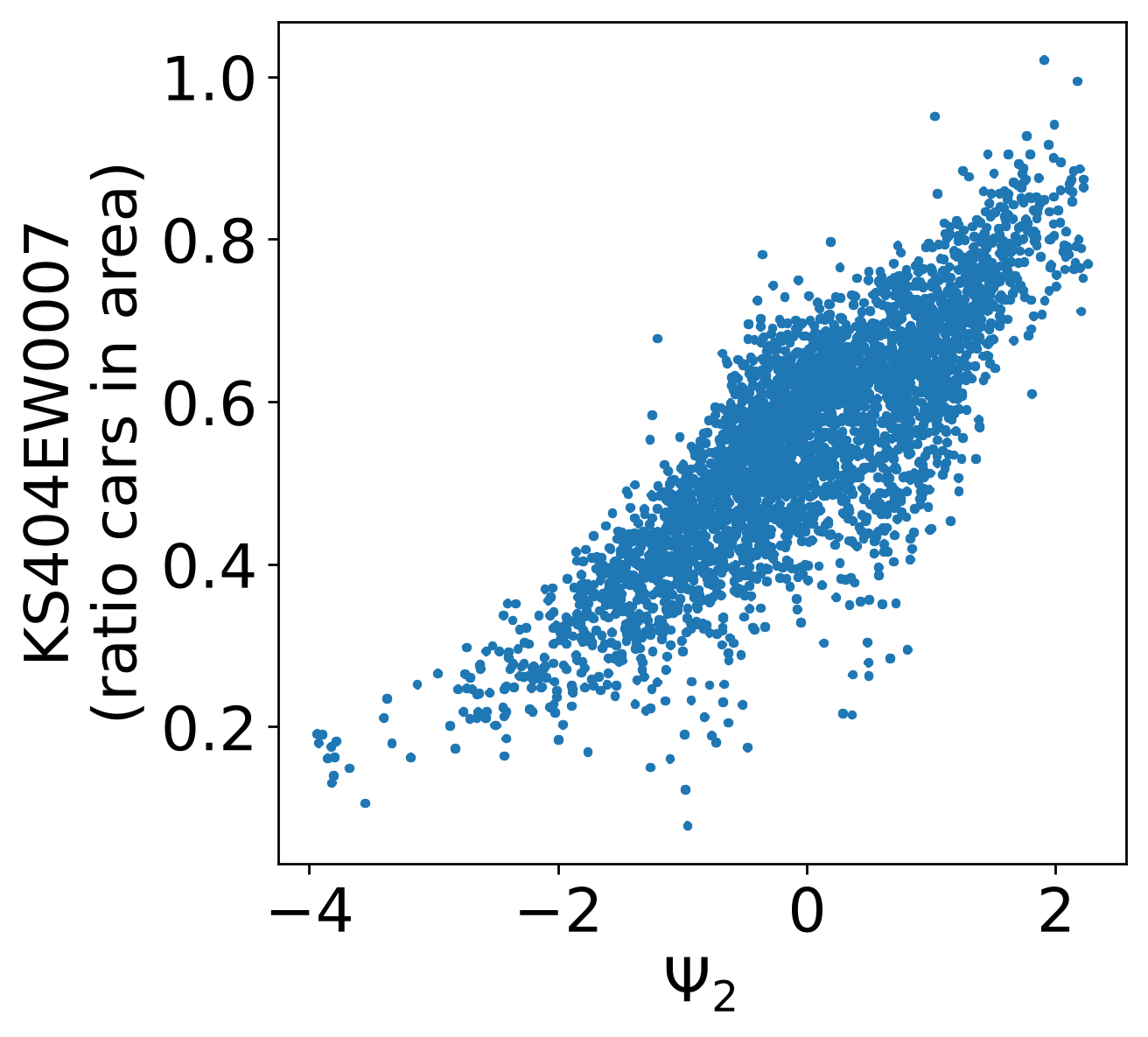}}\\
\subfigure[]{
   \centering
   \label{fig:census101}
   \includegraphics[height = 4.9cm]{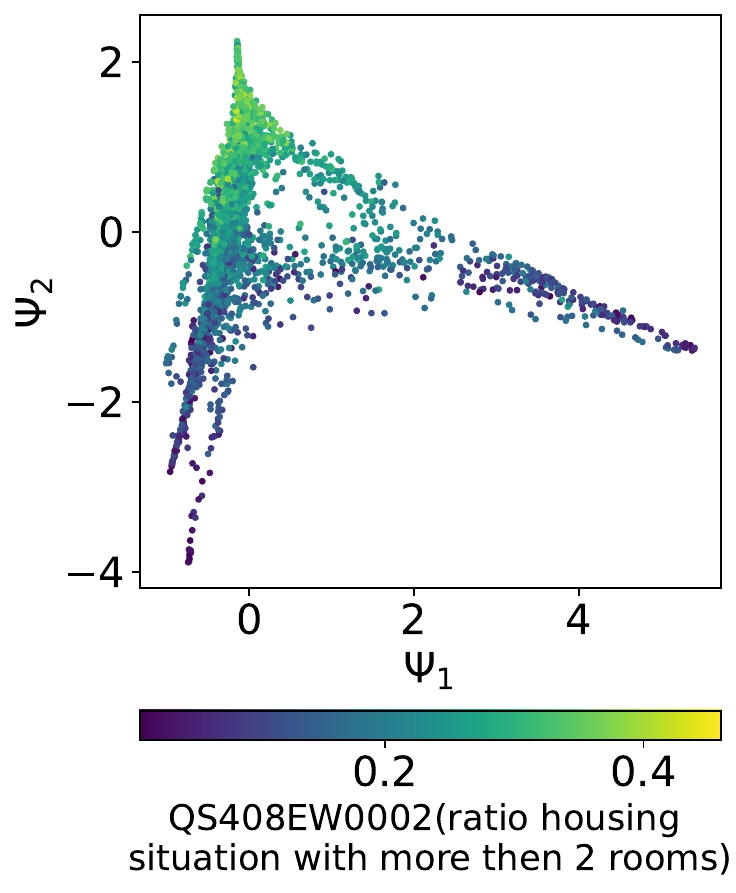}}~
\subfigure[]{
   \centering
   \label{fig:census111}
   \includegraphics[height = 4.9cm]{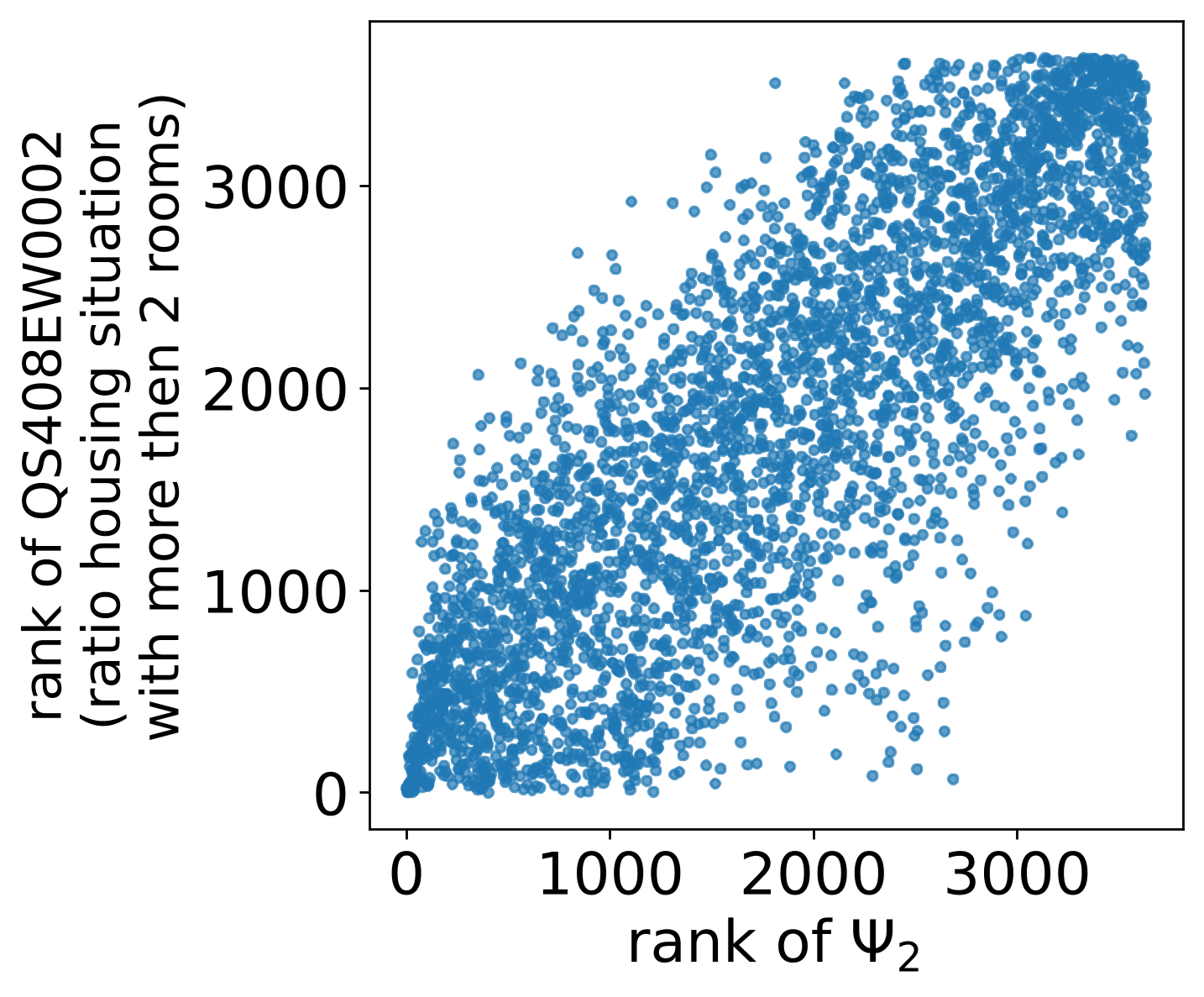}}~
\subfigure[]{
   \centering
   \label{fig:census121}
   \includegraphics[height = 4.9cm]{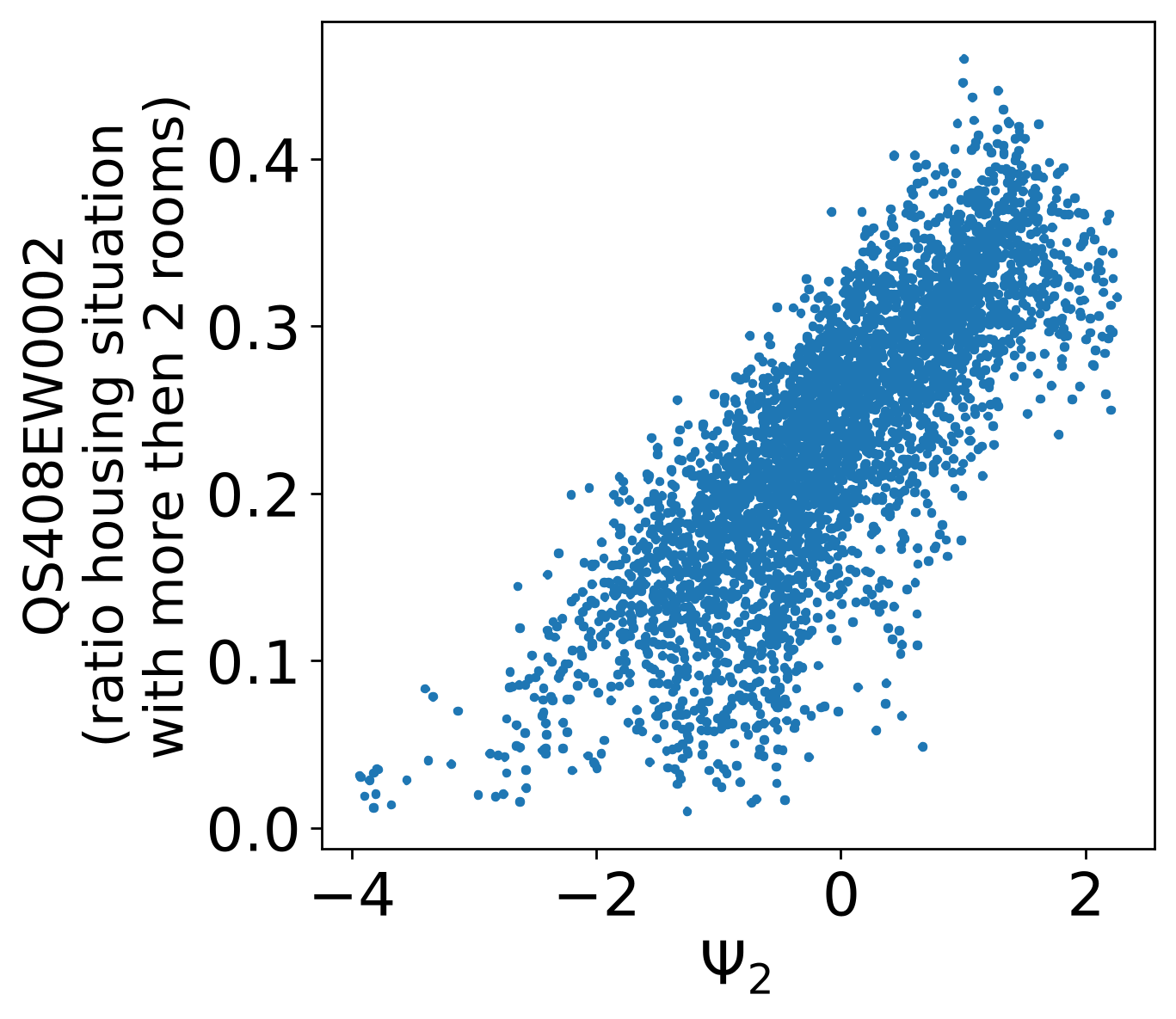}}\\
\subfigure[]{
   \centering
   \label{fig:census13}
   \includegraphics[height = 4.9cm]{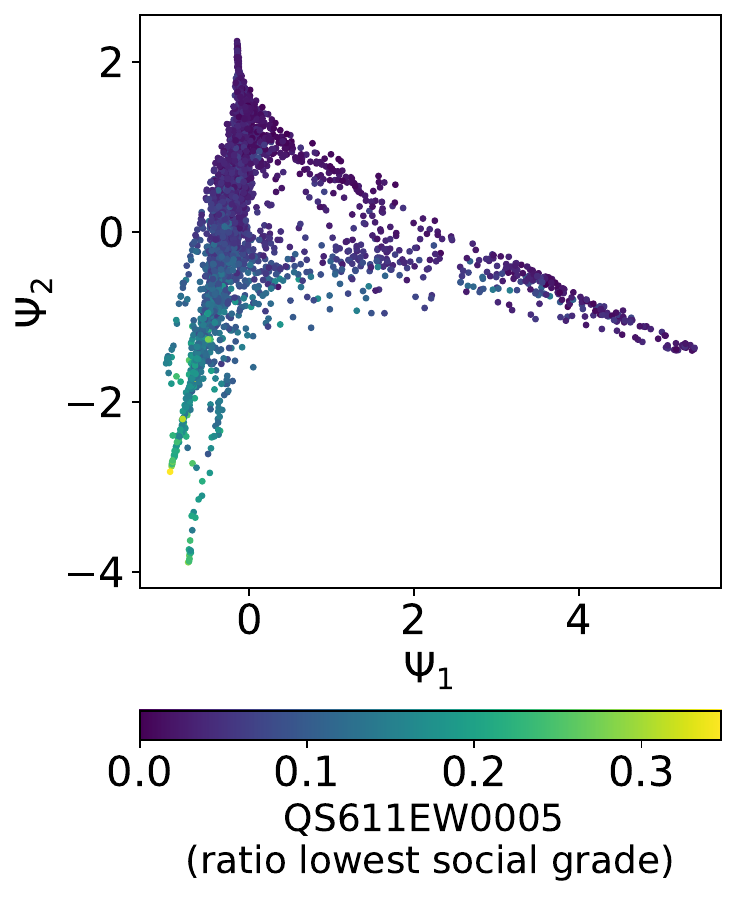}}~
\subfigure[]{
   \centering
   \label{fig:census14}
   \includegraphics[height = 4.9cm]{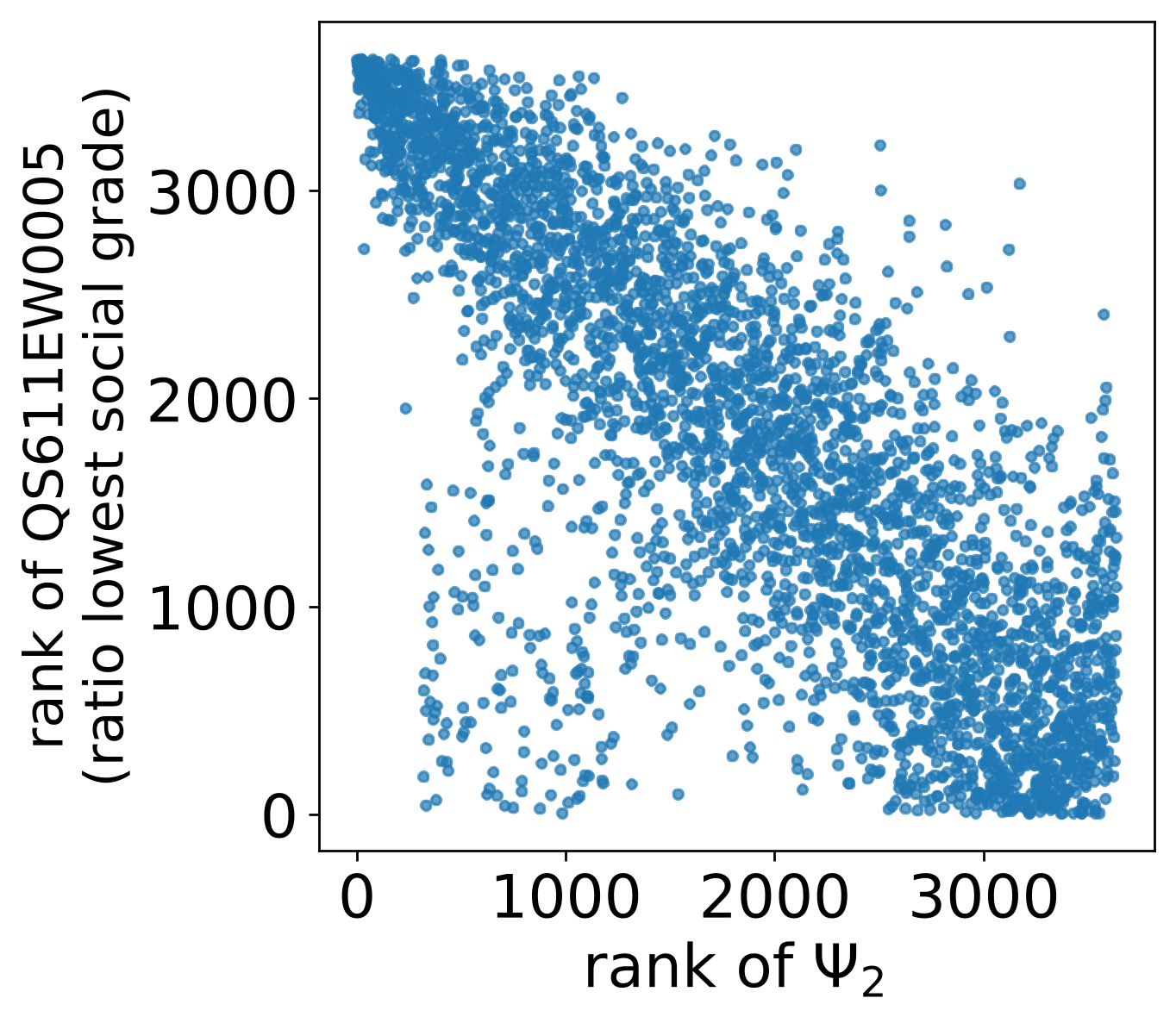}}~
\subfigure[]{
   \centering
   \label{fig:census15}
   \includegraphics[height = 4.9cm]{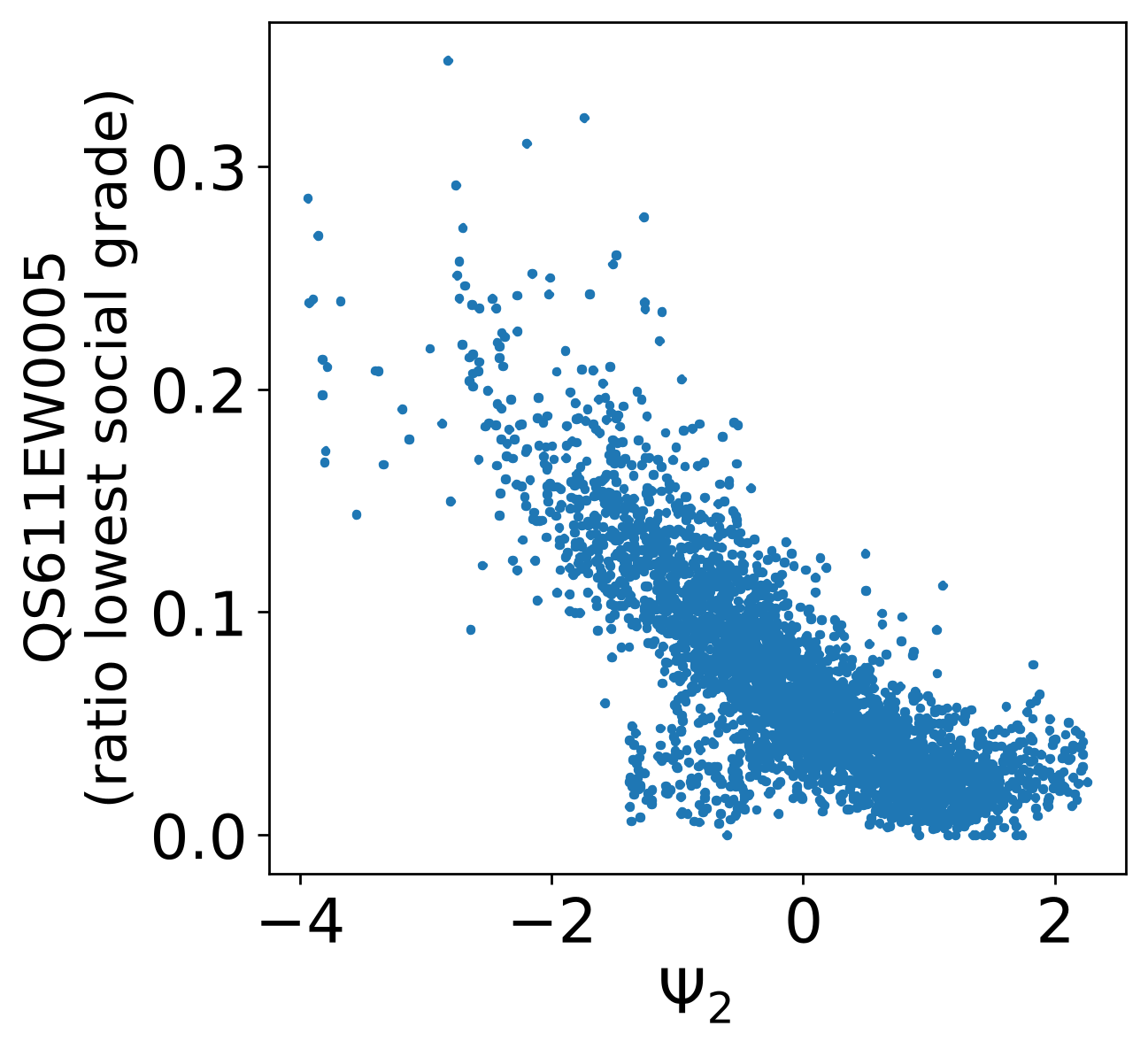}}\\
   \caption{\textbf{Finding interpretation of the second diffusion component $\Psi_2$:} The columns illustrate the relationships with the individual indices \texttt{KS404EW0007} (ratio of cars in the area), \texttt{QS408EW0002} (ratio housing situation with more then 2 rooms), and \texttt{QS611EW0005} (ratio lowest social grade), see table \ref{tab:census_description} for their exact meanings. Each row, respectively, shows the \textit{Diffusion Map} with the first two diffusion components—with a color scale corresponding to the respective index—the rank–rank distribution of the index versus $\Psi_1$ (which can be used to illustrate the Spearman rank correlation), and the index plotted against $\Psi_1$.}
   \label{fig:census_meaning_psi2}
\end{figure}

For instance, the variable \texttt{KS404EW0007}, which measures the number of cars in the examined output areas, shows a strong correlation of $\rho_s = 0.83$, $\rho_p = 0.83$, which can also be seen in figures \ref{fig:census11} and \ref{fig:census12}. In social sciences, it is well established that the number of cars can serve as an indicator of the wealth of citizens \cite{Galobardes2006}. There are also high correlations with the available living space. Specifically, the index \texttt{QS408EW0002}, which represents the number of households with more than two rooms, is highly correlated with $\Psi_2$, showing $\rho_s = 0.81$ and $\rho_p = 0.80$ (see also figure \ref{fig:census101}, \ref{fig:census111}, \ref{fig:census121} for a graphical illustration).\\

The interpretation becomes particularly evident when examining variables that encode the number of individuals with an estimated social status. For the lowest social status (\texttt{QS611EW0005}), we observe a strong negative correlation of $\rho_s = -0.79$ and $\rho_p = -0.80$ (see also figure \ref{fig:census13},\ref{fig:census14},\ref{fig:census15}), further supporting the hypothesis that $\Psi_1$ encodes the financial situation in the output areas.

These findings align with the results of \cite{Barter2019}, who also identified poverty, together with student numbers, as a key variable in describing and structuring the census data.

\subsubsection{Higher diffusion components - revealing ethnic identities and limitations of correlation analysis}
\label{census:higher_components}

\begin{figure}[h]
    \centering
    \includegraphics[width=0.6\linewidth]{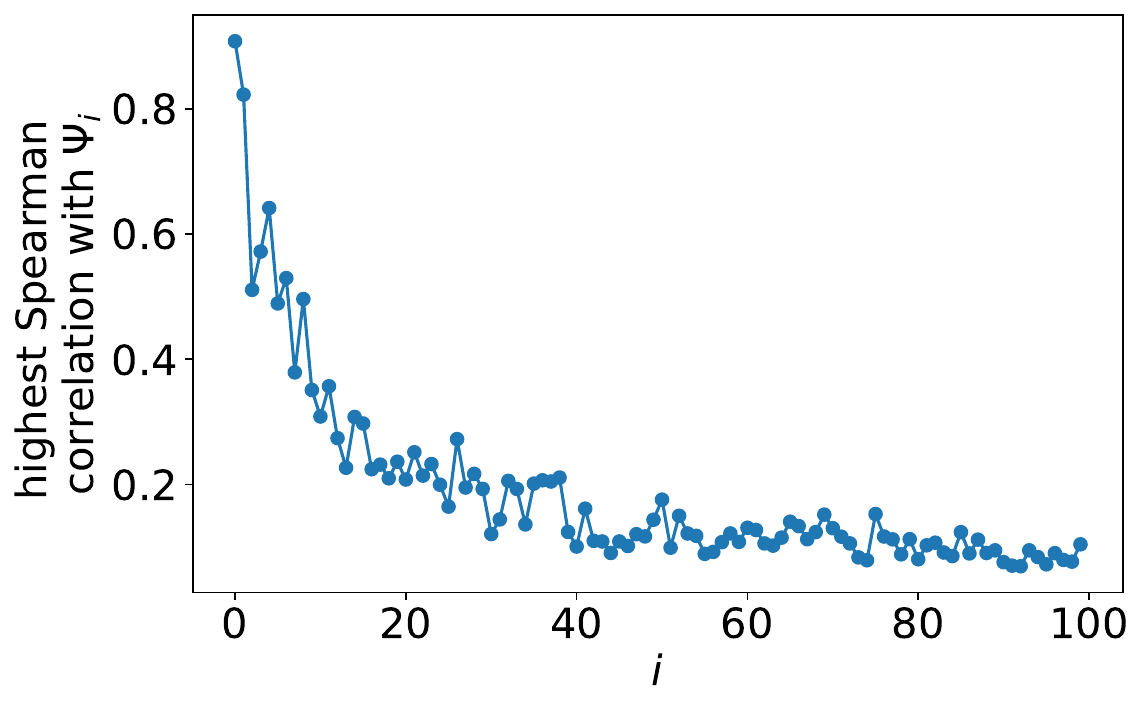}
    \caption{For each of the first 100 diffusion components $\psi_i$, the highest Spearman correlation with one of the dataset indices is plotted. The diffusion parameters are $N=10$, $\epsilon = 5000$.}
    \label{fig:census_highest_correlations}
\end{figure}

From the third diffusion component onward, it becomes increasingly difficult to interpret its meaning using correlation analysis. As shown in figure \ref{fig:census_highest_correlations}, the highest measured Spearman rank correlation decreases for higher components. Since a strong Spearman correlation is crucial for assigning a clear meaning to individual diffusion components, as demonstrated with $\Psi_1$ and $\Psi_2$, this no longer seems achievable for higher diffusion components.

\begin{figure}[H]
\centering
\subfigure[]{
   \centering
   \label{fig:census_psi_3_uk}
   \includegraphics[width = 0.44\textwidth]{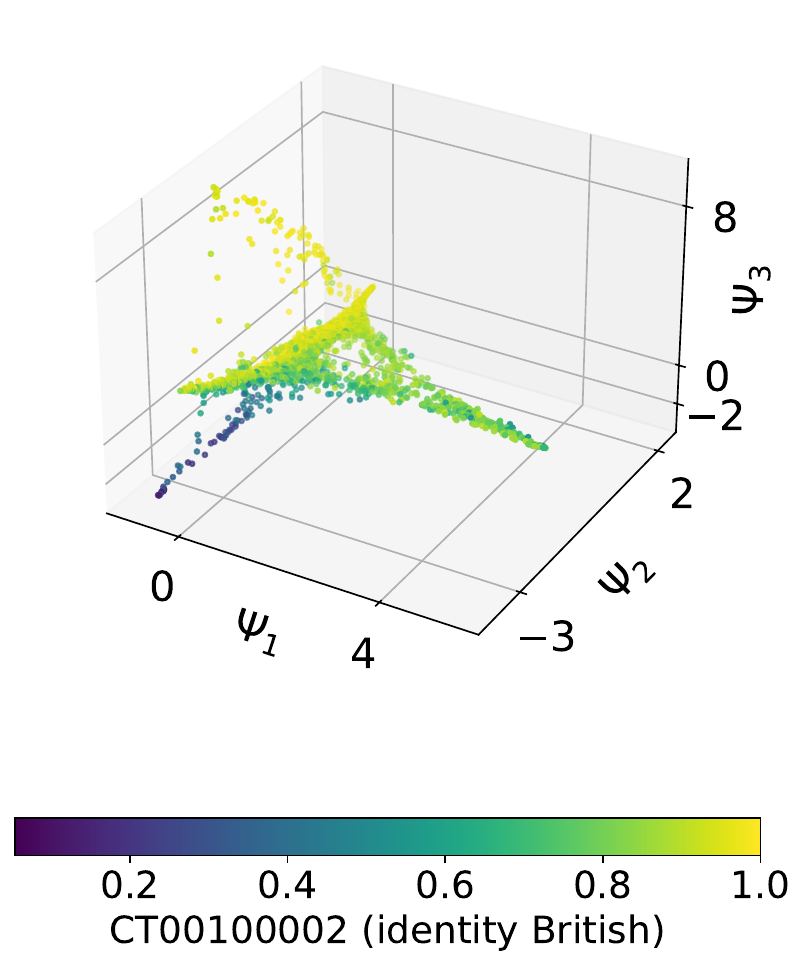}} \hfill
\subfigure[]{
   \centering
   \label{fig:census_psi3_welsh}
   \includegraphics[width = 0.44\textwidth]{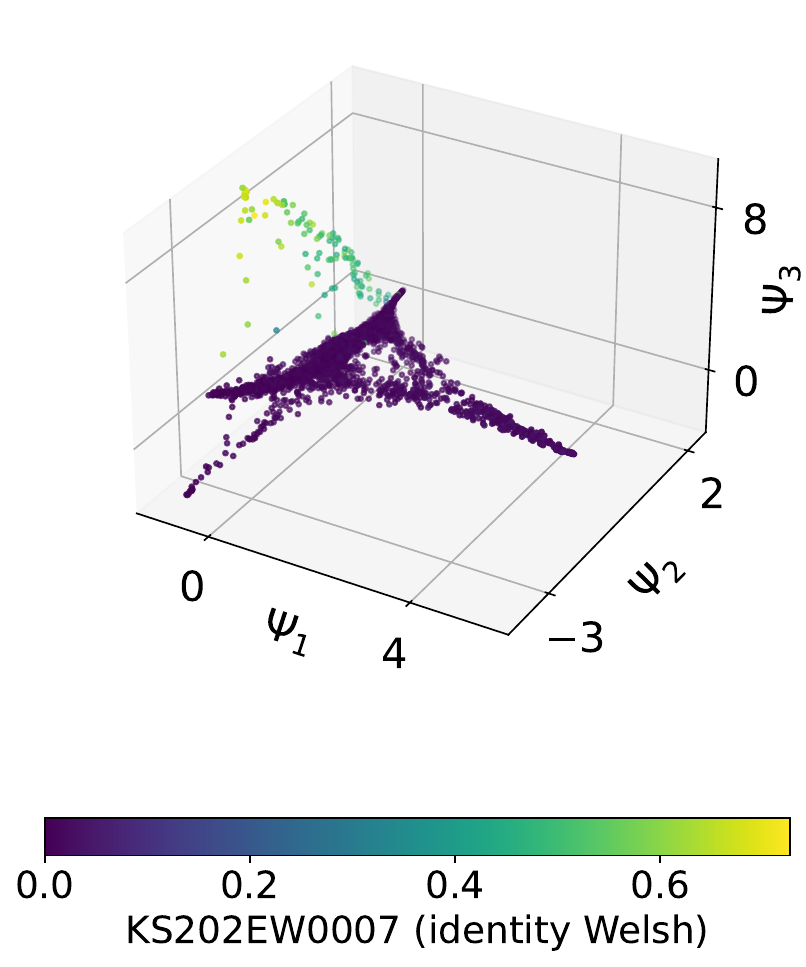}}
\subfigure[]{
   \centering
   \label{fig:census_psi_3_uks}
   \includegraphics[width = 0.38\textwidth]{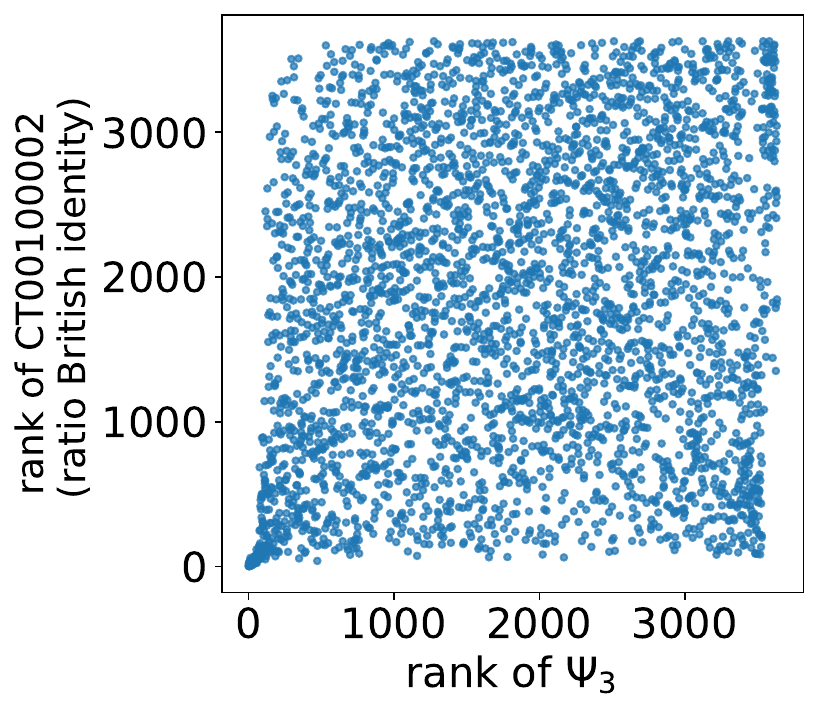}} \hfill
\subfigure[]{
   \centering
   \label{fig:census_psi3_welshs}
   \includegraphics[width = 0.38\textwidth]{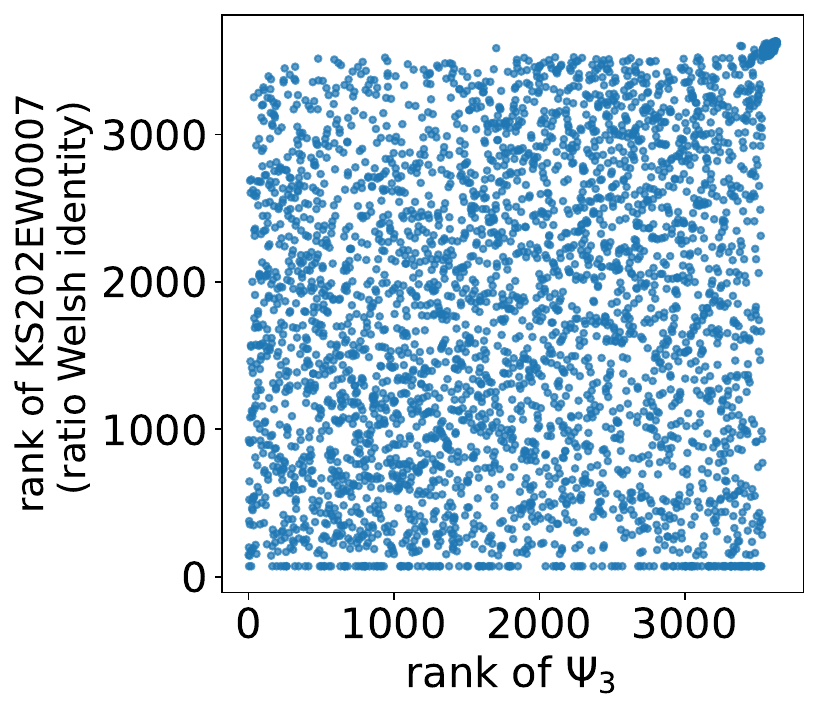}}
\subfigure[]{
   \centering
   \label{fig:census_psi_3_ukp}
   \includegraphics[width = 0.38\textwidth]{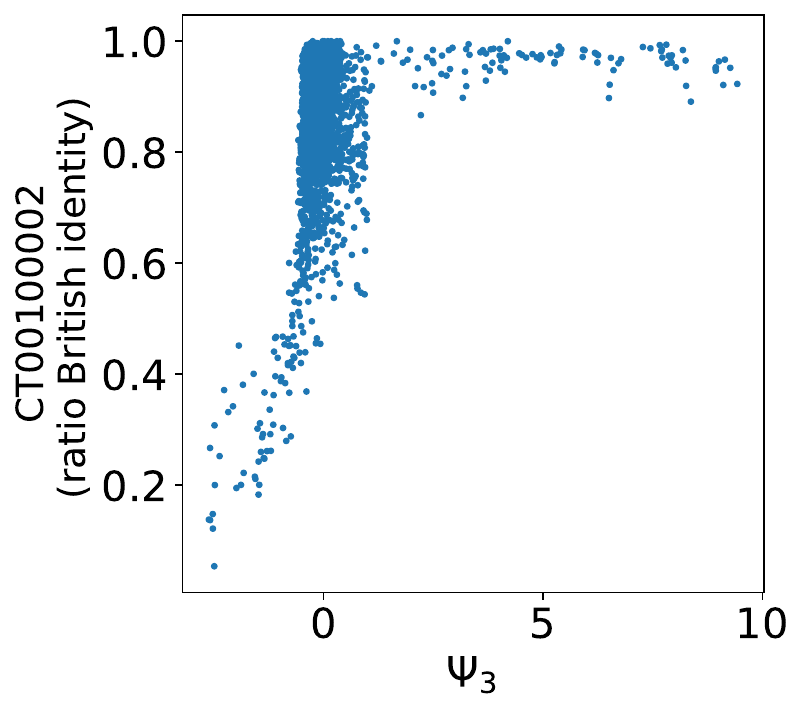}} \hfill
\subfigure[]{
   \centering
   \label{fig:census_psi3_welshp}
   \includegraphics[width = 0.38\textwidth]{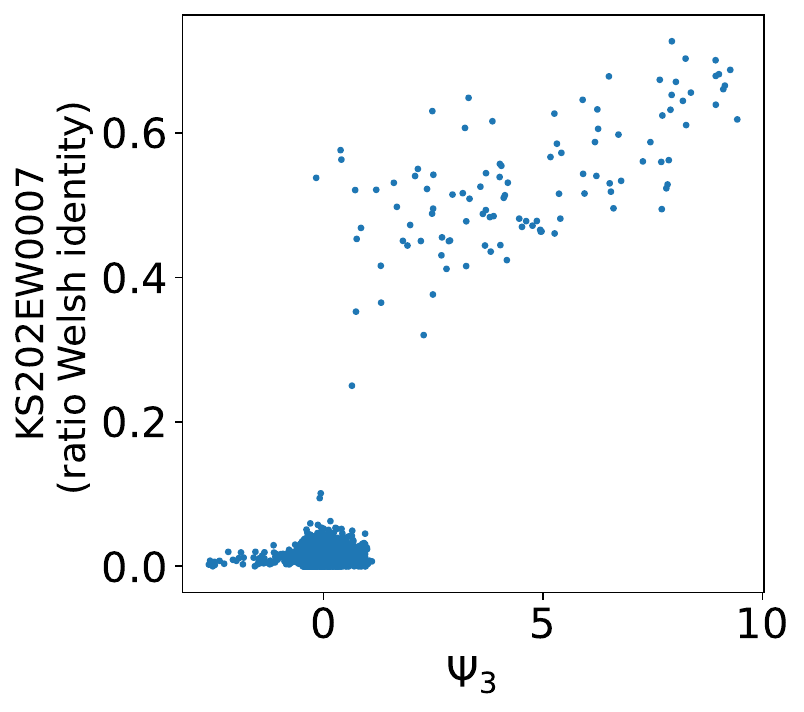}}
   \caption{\textbf{Identifying the meaning of \textit{Diffusion Map} structures in $\Psi_3$:} The plots in the first row show the \textit{Diffusion Map} with the parameters $\epsilon = 5000$ and $N=10$ in the first three diffusion components. The data points are color-coded based on the indices \texttt{CT00100002} (representing the number of people who identify as British) and \texttt{KS202EW0007} (representing the number of people with a Welsh identity). In the second row the rank distributions of both indices according to $\Psi_3$ are shown to visualize the Spearman correlations. In the third row the indices are plotted against $\Psi_3$ for illustrating the Pearson correlation}
   \label{fig:census_meaning_psi3}
\end{figure}

Beyond the holistic interpretation of the different components, smaller features occur and can also be explained, even when no strong Spearman correlation is observed. For instance, one of the peaks visible in the \textit{Diffusion Map} (also when considering only a two-dimensional representation with $\Psi_1$ and $\Psi_2$) corresponds to the proportion of residents who do not identify as British (consider the dark color on one side of the \textit{Diffusion Map} in figure \ref{fig:census_psi_3_uk}). It is possible that these highlighted data points with a very small percentage of British people belong to output areas with refugee accommodations. However, this hypothesis cannot be verified without local knowledge.
The other shown peak is explainable by the (for the data set) unusually high percentage of people with a Welsh identity (consider the light color at the data points with large $\Psi_3$ values in figure \ref{fig:census_psi3_welsh}). 
The reason for this high percentage of Welsh people can be explained by the fact that Bristol is located directly on the border to Wales, and some of the output areas in the dataset belong to Welsh regions. This result also reminds us of the discussions in section \ref{vdem_suffrage} about the discrete variables and their large effect on the \textit{Diffusion Map} result, since here it also seems as if the number of Welsh people is either close to 0 (in the English output areas) or 1 (in the Welsh output areas), which leads to the specific structures in the \textit{Diffusion Map}.

It is important to emphasize that the diffusion component encodes multiple meanings: on one hand, Welsh identity, and on the other, non-British identity. This demonstrates that single diffusion components can carry multiple meanings, which have to be considered during the correlation analysis of components.\\

When analyzing the correlations of these two indices, it also becomes evident how a high Pearson correlation can quickly be misleading, as we previously observed with student populations in comparison to $\Psi_1$. Looking at figures \ref{fig:census_psi_3_uks}, \ref{fig:census_psi3_welshs}, \ref{fig:census_psi_3_ukp}, and \ref{fig:census_psi3_welshp}, we see that both indices describe only a small subset of the data. Thus, $\Psi_3$ and the ratio of British identity are correlated only for small values of $\Psi_3$, while $\Psi_3$ and the ratio of Welsh identity are correlated only for large values of $\Psi_3$. Nevertheless, \texttt{KS202EW0007} (ratio of Welsh identity) exhibits a high Pearson correlation of $\rho_p = 0.87$ despite a much lower Spearman correlation of $\rho_s = 0.21$. Similarly, for \texttt{CT00100002} (ratio of British identity), the Pearson and Spearman correlations are $\rho_p = 0.22$ and $\rho_s = 0.19$, respectively. As already mentioned, the reason why the Pearson correlation cannot be used is that the data would have to be Gaussian distributed without to much outlier. For both indices, however, it is clear that most data points are around 1 for British identity and around 0 for Welsh identity. The data points with British identity at 0 or Welsh identity near 1 are strong outliers. \\

When we compare this result with the literature, we must also point out that the statement by \cite{Barter2019} suggesting that ethnic identities do not appear in the higher diffusion components is not correct for our implementation of the \textit{Diffusion Map} algorithm.

\cite{Barter2019} made this statement based on the fact that ethnic background is only captured by single variables and therefore should not contribute significantly to the \textit{Diffusion Map} result. However, our findings regarding ethnic identities in the highest diffusion components highlight the effect that individual variables can have in altering the appearance of the \textit{Diffusion Map}.\\

\begin{figure}[h]
    \centering
    \includegraphics[width=0.5\linewidth]{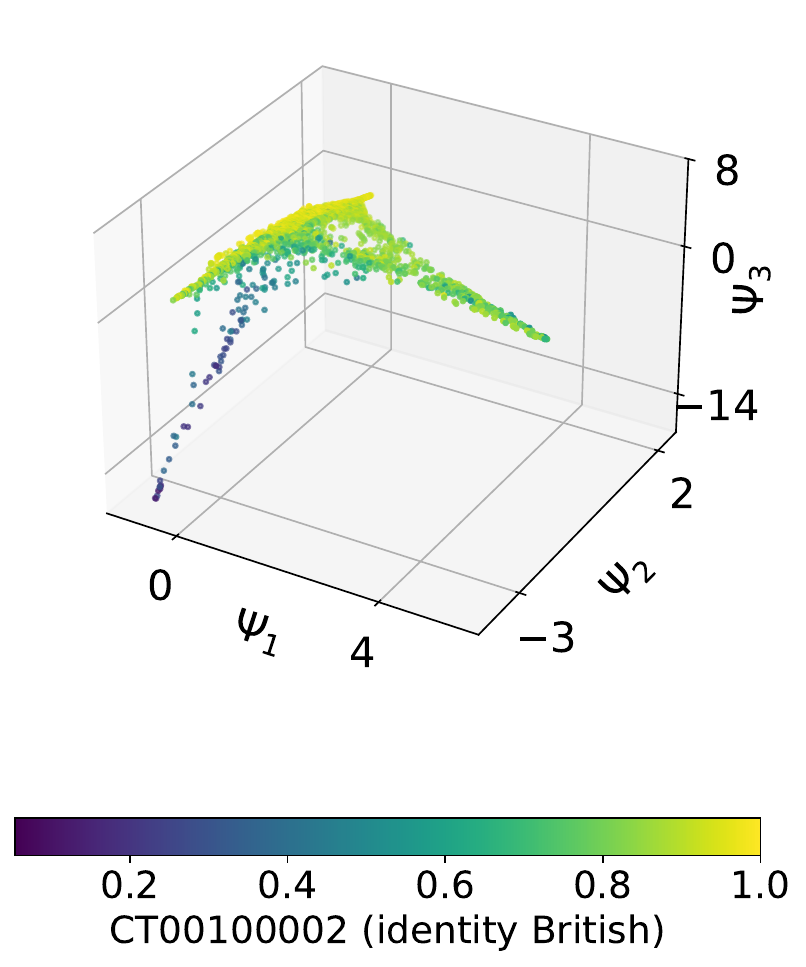}
    \caption{\textbf{Showing the effect of dropping the information about Welsh identity:} for this plot the \textit{Diffusion Map} with $\epsilon=5000$ and $N=10$ was calculated without considering the variable \texttt{KS202EW0007} (ratio Welsh identity). As an effect, the feature that previously encoded the welsh identity is no longer present. The \textit{Diffusion Map} was color-coded using the variable \texttt{CT00100002} (ratio British identity) to reveal that the feature encoding non-British identity has now changed its appearance. The data points of the Welsh output areas are now positioned along the main line of the \textit{Diffusion Map}, but they can not easily visualized in this location, because it is surrounded by other data points.}
    \label{fig:census_without_welsh}
\end{figure}

We tested this in figure \ref{fig:census_without_welsh}: Here, we removed the variable associated with Welsh identity from the dataset, and it became evident that the feature completely disappears. Thus, a single variable can have a huge impact. This result is reminiscent of the effects seen with the suffrage parameter in the V-Dem dataset, which we discussed in section \ref{vdem_suffrage}. It is likely that the reason behind this is that variables with an almost discrete behavior have a particularly strong effect, as the distances between the possible states "output area in England" and "output area in Wales" are large compared to the usual distances of the other variables in the dataset. These larger distances become more significant in the calculation of the \textit{Diffusion Map}.\\

This result also suggests that selecting the right variables beforehand can be crucial. If such a variable is included in the dataset, it determines whether these structures exist at all. We conclude that the \textit{Diffusion Map} can exhibit limited robustness when specific variables are removed from the dataset.

\subsection{Parameter variations: How a different $N$ influence diffusion component meanings}
\label{census_differentN_rotating}

After using the parameter from \cite{Barter2019} to reproduce the results and clarify inaccuracies, we now examine what happens when a different number of nearest neighbors $N$ is used. Already, the use of different $N$ values in the V-Dem dataset in section \ref{vdem_PCA} has shown that much finer structures emerge, such as the features of the V-Dem \textit{Diffusion Map} discussed in section \ref{vdem_interpretation}. The question remains whether such structures also appear in the census dataset given the relatively small number of nearest neighbors of $N=10$. Thus, in section \ref{census:higher_components}, we have discussed the small-scale structures that arise due to ethnic identities. \\

\begin{figure}[H]
    \centering
    \includegraphics[width=0.95\linewidth]{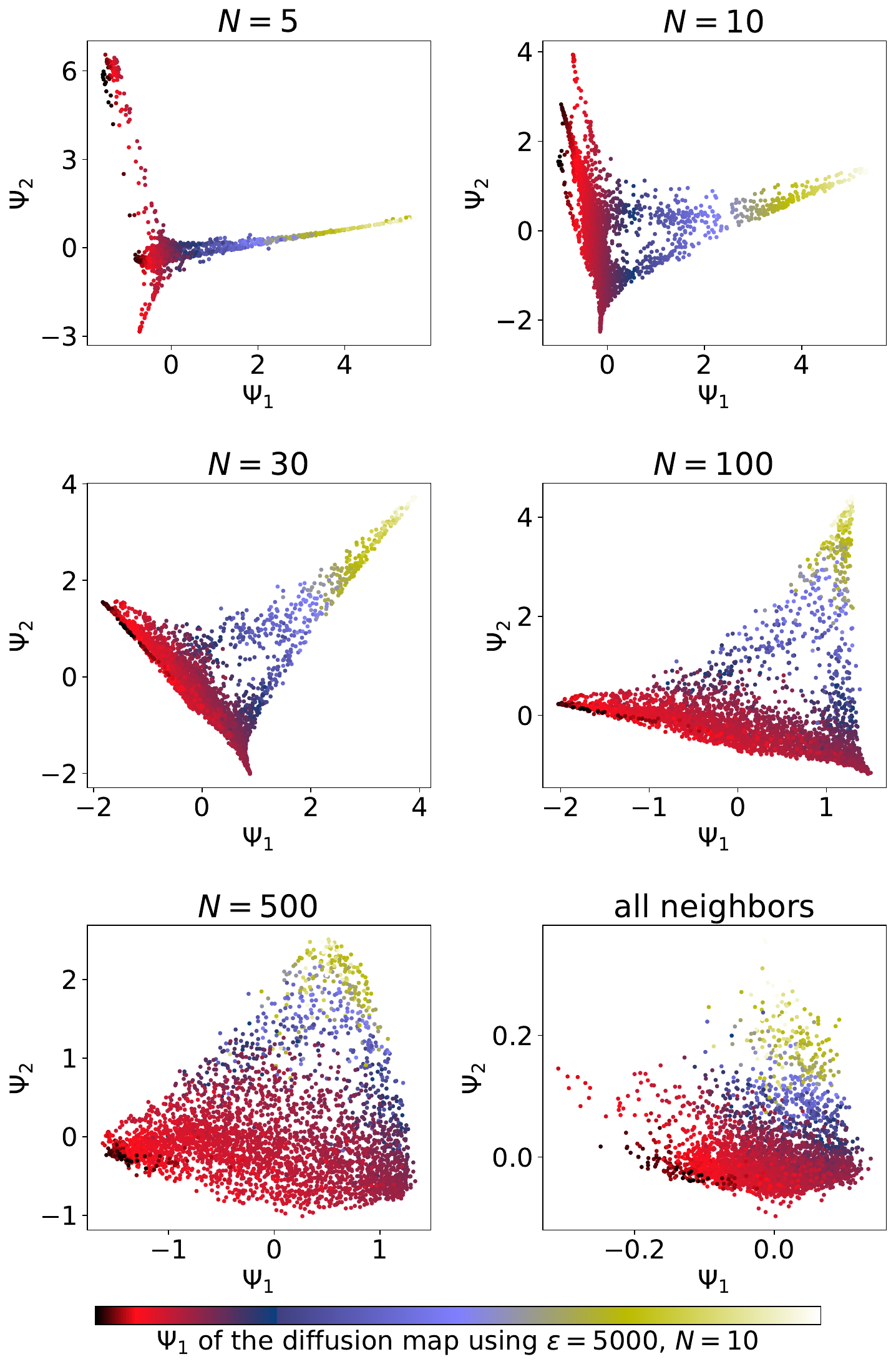}
    \caption{\textbf{Dependence of the \textit{Diffusion Map} on the Neighborhood Parameter $N$:} The plots show the \textit{Diffusion Map} for $\epsilon = 5000$ and varying $N$. The color code is used for better orientation and consistently represents the diffusion component $\Psi_1$ for $N = 10$.}
    \label{fig:census_differentN}
\end{figure}

In order to answer this question, we examine figure \ref{fig:census_differentN}. Here, the \textit{Diffusion Map} for different $N$ are plotted and color coded to facilitate orientation. This figure shows $\Psi_1$ from the comparative \textit{Diffusion Map} that we previously used with $\epsilon=5000$ and $N=10$. We can clearly observe that for large $N>10$, the \textit{Diffusion Map} becomes very broad and no longer exhibits the finer structures that are still visible for $N=10$. The color code also indicates that the meaning of the structure of the \textit{Diffusion Map} remains present for other $N$. A surprising discovery is that the orientation of the \textit{Diffusion Map} changes for different $N$ parameters: the \textit{Diffusion Map} rotates by 90 degrees when moving from $N=5$ to the use of all nearest neighbors. \\

This is problematic because it represents another limitation of the correlation analysis from section \ref{census_exploring_correlation}: for example, in the case of $N=30$, the \textit{Diffusion Map} is no longer aligned parallel to one of the diffusion components. Consequently, it becomes impossible to assign clear meanings to the diffusion components. Because of the rotation, the educational background and the poverty can no longer be definitively considered the first or respectively the second most important explanatory variable. Any claims that the first diffusion components always retain the most crucial information must be made with caution. It is not possible to clearly determine which diffusion component is most important, as its significance may vary depending on different parameters like here the $N$ parameter.
Therefore, the statement from \cite{Barter2019}, that the \textit{Diffusion Map} ranks the discovered variables and that the student number (respectively the educational level) does necessarily have the highest explanatory power of the data set, is not true, as it is described by the second and not the first diffusion component for different values of $N$.\\

\subsection{Comparison to the PCA}
\label{census_comparison_PCA}
\begin{figure}[h]
    \centering
    \includegraphics[width=\linewidth]{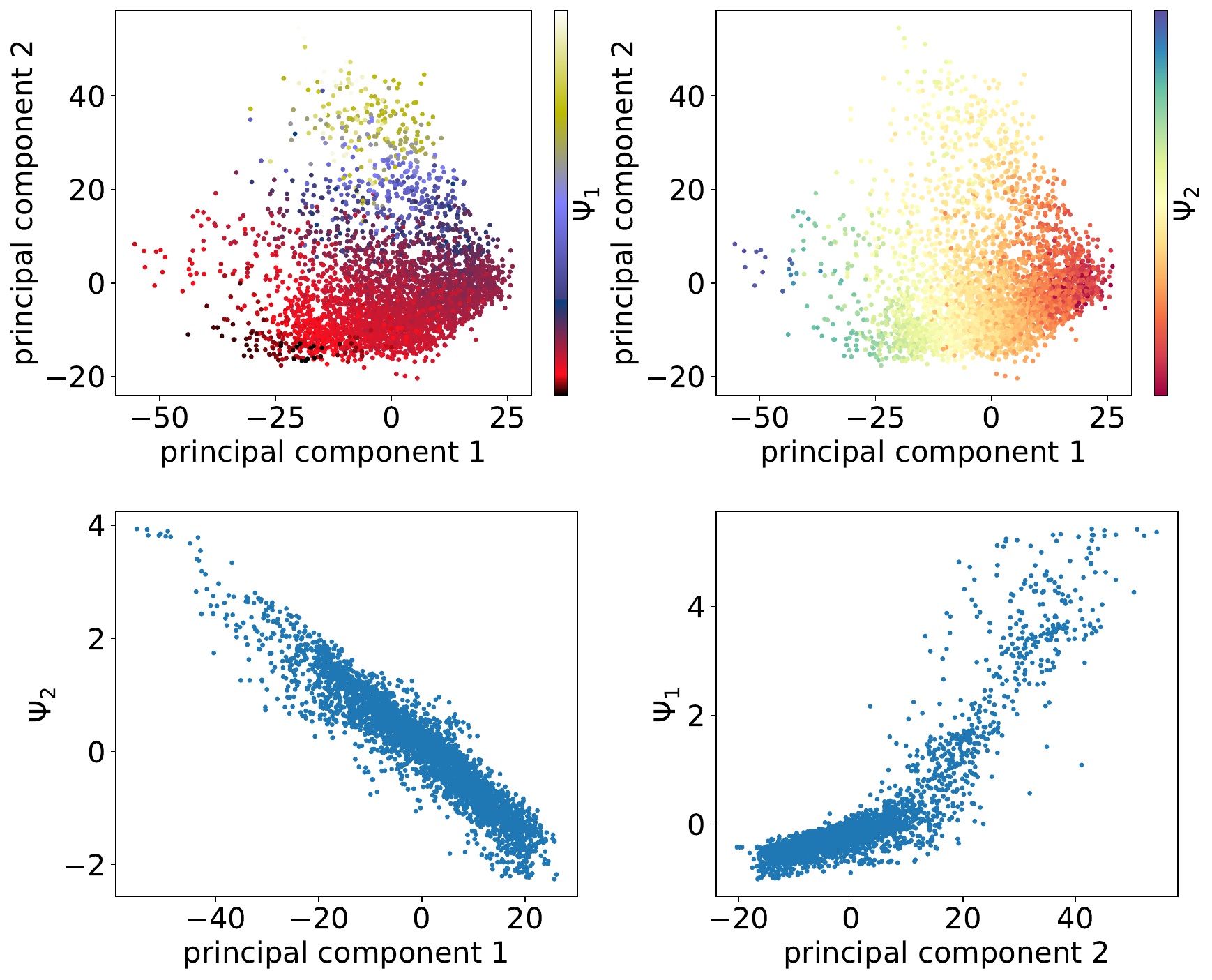}
    \caption{\textbf{Principal components reveals strong correlations to the diffusion components:} The first row presents the principal component analysis of the census dataset, with a color code representing the diffusion components $\Psi_1$ (left) and $\Psi_2$ (right) for \textit{Diffusion Map} parameters $\epsilon = 5000$ and $N=10$. In the second row, the correlated principal components and diffusion components are plotted against each other.}
    \label{fig:census_comparison_PCA}
\end{figure}

From previous analyses, we know that for large neighborhoods, the \textit{Diffusion Map} approximates the PCA. Since we have progressively increased the dominant neighborhood parameter $N$, we now aim to examine whether the \textit{Diffusion Map} has indeed converged to the PCA. To this end, we compare the representation with the PCA, which is shown in figure \ref{fig:census_comparison_PCA}. In the top row, we observe the PCA and note that it is nearly identical to the \textit{Diffusion Map} when considering all nearest neighbors (figure \ref{census_differentN_rotating}, bottom right). For larger values of $\epsilon$, the \textit{Diffusion Map} would further approximate the PCA result.\\

As a color code, we have plotted the relationship with $\Psi_1$ and $\Psi_2$ of our reference \textit{Diffusion Map} with $N=10$ and $\epsilon=5000$. By considering the plots in the bottom row of figure \ref{fig:census_comparison_PCA}, we can clearly observe that the diffusion components are strongly correlated with the principal components. This suggests that the principal components contain nearly the same information as the diffusion components. Consequently, the financial situation in the output areas is described by the first principal component, where the correlation between the first principal component and \texttt{KS404EW0007} (number of cars), which serves as an indicator of financial status, is $\rho_s = 0.85$. The second principal component, in turn, describes the typical educational level in the individual output areas and is strongly correlated with \texttt{QS502EW0008} (number of higher school certificates), with $\rho_s = 0.86$. The financial situation and the educational level therefore describe the greatest variance in the data set.\\

The appearance of the PCA allows us to view a statement from \cite{Barter2019} in a new context: it is stated there that the explanatory variables of students and poverty emerge from complex nonlinear combinations of various other statistics. Since the linear PCA method is also able to explain these two explanatory variables, this suggests that the structures may not actually be nonlinear.

\subsection{Testing for redundant variables}
\label{census_redundant_information}

We now want to address the question of whether the \textit{Diffusion Map} is strongly influenced by redundant variables, as discussed in section for the Swiss roll dataset \ref{swiss_roll_redundant_information}.\\

As \cite{Barter2019} already proposed, the significant influence of individual variables may stem from the fact that the census includes many questions that are answered similarly by a specific group of individuals (as the group of students). This group would then receive special weighting because of the presence of many linearly redundant variables in the dataset, allowing these groups to be easily identified by the \textit{Diffusion Map}.\\

For the example dataset, we have devised an approach to resolve these redundancies. The idea is to first apply PCA to the dataset in order to remove the linearly dependent, redundant variables, and then, through the subsequent application of the \textit{Diffusion Map}, capture the cleaned nonlinear structure. \\

We want to specifically examine the question of whether the student group, as mentioned by \cite{Barter2019}, gains significance due to many redundant variables, and investigate whether the students lose significance in the \textit{Diffusion Map} after applying the proposed procedure and resolving these redundancies.
First, we checked how many variables have a strong correlation with \texttt{KS501EW0010} (ratio of students) and found 42 variables with a Spearman correlation $\rho_s > 0.75$.

\begin{figure}[H]
\centering
\subfigure[Spectrum of the PCA. The inset show a zoom of the explained variance of the first principal components.]{
   \centering
   \label{fig:census_redundant1}
   \includegraphics[width = 0.65\textwidth]{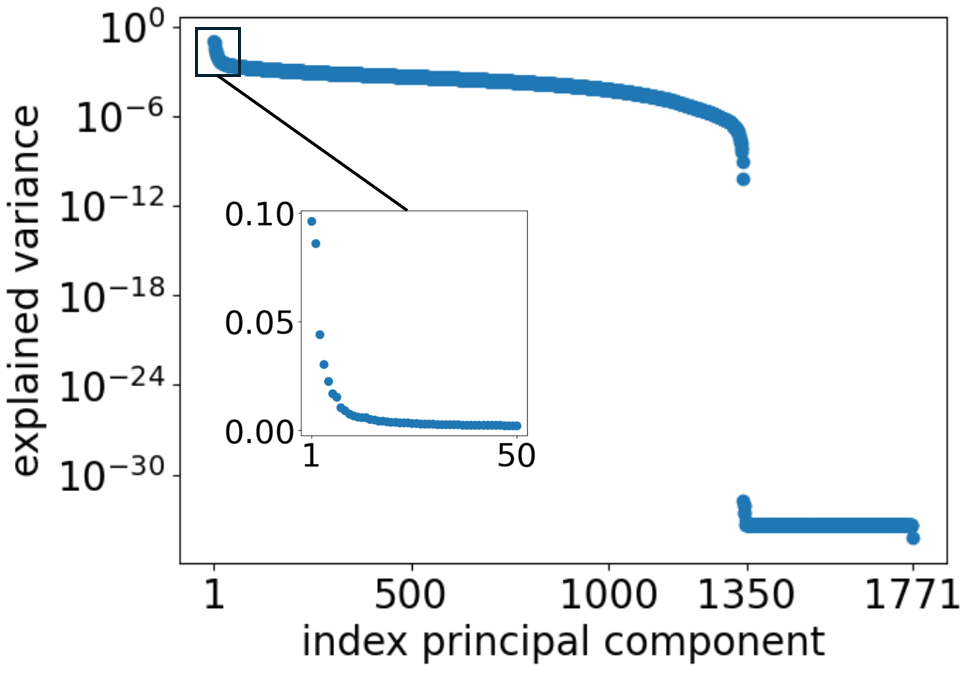}}\\
\subfigure[\textit{Diffusion Map} result on the first 100 principal components. Each column have a specific color code, which shows different variables of the dataset. ]{
   \centering
   \label{fig:census_redundant2}
   \includegraphics[width = \textwidth]{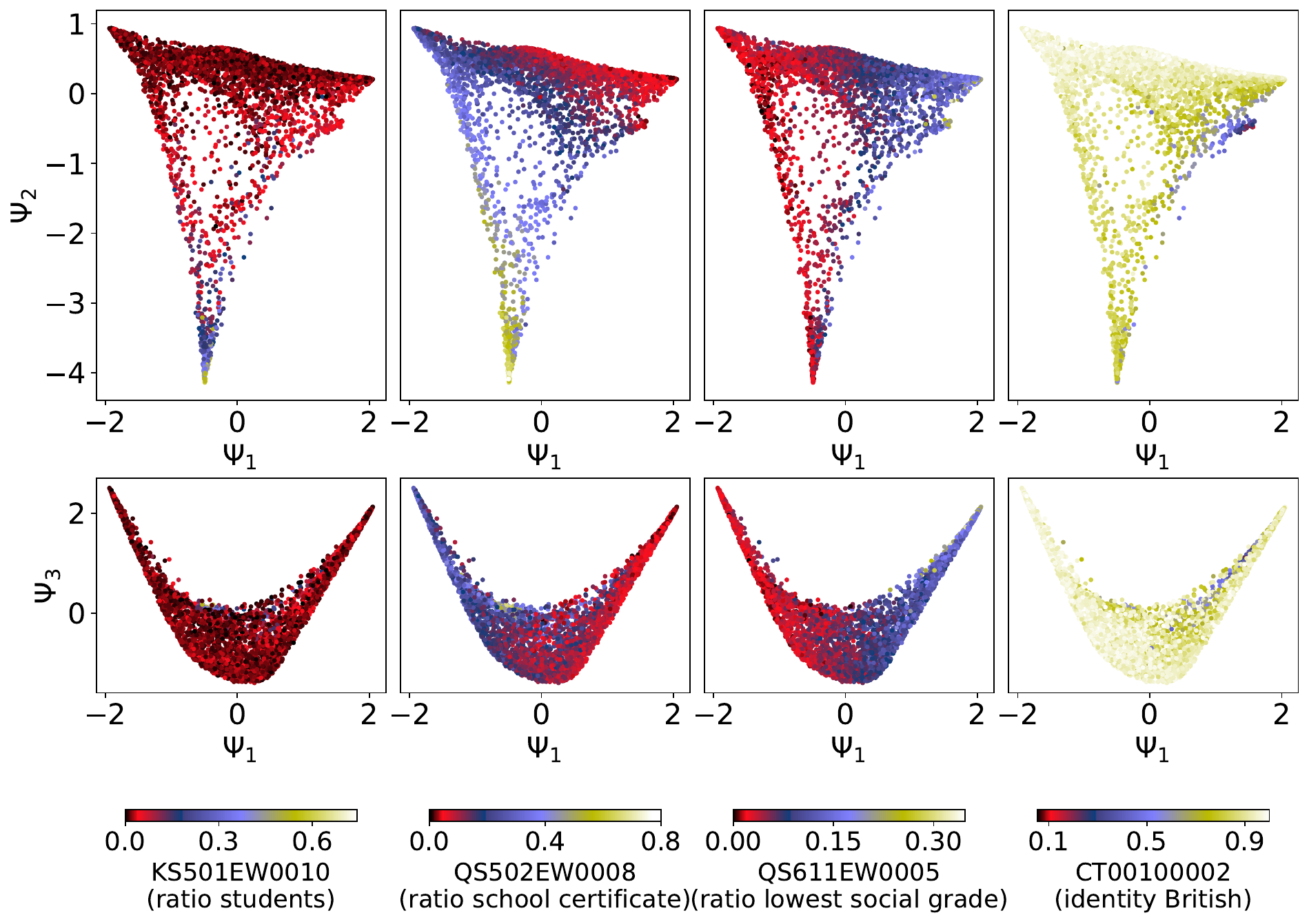}}
   \caption{\textbf{Effect on the \textit{Diffusion Map} when preprocessing the data using PCA:} the \textit{Diffusion Map} is applied to the standardized first 10 principal components. The \textit{Diffusion Map} parameters are $\epsilon = 1000$ and $N = 100$. The PCA is intended to eliminate the effect of redundant variables (see section \ref{swiss_roll_redundant_information}).}
   \label{fig:census_redundant}
\end{figure}

Next, we compute the PCA and examine the eigenvalue spectrum, which provides the explained variance of the individual principal components (see figure \ref{fig:census_redundant1}). There, we observe a sharp decline in explained variance at the beginning, with another drop occurring around the 1350th principal component.

As a first attempt, we will now consider only the first 10 principal components as the "new dataset." Among these principal components, there is only one with a high correlation, $\rho_s = 0.85$.

The \textit{Diffusion Map} applied to the PCA then gives us the results shown in figure \ref{fig:census_redundant2}. Here, we observe that students and education levels are still described by $\Psi_2$. The social situation of individuals is described by $\Psi_1$. What is new, however, is that there is now also a visible relationship between education and social situation, represented by $\Psi_1$. The British identity now has only a minor influence on the map, while the Welsh identity is spread across the entire \textit{Diffusion Map} and is not clearly shown. 

What is new here is that $\Psi_3$ now exhibits the typical polynomial form and no longer carries any new information (no variables with a Spearman correlation of $\rho_s > 0.7$ can be found for this component).

We can conclude that education and social situation are still marked as the most important natural parameters, even after all linearly redundant variables have been removed. Therefore, the reason for the importance of these variables must be something other than redundancy.

The variable related to British identity is no longer as prominent, which suggests that there may be many redundant variables contributing to its importance. This could be due to the fact that a wide range of ethnic identities is queried, and it is likely that many non-British individuals reside in the same output areas, causing these variables to be nearly identical.

\subsection{Insight gained from applying the \textit{Diffusion Map} to census data}
\label{census_conclusion}

We have learned that education and wealth are important explanatory variables. Through correlation analysis, it is sometimes possible to assign meaning to the individual diffusion components (section \ref{census_exploring_correlation}).

However, in certain cases, the diffusion components may have multiple interpretations, or the \textit{Diffusion Map} might become spatially misaligned, complicating the analysis. Furthermore higher-level components no longer seem to have a clear meaning. (section \ref{census:higher_components}).

We have also shown that the application of the \textit{Diffusion Map} leads to many challenges (e.g., finding the right parameters, see section \ref{census_differentN_rotating}), and the analysis is often complex and requires careful consideration. As a result, the application of this method can easily lead to misunderstandings, as evidenced by our comparison with the results from \cite{Barter2019}. 

We have also learned that the \textit{Diffusion Map} results can be strongly influenced by the inclusion of single variables (section \ref{census:higher_components}), which highlights the importance of variable selection for the outcome of the analysis.\\

These difficulties in the application of the method raise the question of what applications the \textit{Diffusion Map} can be used for in the context of similar datasets. This question is emphasized because it seems that the easier PCA method can also find the most important explanatory variables in this case (section \ref{fig:census_comparison_PCA}).

%% file: Chapter/44_germany.tex
\section{Regional data of the districts in Germany}
\label{regionaldata_germany_section}

In this section, we will look at a smaller data set of regional statistics from Germany, which can be found at \cite{regionaldata}. We will have a look at the results of the \textit{Diffusion Map} and the PCA and compare to which extent both embeddings can describe the data set.

\subsection{The dataset and motivation for the analysis}
\label{districts_dataset}
We use the \textit{Diffusion Map} to examine 369 rural districts ('Landkreise') and urban districts ('kreisfreie Städte' and 'Stadtverbünde') in Germany. Although this administrative definition is not entirely accurate and there are several 'Landkreise' with large cities, we will refer to 'Landkreise' as rural districts and 'kreisfreie Städte' and 'Stadtverbünde' as urban districts in this section.

We will use 16 variables from different areas of statistics as variables in order to represent a wide spectrum of society. Most of these variables are available in absolute numbers, so they are normalized by the number of inhabitants. See \ref{tab:variables_regionaldata} for the list of variables used. \\

The aim is to find the most important properties of the dataset and possible natural parameters of this German district data. As an initial hypothesis, we expect a significant difference between districts located in the former East Germany and those in the western states, as there are still substantial disparities between East and West, for example, in terms of wages, home ownership, and wealth (see, e.g., \cite{ost_west2024}). A second distinction should emerge between rural and urban areas, given the many differences in unemployment rates, housing prices, income levels, available services, and various other statistics (see, e.g., \cite{Pateman2010,Fuest2019}). In the German context, some urban-rural disparities—such as those in productivity—are known to be even greater than those between East and West \cite{Belitz2019}, and these urban-rural differences continue to grow \cite{Fuest2019}. 
With this preliminary consideration, we will now examine whether the \textit{Diffusion Map} can detect these differences and what contribution the method can make to understanding these patterns.

\begin{table}[h]
    \footnotesize
    \centering
    \begin{tabular}{l}
        wages- and income tax per inhabitant \\
        unemployed per inhabitant \\
        marginally employed per inhabitant \\
        social welfare benefits per inhabitant \\
        insolvencies per inhabitant \\
        amount of building land sales per inhabitant \\
        construction of living space per inhabitant\\
        cars per inhabitant \\
        general higher education qualification per school leaving \\
        public dept per inhabitant \\  
        full time equivalents of public service per inhabitant \\
        youth quotient \\
        old age quotient \\        
        asylum seekers per inhabitant \\
        naturalizations per inhabitant \\
        immigration per inhabitant \\
        emigration per inhabitant \\
    \end{tabular}
    \caption{Used variables from the regional statistics dataset \cite{regionaldata}}
    \label{tab:variables_regionaldata}
\end{table}

\subsection{Identifying natural parameters of german Districts: rural-urban and east-west differences}
\label{germany_natural_parameters}

\begin{figure}[H]
    \centering
    \includegraphics[width=\linewidth]{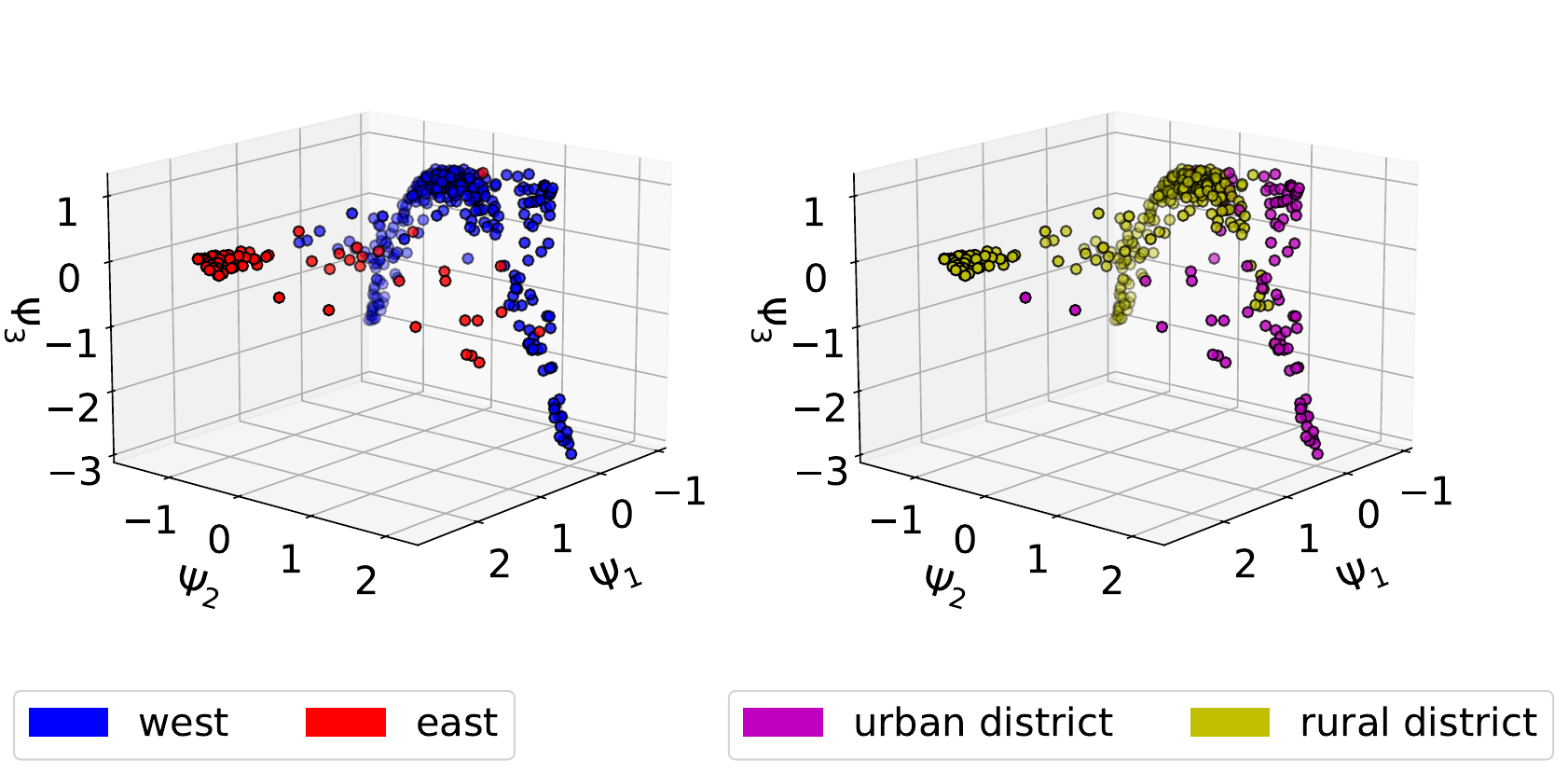}
    \caption{\textit{Diffusion Map} of the regional statistics of german districts. The data was standardized and for the calculation of the \textit{Diffusion Map} we used $\epsilon = 50$ and 10 nearest neighbors. The color of the data points indicates whether the district is located in the territory of former East Germany or West Germany (left plot) or whether the district is classified as rural or urban (right plot).}
    \label{fig:germany_diffmap}
\end{figure}

After optimizing the neighborhood parameters, we obtain the \textit{Diffusion Map}, which is depicted in figure \ref{fig:germany_diffmap}. Our analysis will be confined to the first three dimensions. Examining the color-coded data points on the left, we observe that the first diffusion component, $\Psi_1$, effectively distinguishes between East and West German districts: western districts exhibit low values of $\Psi_1$, whereas eastern districts display high values. However, some outliers are present, which are eastern districts with unexpectedly low $\Psi_1$ values.

By incorporating information from the plot on the right in figure \ref{fig:germany_diffmap}, we can conclude that these outliers predominantly correspond to urban districts. This observation aligns with our initial assessment in Section \ref{districts_dataset}, where we hypothesized that differences between urban and rural areas may surpass those between East and West Germany. Therefore, these eastern urban data points exhibit greater similarity to urban districts than to other eastern districts and are consequently categorized within the urban group.

The three strongest correlations with $\Psi_1$ are found in the following variables:
\begin{itemize}
    \item Unemployment per inhabitant ($\rho_s = 0.76$)
    \item Construction of living space per inhabitant ($\rho_s = -0.62$)
    \item Wages and income tax per inhabitant ($\rho_s = -0.683$)
\end{itemize}
These correlations highlight the persistent socioeconomic disparities between East and West Germany, which lead to the formation of two distinct clusters in the data.\\

Moreover, the \textit{Diffusion Map} also differentiates between urban and rural districts. Urban districts exhibit high values of $\Psi_2$, whereas rural districts correspond to lower values.

The three strongest correlations with $\Psi_2$ are as follows:
\begin{itemize}
    \item Social welfare benefits per inhabitant ($\rho_s = 0.85$)
    \item Asylum seekers per inhabitant ($\rho_s = 0.84$)
    \item Unemployment per inhabitant ($\rho_s = 0.74$)
\end{itemize}

A notable pattern emerges in the $\Psi_2$-$\Psi_3$ plane, resembling a parabolic dependence. This suggests that $\Psi_3$ in this dataset may only have a functional parabolic relationship with $\Psi_2$.\\

In section \ref{sorting_ability}, we previously discussed the potential emergence of clustered data points and their meaningful interpretation despite cluster formation. The \textit{Diffusion Map} result may thus reflect the ability to separate east and west German districts due to their remaining structural differences. Simultaneously, at least the western districts appear to be meaningfully arranged according to their similarities.\\

In summary, it can be stated that the East-West difference and the urbanization of the districts could be considered natural parameters of the dataset.

\subsection{Comparison to the PCA}
\label{germany_comparison_PCA}

\begin{figure}[H]
    \centering
    \includegraphics[width=\linewidth]{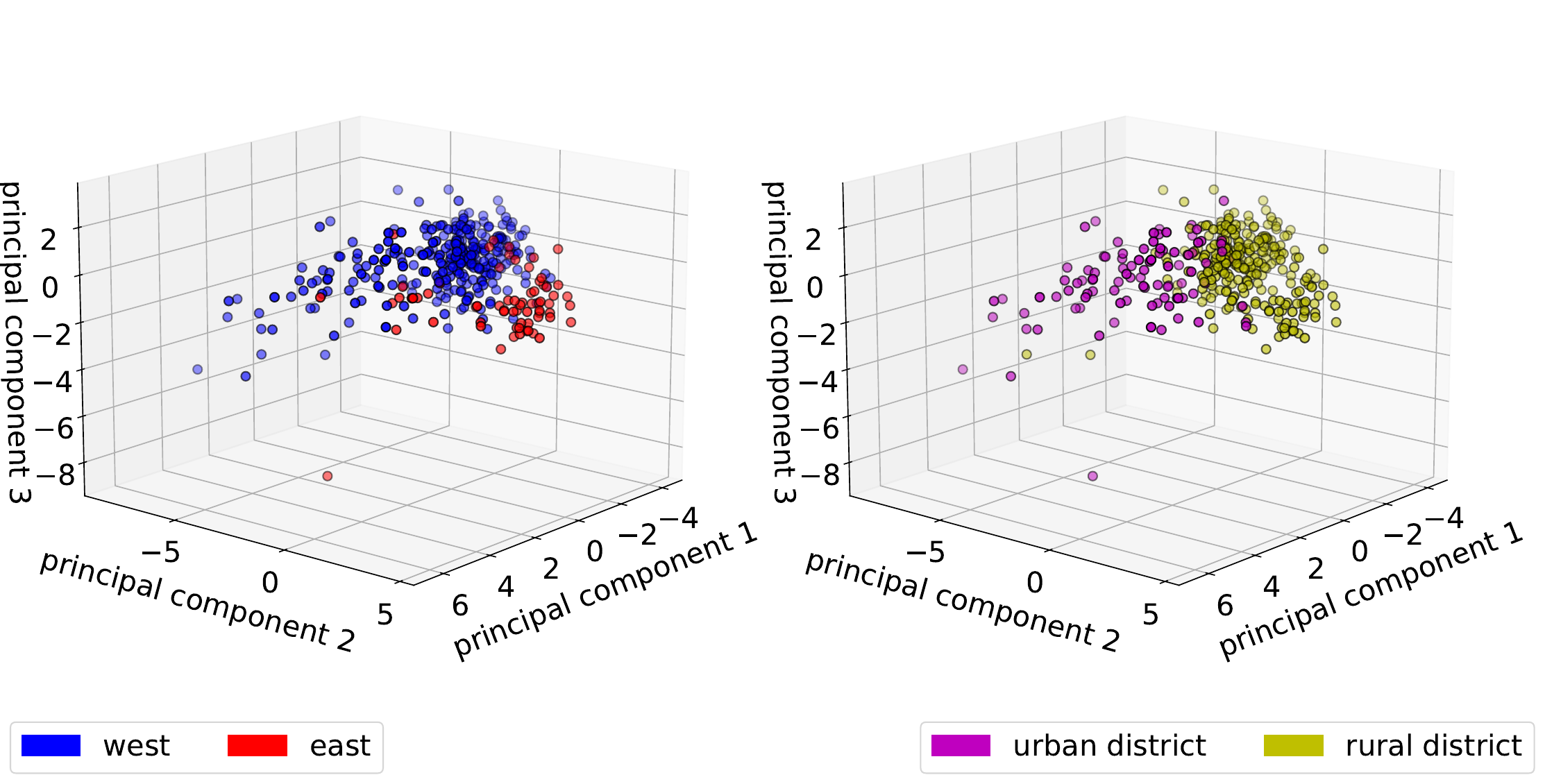}
    \caption{Principal component analysis of the standardized regional statistics of German districts.}
    \label{fig:germany_diffmap2}
\end{figure}

We now take a look at figure \ref{fig:germany_diffmap2} to assess whether the same patterns can be found in the PCA results. As observed in the left plot, principal component 2 organizes the districts according to an East-West gradient, while the right plot demonstrates that principal component 1 effectively differentiates between urban and rural districts. Upon examining the correlations between the principal components and the diffusion components, it becomes apparent that principal component 1 exhibits a strong correlation with $\Psi_2$, with $\rho_s = 0.92$. Furthermore, principal component 2 shows a relatively strong correlation with $\Psi_1$, with $\rho_s = 0.79$. The arrangement of the data points is therefore very similar to the \textit{Diffusion Map}, although no distinct clusters can be seen in the PCA result.

\subsection{Usefulness of the \textit{Diffusion Map} for the dataset}
It can be concluded that the \textit{Diffusion Map} was able to arrange the data according to our hypothesis. The significant differences due to the east-west and the urban-rural differences are clearly visualized. 
The key difference compared to the PCA results lies in the fact that the \textit{Diffusion Map} is capable of meaningfully clustering very similar data. However, the PCA is also effective in sorting the data points according to the main features (urban/rural, East/West).
Finally, the question arises of what further conclusions can be drawn from the \textit{Diffusion Map} to better understand the data or to get more insights, beyond the facts we already know. Perhaps additional insights could be gained by examining individual data points or outliers.

%% file: Chapter/5Outlook.tex
\label{4discussion_outlook}


\section{Summary}
In this study, we have examined the \textit{Diffusion Map} method in general and its applications to social science data in particular. To do so, we analyzed the relevant literature, explored its behavior using simple example datasets, and ultimately applied it to three social science datasets. In this section, we outline the key insights gained from our investigation of the method.

\subsection{What can influence the \textit{Diffusion Map} result?}
One key finding is that we have identified several factors that can significantly influence the outcome of the \textit{Diffusion Map} for individual datasets. Awareness of these factors is essential to avoid misapplications and misinterpretations.

These factors include:
\begin{enumerate}
    \item The \textbf{choice of different} neighborhood \textbf{parameters} ($N$, $\epsilon$) has a significant impact on the results. Existing methods provide only an initial guess and do not find a definitively optimal value (sections \ref{finding_right_epsilon}, \ref{finding_N}, \ref{vdem_PCA}, \ref{census_exploring_correlation}, \ref{census_differentN_rotating}). When simultaneously using an $\epsilon$-neighborhood and a small number of nearest neighbors, the computational effort is reduced due to the sparse matrix, but the $N$ parameter can become dominant, revealing fine-scale structures (as observed in sections \ref{vdem_PCA}, \ref{census_motivation}). Additionally, for all datasets, we observed that for very large neighborhoods, the \textit{Diffusion Map} converges towards PCA. The parameter $t$, however, does not affect the structure of the \textit{Diffusion Map} (section \ref{time_parameter}).
    \item \textbf{Scaling individual variables} affects their influence on the \textit{Diffusion Map} results, as it impacts distance calculations. Since normalizations also alter the scale of individual variables, they consequently influence the results as well (section \ref{normalization}). The effect of scaling presents the hypothetical possibility of increasing the influence of more important variables on the embedding through upscaling while reducing the impact of less relevant variables through downscaling.  
    \item When multiple variables containing the same information, i.e. when \textbf{redundant variables} are present in the dataset, this alters the distance calculations for the \textit{Diffusion Map}, giving more weight to these duplicated variables. Applying PCA beforehand can resolve this issue for linear redundancies for simple datasets (section \ref{swiss_roll_redundant_information}). However, for more complex datasets, this effect remains unclear and could not be conclusively demonstrated (e.g., section \ref{census_redundant_information}).  
    \item The \textbf{orientation of \textit{Diffusion Maps} can change} depending on the choice of the neighborhood parameter, which in turn alters the meaning of the diffusion components (section \ref{census_differentN_rotating}).   
\end{enumerate}
The preprocessing of data has not been widely discussed in the literature, likely because initial experiments with the \textit{Diffusion Map} often use test datasets like the Swiss roll, which yield clear results, lack redundant variables, and feature equally important dimensions.\\

Nevertheless, our work also demonstrates that it is useful to use simple example datasets, specifically designed to address different aspects, in order to better understand the effects and expose misinterpretations of the method. However, it remains important to investigate more complex datasets to truly assess how effectively the method can handle them \cite{Strange2014}.\\

Caution is also needed, as the definition of the method - also aside from the use of alternative kernels as similarity measures (e.g., in \cite{Barter2019,Zeng2024} the kernel $K_{i,j} = 1/\mathcal{D}_{i,j}$ or in \cite{Xiu2023,Xiu2024} the Spearman correlation) - can vary. As we noted in section \ref{different_normalizations}, the literature is also not unified on the definition for the Laplacian matrix. Furthermore, a different normalization of the eigenvectors can influence the results (sections \ref{spectral_decomposition} and \ref{counterexample_normalisation}).

\subsection{Capabilities and Limitations of the \textit{Diffusion Map}}

We have observed that, for all the test and real-world datasets analyzed, the \textit{Diffusion Map} is able to identify important explanatory/natural variables (sections \ref{swiss_roll}, \ref{vdem_interpretation}, \ref{census_exploring_correlation}, \ref{germany_natural_parameters}). This contributes to a better understanding of these datasets.\\

In section \ref{sorting_ability}, we explored how the \textit{Diffusion Map} is able to meaningfully order data points and what conditions must be met for this sorting ability. In order for the \textit{Diffusion Map} to meaningfully arrange two points, it seems necessary that there are sufficiently similar points between them. Otherwise, the points may belong to clusters, but the position of these clusters relative to each other may lose its significance (section \ref{clustering}). It is important to note that when applying the \textit{Diffusion Map} to both discrete and continuous variables, the discrete variables are considered more significant. This is because discrete variables have greater distances between their possible values, and since the \textit{Diffusion Map} relies on these distances, they gain more importance (see section \ref{vdem_suffrage}). In the analysis of the \textit{Diffusion Map} for democracy data in section \ref{vdem_quality_of_election}, we showed that gaps in the dataset can produce features that are more difficult to understand.

Overall, it was shown that the linear PCA method was also able to meaningfully order the data according to the most important explanatory variables for the datasets (sections \ref{vdem_PCA}, \ref{census_comparison_PCA}, \ref{germany_comparison_PCA}). However, the \textit{Diffusion Map} also enabled the detection and analysis of smaller structures. Additionally, we observed that PCA could not cluster the data in the case of the German district dataset (section \ref{germany_comparison_PCA}).

The clustering ability can be traced through the \textit{Diffusion Map} property (section \ref{diffusion_property}) and has been described in several publications (see section \ref{clustering}).\\

We have seen that it can be challenging to assign importance or meaning to individual diffusion components. We observed that components may exhibit functional dependencies on each other. Polynomial dependencies within the datasets were frequently observed (sections \ref{typical_shapes}, \ref{vdem_one_dimensional_curve}, \ref{germany_natural_parameters}), which were first described for one-dimensional datasets by \cite{Nadler2007} and were also derived in our work (section \ref{typical_shapes}). This can lead to misinterpretation, for example, when assuming that the parabolic shape of the dependence of two diffusion components (for example $\Psi_2(\Psi_1)$) has a specific meaning.

These behaviors and dependencies are not visible in the \textit{Diffusion Map} spectrum (section \ref{swiss_roll_PCA_spectrum}), nor can they be predicted by other methods. Another issue is that, unlike PCA, the importance of individual \textit{Diffusion Map} components cannot be derived from the eigenvalue spectrum (section \ref{swiss_roll}). It is also possible that important natural parameters do not appear in the first diffusion components, but rather emerge in later components, as earlier components may only reflect polynomial dependencies (section \ref{swiss_roll}). As mentioned above, neighborhood parameters can affect which information is encoded by which diffusion component, as the \textit{Diffusion Map} exhibits different orientations in space depending on the parameters, further complicating the analysis (section \ref{census_differentN_rotating}).

To identify natural parameters or explanatory variables in a \textit{Diffusion Map}, one can resort to correlation analysis using Spearman correlation of variables in or outside the data set with the individual diffusion components. (section \ref{census_exploring_correlation}). However, diffusion components can also encode multiple meanings, as we observed in section \ref{fig:census_highest_correlations} through the encoding of various ethnic identities. In doubt, individual structures, such as specific peaks or patterns in the \textit{Diffusion Map}, must be examined individually.\\

In conclusion, it can be stated that the \textit{Diffusion Map} is not easy to handle, and its structure can lead to incorrect interpretations. More work is needed to better describe the behavior of the \textit{Diffusion Map} and to point out potential misunderstandings in the analysis.

\section{Recommendation for further studies}
This leads us to the potential next steps that should be taken to improve the use of the \textit{Diffusion Map}.

\subsection{Conquest the limitations of this work}

In this work, we primarily aimed to better understand the \textit{Diffusion Map} and to uncover potential misunderstandings of the method through visual examples, qualitative descriptions, and counterexamples. These analyses are only the beginning, and further theoretical investigations and quantifications of the aforementioned effects are needed.

There is an urgent need for techniques that can make clear statements about the descriptive importance of individual diffusion components. So far, there have only been statements regarding the accuracy of the \textit{Diffusion Map} in terms of the diffusion property (section \ref{diffusion_property}). However, it does not address questions such as when natural parameters of the manifold should be identified, particularly when they appear only after less significant diffusion components, as observed with the width of the Swiss Roll, which was only encoded in the fourth diffusion component. (section \ref{swiss_roll}).\\

Better techniques are also needed to determine the correct neighborhood parameters for each dataset. An idea involving variable bandwidths and neighborhoods within the \textit{Diffusion Map} is presented in \cite{Berry2016}, but was not further explored in this work.\\

Also, the relationship between the \textit{Diffusion Map} and the transition of \textit{Diffusion Map} results to PCA with increasing neighborhood sizes must also be quantitatively described and explained.

\subsection{Analysis framework for spectral non-linear methods}

Overall, there is a lack of a well-established analytical framework for applying nonlinear dimensionality reduction methods to complex data, those from social sciences.

\paragraph*{Preprocessing data}~\\
Most scientific studies (after possibly normalizing the data) proceed directly to applying dimensionality reduction techniques to the dataset under consideration. However, a more critical question should be addressed: How should the data be preprocessed to yield meaningful and interpretable results?

For datasets where individual variables have a similar significance or structure — such as spatial coordinates or pixel data — extensive preprocessing may not be necessary. However, in the case of social science data, this is not the case.

It seems most reasonable to develop a theory before applying the \textit{Diffusion Map}, outlining how the concept (for example 'democracy') described by the data relates to the variables (i. e. the properties of 'democracy') in the dataset (This must be done by experts in the corresponding field.). Based on this idea, we propose using the following questions to prepare the dataset for application with the \textit{Diffusion Map} method:

\begin{enumerate}
     \item How is the concept described or influenced by the variables in the dataset? Are there \textbf{more important and less important variables}, and can this be accounted through scaling these variables?  
    \item Are there potential \textbf{redundant variables} that are either intentionally included (because two independent aspects of the concept convey the same information) or should they be removed?  
    \item Does the dataset contain both \textbf{continuous and discrete variables}, or are there large gaps in some variables? Should these variables be treated separately to prevent them from having an excessive impact on the results?
    \item Has the dataset to be \textbf{normalized} to establish an equivalence in contributing to the \textit{Diffusion Map} result for the single variables, or does the original scale already reflect the importance of the variable. (This could be the case, if all variables were for example spatial coordinates and one coordinate remained nearly constant across all data points, it could be problematic if this variable were artificially assigned greater importance, when all variables are scaled to the range $[0,1]$ before applying the \textit{Diffusion Map}.)?
\end{enumerate}

\paragraph*{Unifying spectral methods}~\\
As discussed in section \ref{related_methods_and_history}, there have already been initial proposals to unify various dimensionality reduction methods within a kernel PCA framework.
This is a promising approach, as there currently exist too many methods with highly similar functionalities, making it increasingly unclear what their actual differences are. Even the \textit{Diffusion Map} method is essentially just the probabilistic interpretation of spectral embedding \cite{Nadler2005,Nadler2007}.

However, the idea of a unifying framework does not seem to be gaining traction within the scientific community. This may be due to various review articles that either overlook or introduce this perspective too late, failing to emphasize the many commonalities between methods (e.g., \cite{Izenman2012,Ghojogh2023,Meila2024}).\\

As a result, there is likely a lack of clarity within the scientific community regarding the advantages and disadvantages of individual methods, as well as which computational steps in these methods contribute to their strengths and weaknesses. For instance, there are too few quantitative comparisons of these methods \cite{Strange2014}, despite the existence of initial studies proposing criteria for evaluating dimensionality reduction techniques \cite{Lee2009,Lee2010a,Meng2011,Fernandez2013}.

This raises the fundamental questions: what criteria must a dimensionality reduction method meet to be considered effective? Should local or global structures be preserved as much as possible, or should they be restructured to facilitate deeper insights (as is the case with clustering methods)? Should the new dimensions of the embedding carry a specific interpretable meaning, or at least have a clearly defined importance (as in PCA, where significance is quantitatively determined by eigenvalues and explained variance)? Alternatively, should the method prioritize preserving pairwise distances in the new embedding space, as seen in \textit{Diffusion Maps}? Is accuracy according to a predefined metric (such as diffusion distance in the case of the \textit{Diffusion Map}) the most important criterion? To what extent should the method be able to identify the natural parameters or key explanatory variables underlying the dataset? These questions need to be answered in future studies.

\paragraph*{From single methods to a full toolbox: rethinking dimensionality reduction}~\\
A next step in the discourse on dimensionality reduction methods should be a shift away from viewing them as complete methods and increasingly complex concepts, without properly explaining the origin of the methods, and instead toward thinking of them as tools within a toolbox: at the beginning of an analysis, one could ask: is it sufficient for my purposes to examine the eigenvectors of the Euclidean distance matrix (as it is the case for classical multi-dimensional scaling)? No? Then try using the eigenvectors of the geodesic distance matrix (as for the Isomap method). That doesn’t work either? Then one could use local kernels (as in \textit{Diffusion Maps} or Laplacian Eigenmaps). Perhaps one also only considers the $N$ nearest neighbors.\\

By understanding the existing methods as a sequence of tools, comprehension would certainly become easier. These tools, along with their effects and potential misunderstandings, should be systematically collected in a catalog and made accessible to everyone through simple illustrative examples.

%% file: Chapter/6Appendix.tex
\section*{Improvement of normalization to maintain the \textit{Diffusion Map} property}
\addcontentsline{toc}{section}{Improvement of normalization to maintain the \textit{Diffusion Map} property}


In sections \ref{idea_algorithm} and \ref{probabilistic_interpretation}, it was noted that correct normalization is important to ensure the diffusion property. In private conversations with Paula Pirker-Díaz and Maximilian Graf, we found out that the normalization of the length of the eigenvectors is important so that the diffusion property is fulfilled. It should be noted that the normalization from \cite{Nadler2005} does not seem to work properly. In this section, we take a closer look at this problem and a possible solution. We emphasize (using two examples) that for the calculation of the \textit{Diffusion Map} the definition of the eigenvectors given in equation \ref{norm_eigenvec} should be used to fulfill the \textit{Diffusion Map} property. \\
For the purpose of completeness, the entire argumentation has been inserted here, although some definitions are repeated.\\

In \cite{Nadler2005} is introduced as a theorem that \textit{''the diffusion distance is equal to Euclidean distance in the \textit{Diffusion Map} space with all ($n-1$) eigenvectors''}. The diffusion distance is defined as:

\begin{equation}
\label{diff_dist}
D_t^2(x_0,x_1) = \sum_y (p(t,y|x_0)-p(t,y|x_1))^2 \frac{1}{\phi_0(y)},    
\end{equation}

where $p(t, y | x) = \phi_0(y) + \sum_{j \geq 1} \psi_j(x)\lambda_j^t\phi_j(y)$. Then, according to the mentioned theorem, it must coincide with:

\begin{equation}
D_t^2 = ||\Psi_t(x_0) - \Psi_t(x_1)||^2 = \sum_{j \geq 1} \lambda_j^{2t}(\psi_j(x_0)-\psi_j(x_1))^2
\end{equation}

Here we check this equality in two different ways for two simple examples. On the one hand, following the equations described in \cite{Nadler2005} and on the other hand, with our alternative definition of the eigenvectors $\phi_j$ and $\psi_j$. \\

\paragraph*{1. Symmetric example} Let's take $x_1 = 0$ and $x_2 = \sqrt{2\log(2)}$ as data points in the $\mathbb{R}$ feature space. \\

We use the Gaussian kernel 
\begin{equation}
    \label{kernel}
    K_{ij} = k(\mathbf{x_i},\mathbf{x_j}) = \exp{(-\frac{||\mathbf{x_i}-\mathbf{x_j}||^2}{2\epsilon})}
\end{equation} with width $\epsilon = 1$ to calculate the pairwise similarity matrix $K$. Consequently, we get:

$$K = 
\begin{pmatrix}
1 & \frac{1}{2}\\
\frac{1}{2} & 1
\end{pmatrix}
$$

Now we normalize each row of the matrix $K$ so that the sum of all entries equals 1. In other words, we create a diagonal normalization matrix $D_{ii} = \sum_j K_{ij}$. The normalized matrix is then $M = D^{-1}K$, which leads us to the next result:

$$M = 
\begin{pmatrix}
\frac{2}{3} & \frac{1}{3}\\
\frac{1}{3} & \frac{2}{3}
\end{pmatrix}
$$

If our matrix was not symmetric, we would compute the symmetric matrix
\begin{equation}
    M_S = D^{1/2} M D^{-1/2} = D^{-1/2}KD^{-1/2},
\end{equation}
characterized by its eigenvalues $\lambda_j$ and the corresponding eigenvectors $v_j$. For this particular case, $M_S = M$. \\

Now we calculate the eigenvalues with their normalized eigenvectors of length $1$:
\begin{align*}
\begin{cases}
    \lambda_0 &= 1\\
    \lambda_1 &= \frac{1}{3}\\
\end{cases}
    ~\\
\begin{cases}
    v_0 &= \begin{pmatrix}
        \frac{1}{\sqrt{2}}\\
        \frac{1}{\sqrt{2}}
    \end{pmatrix}\\
    v_1 &= \begin{pmatrix}
        -\frac{1}{\sqrt{2}}\\
        \frac{1}{\sqrt{2}}
    \end{pmatrix}\\
\end{cases}
\end{align*}

\begin{center}

\begin{minipage}[t]{0.47\textwidth}
\scriptsize
\underline{\textbf{Calculation following \cite{Nadler2005}}}\\
In \cite{Nadler2005}, the relationship between the eigenvectors of $M_S$ and $M$ is directly used: 
\begin{align}
     \phi_j &= v_jD^{1/2}\\ \psi_j &= v_jD^{-1/2}    
\end{align}

For this example we get the following vectors:
\begin{align*}
    \begin{cases}
    \phi_0 &= \begin{pmatrix}
        \sqrt{3/4}\\
        \sqrt{3/4}        
    \end{pmatrix}\\
    \phi_1 &= \begin{pmatrix}
        -\sqrt{3/4}\\
        \sqrt{3/4}        
    \end{pmatrix}\\
    \end{cases}
    ~\\
    \begin{cases}
    \psi_0 &= \begin{pmatrix}
        \sqrt{1/3}\\
        \sqrt{1/3}
    \end{pmatrix}\\
    \psi_1 &= \begin{pmatrix}
        -\sqrt{1/3}\\
        \sqrt{1/3}
    \end{pmatrix}
    \end{cases}
\end{align*}
The \textit{Diffusion Map} is defined as:
\begin{equation}
    \Psi_t(x) = (\lambda_1^t\psi_1(x), \lambda_2^t\psi_2(x), ..., \lambda_k^t\psi_k(x))
\end{equation}
Here we set the number of time steps of the diffusion process to $t=1$.

The Euclidean distance in the diffusion space in our example is:
\begin{align*}
    D_t^2 &= ||\Psi_t(x_0) - \Psi_t(x_1)||^2\\
    &= ||-1/3\sqrt{1/3} - 1/3\sqrt{1/3}||^2 = \boxed{\frac{4}{27}}
\end{align*}

On the other hand, the diffusion distance in the original feature space is given by:
\begin{equation}
\label{diff_dist2}
D_t^2(x_0,x_1) = \sum_y (p(t,y|x_0)-p(t,y|x_1))^2 \frac{1}{\phi_0(y)}    
\end{equation}
$p(t,y|x)$ is the probability of being at $y$ after $t$ steps after starting at $x$. 

For this case the required probabilites are given by:
\begin{align*}
    p(t=1,x_0|x_0) &= \phi_0(x_0) + \psi_1(x_0)\lambda_1\phi_1(x_0) \\
    &= 1/6+\sqrt{3}/2\\
    p(t=1,x_0|x_1) &= \sqrt{3}/2 - 1/6\\
    p(t=1,x_1|x_0) &= \sqrt{3}/2 - 1/6\\
    p(t=1,x_1|x_1) &= 1/6+\sqrt{3}/2\\
\end{align*}
Therefore the diffusion distance is given by:
\begin{align*}
    D_{t=1}^2 &= (p(t,x_0|x_0)-p(t,x_0|x_1))^2 \frac{1}{\phi_0(x_0)}\\ 
    &+ (p(t,x_1|x_0)-p(t,x_1|x_1))^2 \frac{1}{\phi_0(x_1)} \\
    &= 1/9*\sqrt{4/3}+1/9*\sqrt{4/3} = \boxed{\frac{4\sqrt{3}}{27}}
\end{align*}
Consequently, following the described steps we obtain that the Euclidean distance in the diffusion space is not equal to the diffusion distance in the feature space (theorem not fulfilled).
\end{minipage}
\hfill\vline\hfill
\begin{minipage}[t]{0.47\textwidth}
\scriptsize
\underline{\textbf{Adjusted calculations}}\\
To ensure equality of both distances we change the definition of the eigenvectors $\phi_j$ and $\psi_j$ to:
\begin{align}
    \label{norm_eigenvec}
    \phi_j &= \frac{1}{\sqrt{tr(D)}} v_j D^{1/2}\\
    \psi_j &= \sqrt{tr(D)} v_j D^{-1/2}  
\end{align}

We get the vectors:
\begin{align*}
    \phi_0 &= \begin{pmatrix}
        1/2\\
        1/2       
    \end{pmatrix}\\
    \phi_1 &= \begin{pmatrix}
        -1/2\\
        1/2        
    \end{pmatrix}\\
    ~\\
    \psi_0 &= \begin{pmatrix}
        1\\
        1
    \end{pmatrix}\\
    \psi_1 &= \begin{pmatrix}
        -1\\
         1
    \end{pmatrix}
\end{align*}

With the definition of the \textit{Diffusion Map} the Euclidean distance in diffusion space is given by:
\begin{align*}
    D_{t=1}^2 &= ||\Psi_t(x_0) - \Psi_t(x_1)||^2 \\
    &= ||1/3+1/3||^2 = \boxed{\frac{4}{9}}
\end{align*}

For our case the needet probabilites are given by:
\begin{align*}
    p(t=1,x_0|x_0) &= \phi_0(x_0) + \psi_1(x_0)\lambda_1\phi_1(x_0) \\
    &= 2/3\\
    p(t=1,x_0|x_1) &= 1/3\\
    p(t=1,x_1|x_0) &= 1/3\\
    p(t=1,x_1|x_1) &= 2/3\\
\end{align*}
Then, the diffusion distance is given by:
\begin{align*}
    D_{t=1}^2 = 1/9*2+1/9*2 = \boxed{\frac{4}{9}} 
\end{align*}

Therefore the theorem is fulfilled with the alternative definition of the $\psi_j$ and $\phi_j$.

\end{minipage}
\end{center}

\paragraph*{2. Non symmetric example}
\small
Following the same procedure, we have analyzed an additional case consisting of a 3 points dataset:

\[
\begin{cases} x_1=(1,2,3) \\ x_2=(1,1,0) \\ x_3=(5,0,4) \end{cases}
\]

The Gausian kernel matrix with width $\epsilon=8$ leads to the pairwise similarity matrix:

$$L = 
\begin{pmatrix}
1 & 0.5353 & 0.2692\\
0.5353 & 1 & 0.1271\\
0.2692 & 0.1271 & 1
\end{pmatrix}
$$

Being $D_{i,i} = \sum_j L_{i,j}$,

$$M=D^{-1}L = 
\begin{pmatrix}
0.5542 & 0.2966 & 0.1492\\
0.3219 & 0.6015 & 0.0765\\
0.1928 & 0.0911 & 0.7162
\end{pmatrix}
$$

Let's compute the symmetric matrix:

$$M_S=D^{-1/2}LD^{-1/2} = 
\begin{pmatrix}
0.5542 & 0.3091 & 0.1696\\
0.3091 & 0.6015 & 0.0835\\
0.1696 & 0.0835 & 0.7162
\end{pmatrix},
$$

whose eigenvalues and normalized eigenvectors are:
\vspace{.5cm}
\begin{equation*}
\begin{cases}
    \lambda_0 &= 1\\
    \lambda_1 &= 0.6151\\
    \lambda_2 &= 0.2567\\
\end{cases}
\end{equation*}

\vspace{1cm}

\begin{equation*}
\begin{cases}
    v_0 &= \begin{pmatrix}
        -0.6091\\
        -0.5847\\
        -0.5358
    \end{pmatrix}\\
    \\[7pt]
    v_1 &= \begin{pmatrix}
        -0.2459\\
        -0.5031\\
        0.8285
    \end{pmatrix}\\
    \\[7pt]
    v_2 &= \begin{pmatrix}
        -0.7539\\
        0.6364\\
        0.1627
    \end{pmatrix}\\
\end{cases}
\end{equation*}

\begin{center}
\begin{minipage}[t]{0.47\textwidth}
\scriptsize
\underline{\textbf{Calculation following \cite{Nadler2005}}}

\begin{align*}
     \phi_j &= v_jD^{1/2}\\ \psi_j &= v_jD^{-1/2}    
\end{align*}

We get the vectors:
\begin{tiny}
\begin{equation*}
    \phi_0 = \begin{pmatrix}
        0.8182\\
        0.7538\\
        0.6332 
    \end{pmatrix},\hspace{0.2cm}
    \phi_1 = \begin{pmatrix}
        -0.3303\\
        -0.6487\\
        0.9790   
    \end{pmatrix},\hspace{0.2cm}
    \phi_2 = \begin{pmatrix}
        -1.0128\\
        0.8206\\
        0.1922  
    \end{pmatrix}\\
\end{equation*}
\begin{equation*}
    \psi_0 = \begin{pmatrix}
        -0.4535\\
        -0.4535\\
        -0.4535
    \end{pmatrix},\hspace{0.2cm}
    \psi_1 = \begin{pmatrix}
        -0.1831\\
        -0.3902\\
        0.7011
    \end{pmatrix},\hspace{0.2cm}
    \psi_2 = \begin{pmatrix}
        -0.5613\\
        0.4936\\
        0.1376
    \end{pmatrix}
\end{equation*}
\end{tiny}
\vspace{.35cm}

Euclidean distance in the diffusion space:
\begin{align*}
    D_{t=1_{i,j}}^2 &= ||\Psi_t(x_i) - \Psi_t(x_j)||^2 \\
    &= \boxed{\begin{pmatrix}
   0 & 0.0896 & 0.3281\\
   0.0896 & 0 & 0.4591\\
   0.3281 & 0.4591 & 0
   \end{pmatrix}}
\end{align*}

Diffusion distance defined as 
$$D_t^2 = \sum_k (p(t, x_k | x_i)-p(t, x_k | x_j))^2/\Phi_0^k,$$ 
where 
$$p(t, x_k | x_i) = \phi_0(x_k) + \sum_{j \geq 1} \psi_j(x_i)\lambda_j^t\phi_j(x_k),$$

\begin{equation*}
D^2_{t=1, i,j} = 
   \boxed{\begin{pmatrix}
   0 & 0.1976 & 0.7235\\
   0.1976 & 0 & 1.0124\\
   0.7235 & 1.0124 & 0
   \end{pmatrix}}
\end{equation*}

Again, the Euclidean distance in diffusion space is not equal to the diffusion distance in feature space (theorem not fulfilled).

\end{minipage}
\hfill\vline\hfill
\begin{minipage}[t]{0.47\textwidth}
\scriptsize
\underline{\textbf{Adjusted calculations}}

\begin{align*}
    \phi_j &= \frac{1}{\sqrt{tr(D)}} v_j D^{1/2}\\
    \psi_j &= \sqrt{tr(D)} v_j D^{-1/2}  
\end{align*}

\begin{tiny}
We get the vectors:
\begin{equation*}
    \phi_0 = \begin{pmatrix}
        0.371\\
        0.3418\\
        0.2871 
    \end{pmatrix},\hspace{0.2cm}
    \phi_1 = \begin{pmatrix}
        -0.1498\\
        -0.2941\\
        0.4439   
    \end{pmatrix},\hspace{0.2cm}
    \phi_2 = \begin{pmatrix}
        -0.4592\\
        0.3721\\
        0.0872  
    \end{pmatrix}\\
\end{equation*}
\begin{equation*}
    \psi_0 = \begin{pmatrix}
        -1\\
        -1\\
        -1 
    \end{pmatrix},\hspace{0.2cm}
    \psi_1 = \begin{pmatrix}
        -0.4037\\
        -0.8605\\
        1.5462
    \end{pmatrix},\hspace{0.2cm}
    \psi_2 = \begin{pmatrix}
        -1.2378\\
        1.0885\\
        0.3036 
    \end{pmatrix}
\end{equation*}
\end{tiny}
\vspace{.35cm}

Euclidean distance in the diffusion space:
\begin{align*}
    D_{t=1_{i,j}}^2 &= ||\Psi_t(x_i) - \Psi_t(x_j)||^2 \\
    &= \boxed{\begin{pmatrix}
   0 & 0.4357 & 1.5955\\
   0.4357 & 0 & 2.2326\\
   1.5955 & 2.2326 & 0
   \end{pmatrix}}
\end{align*}

Diffusion distance defined as 
$$D_t^2 = \sum_k (p(t, x_k | x_i)-p(t, x_k | x_j))^2/\Phi_0^k,$$ 
where 
$$p(t, x_k | x_i) = \phi_0(x_k) + \sum_{j \geq 1} \psi_j(x_i)\lambda_j^t\phi_j(x_k),$$

\begin{equation*}
D^2_{t=1, i,j} = 
   \boxed{\begin{pmatrix}
   0 & 0.4357 & 1.5955\\
   0.4357 & 0 & 2.2326\\
   1.5955 & 2.2326 & 0
   \end{pmatrix}}
\end{equation*}

As seen in the previous example, the theorem is fulfilled with the alternative definition of $\psi_j$ and $\phi_j$.

\end{minipage}
\end{center}

\vspace{1cm}

\small
It is worth mentioning that cases with more than 1000 data points have been tested (including the V-Dem dataset with 12296 data points), and the analogous result has been obtained for all cases. This discrepancy may also be due to different length definitions of the eigenvectors, although length $1$ seems to make the most sense.